\documentclass[aps,amsmath,amssymb, notitlepage,nofootinbib, twocolumn,longbibliography
    ]{revtex4-1}

\usepackage[normalem]{ulem}

\usepackage[dvipsnames]{xcolor}
\definecolor{colorforS}{rgb}{0.858, 0.188, 0.8}
\definecolor{colorforS}{rgb}{0.1, 0.5, 0.3.}
\definecolor{colorforU}{rgb}{0.05, 0, 1}
\definecolor{colorforW}{rgb}{0.9, 0, 0}
\def\colZ{white}
\def\colW{colorforW!70}
\def\colU{colorforU!70}
\def\colI{gray}
\def\colS{colorforS!70}
\def\minsize{18}
\def\innersep{2}

\usepackage{hyperref}
\hypersetup{
    colorlinks,
    linkcolor={blue!70!black},
    citecolor={blue!75!},
    urlcolor={blue!80!black}
}

    \usepackage{tikz}

    \usetikzlibrary{decorations.markings}
    \usepackage{graphicx}
    \usepackage{xcolor}
    \usepackage{bbold}

    \newcommand{\bit}{\begin{itemize}}
    \newcommand{\eit}{\end{itemize}}

    \newcommand{\f}{\frac}
    \renewcommand{\>}{\right\rangle}
    \newcommand{\<}{\left\langle}
    \newcommand{\ba}{\begin{align}}
    \newcommand{\ea}{\end{align}}
    \newcommand{\be}{\begin{equation}}
    \newcommand{\ee}{\end{equation}}
    \newcommand{\bi}{\begin{itemize}}
    \newcommand{\ei}{\end{itemize}}
    \newcommand{\lf}{\left(}
    \newcommand{\ri}{\right)}
    \newcommand{\dd}{\mathrm{d}}

    \newcommand{\tr}{\operatorname{tr}}

    \newcommand{\measO}{\mathcal{O}}

    \def\tr{\text{tr}}

    \def\+{\dagger}

\begin{document}

    \newcommand{\bra}[1]{\< #1 \right|}
    \newcommand{\ket}[1]{\left| #1 \>}

        \newcommand{\bbra}[1]{\<\< #1 \right| \right|}
    \newcommand{\kket}[1]{\left|\left| #1 \>\>}
    
           \newcommand{\bbraket}[2]{\<\< #1 || #2 \>\>}

    \newcommand{\fpW}{\protect\tikz[baseline=-2.5pt]
  \protect\node[minimum width=0,inner sep=0.2] (u) at (0,0) [draw, circle, fill=\colW] {\footnotesize $\mathcal{W}$};}.

 \newcommand{\fpFK}{\protect\tikz[baseline=-2.5pt] \protect\node[minimum width=0,inner sep=0.5] (u) at (0,0) [draw, circle, fill=white] {\tiny$\mathcal{FK}$};}

 \newcommand{\fpS}{\protect\tikz[baseline=-2.5pt]
  \protect\node[minimum width=0,inner sep=0.5] (u) at (0,0) [draw, circle, fill=\colS] { $\mathcal{S}$};}

\newcommand{\fpU}
{\protect\tikz[baseline=-2.5pt]
  \protect\node[minimum width=0,inner sep=0.5] (u) at (0,0) [draw, circle, fill=\colU] { $\mathcal{U}$};}

\newcommand{\fpL}
{\protect\tikz[baseline=-2.5pt]
  \protect\node[minimum width=0,inner sep=0.5] (i) at (0,0) [draw, circle, fill=\colI] { $\mathcal{L}$};}

\newcommand{\fpZ}
{\protect\tikz[baseline=-2.5pt]
  \protect\node[minimum width=0,inner sep=0.5] (i) at (0,0) [draw, circle,] {$0$};}

    \title{Bayesian critical points in classical lattice models}
\author{Adam Nahum}
\affiliation{Laboratoire de Physique de l’\'Ecole Normale Sup\'erieure, CNRS, ENS \& Universit\'e PSL, Sorbonne Universit\'e, Universit\'e Paris Cit\'e, 75005 Paris, France}
\author{Jesper Lykke Jacobsen}
\affiliation{Laboratoire de Physique de l’\'Ecole Normale Sup\'erieure, CNRS, ENS \& Universit\'e PSL, Sorbonne Universit\'e, Universit\'e Paris Cit\'e, 75005 Paris, France}
\affiliation{Sorbonne Universit\'e, \'Ecole Normale Sup\'erieure, CNRS, Laboratoire de Physique (LPENS), F-75005 Paris, France}
    \date{\today}

\begin{abstract} 
The Boltzmann distribution encodes our subjective knowledge of the configuration in a classical lattice model, given only its Hamiltonian. If we acquire further information about the configuration from measurement, our knowledge is updated according to Bayes' theorem. We examine the resulting ``conditioned ensembles'', finding that they show many new phase transitions and new renormalization-group fixed points. (Similar conditioned ensembles also describe ``partial quenches'' in which  some of the system's degrees of freedom are instantaneously frozen, while the others continue to evolve.) After describing general features of the replica field theories for these problems, we analyze the effect of measurement on illustrative critical systems, including: critical Ising and Potts models, which show surprisingly rich phase diagrams, with RG fixed points at weak, intermediate, and infinite measurement strength; 
various models involving free fields, XY spins, or flux lines in 2D or 3D; and 
geometrical models such as polymers or clusters. We make connections with quantum dynamics, in particular with ``charge sharpening'' in 1D, by giving a formalism for measurement of classical stochastic processes: e.g. we  give a purely hydrodynamic derivation of the known effective field theory for charge sharpening. We discuss qualitative differences between RG flows for the above measured systems, described by $N\to 1$ replica limits, and those  for disordered systems, described by $N\to 0$ limits. In addition to discussing measurement of critical states, we give a unifying treatment of a family of inference problems for non-critical states. These are  related to the Nishimori line in the phase diagram of the random-bond Ising model, and are relevant to various quantum error correction problems. We describe distinct physical interpretations of conditioned ensembles and note interesting open questions.
\end{abstract}

        \maketitle

 \noindent

 \section{Introduction}

Conventional expectation values and correlators express our knowledge of the state of a classical system --- 
for example, a lattice magnet at its critical point --- based only on its Hamiltonian. If we obtain further information about the state from measurements, 
our knowledge and our conditional estimates of correlators are updated.  
This response of nontrivial classical ensembles  to measurement contains universal physics that cannot be simply inferred from the correlators of the  initial (unconditioned) ensemble.

Let $S$ represent a configuration in a classical statistical ensemble: 
for example, an equilibrated  configuration of spins in the lattice magnet or of fields in a lattice field theory.  Imagine that we acquire {partial} information about $S$ by making a set of measurements of some local observable $\mathcal{O}$ (extensively throughout the spatial configuration), whose outcomes we denote by $M$. Since the measurements may be incomplete or imprecise, their outcomes do not in general fully reveal the configuration $S$. 
However, we may ask to what extent the measurements  reveal the ``long wavelength'' information in the configuration.
As a simple example, it is perhaps intuitive  that even very imprecise measurement of the microscopic spins in the critical Ising model is sufficient to completely fix the configuration of ``block spins'' on very large scales:
in this case, the measurement ``strength'' effectively renormalizes to infinity. In other models, or for measurement of other local observables,
this may not be the case. 
We may either effectively obtain no information about large scales, 
or we may  end up in an intermediate situation where measurement changes our estimates of long-distance correlations in a nontrivial way, without fully revealing the configuration. Formally, the question is about the nature of the distribution $P(S|M)$ of the degrees of freedom~$S$, conditioned on a typical set of measurement outcomes~$M$. For example, are the fluctuations of $S$ in the conditioned ensemble short-range correlated, or critical, or long-range ordered, or something else?

A closely related problem arises if we imagine that configurations $S$ are being sampled using a Monte Carlo Markov chain, and at some time we decide to freeze (``quench'') the values of a subset of the local degrees of freedom.
As an example, we may imagine that our magnet is made up of two species of spins, and the second species is quenched. Let the quenched degrees of freedom be denoted $M$.
Now, as the unquenched degrees of freedom continue to evolve, they explore the distribution $P(S|M)$. 
In this context, we ask whether  long-wavelength fluctuations are frozen out by the ``partial quenching'', or whether they remain nontrivial. 

These two kinds of problem are closely related --- indeed, viewing  
the measurement outcomes as additional degrees of freedom that are coupled to the initial physical ones shows that formally the first type of problem can be mapped to the second type.

Finally,  conditioned ensembles are relevant to the real-space renormalization group (RG) in, say, a lattice spin model.
In a single step of the real-space RG transformation \cite{kadanoff2000statistical}, ``block spins'' $M$ are defined in terms of the microscopic spins $S$. 
For the renormalized couplings to remain short-range, 
what we require, loosely speaking, is that conditioning on the block spins renders the microscopic spins short-range correlated \cite{kennedy1993some,ould1997effect}.
Formally this is another problem of the above type, with the block spin values acting as the measurements. 
Therefore the properties of the  conditioned ensemble in this case are relevant to understanding when a  real-space RG procedure works, and when it might break down.

We will begin to explore measured critical ensembles,  revealing several new kinds of phase transitions and fixed points, and analyzing the general features of the RG flows.
(We will also revisit some older problems that can be mapped to Bayesian inference problems for uncorrelated degrees of freedom, and which are relevant to error correction.)

Before describing our aims in this paper, let us broaden our focus away from critical lattice models to review some other kinds of  Bayesian inference problems that have arisen in the literature. First, there is a significant body of work on connections between 
Bayesian inference  or error correction and the  statistical physics of spin glasses  \cite{zdeborova2016statistical,nishimori2001statistical,ricci2019typology,gamarnik2022disordered,dennis2002topological}.   
In particular, the random-bond Ising model on the Nishimori line \cite{nishimori1980exact} is closely related to a Bayesian inference (reconstruction) problem for noisy uncorrelated images, or equivalently for configurations in the infinite-temperature Ising model \cite{sourlas1994spin,
iba1999nishimori}.
The formal treatment of such problems using the replica trick also has analogies with methods for exploring free-energy landscapes in structural glasses, by coupling together several replicas of the system \cite{monasson1995structural,franz1995recipes,franz1997phase}.
Finally, the vulcanization (cross-linking)  process for rubber
becomes, in an idealized limit, a partial quenching process of a certain type \cite{deam1976theory,goldbart2004sam,goldbart2000random}. 
At the formal level,  these problems, and  the problems we will study here, have in common the  applicability of an ``${N\to 1}$'' replica limit (as opposed to the ${N\to 0}$ limit used for averaging over quenched disorder). 
However, the physical questions and relevant effective theories differ widely between the above examples and again differ widely
from the theories we will consider here. (For example, a key focus of the Bayesian inference literature is on finding efficient algorithms.)

In a different domain, there are connections  with problems studied in the \textit{quantum} literature, and these provide a more direct inspiration for this work \cite{skinner2019measurement,
li2018quantum,agrawal2022entanglement,garratt2023measurements}.   
There are physical and formal analogies
(for example in terms of replica symmetry)
with phase transitions and critical states induced by repeated quantum measurement   \cite{skinner2019measurement,
li2018quantum,
chan2019unitary,
gullans2020dynamical,
nahum2020entanglement,
jian2020measurement,bao2020theory,nahum2021measurement,li2021conformal,sang2021measurement,lavasani2021measurement,
buchhold2021effective,
nahum2023renormalization,
jian2023measurement,fava2023nonlinear,poboiko2023theory,fava2024monitored,garratt2023measurements,agrawal2022entanglement}.
(Of course, while quantum measurements have an ``objective'' effect on the wavefunction, classical measurements only update our subjective knowledge of the system.)
Some recent strands of work in the quantum literature should be specifically highlighted in the present context.
Ref.~\cite{garratt2023measurements} began exploring the effect of measurements on quantum-critical ground states \cite{garratt2023measurements,parameswaranjournalclub2022,yang2023entanglement,
weinstein2023nonlocality,lee2023quantum,murciano2023measurement,ma2023exploring,patil2024highly,baweja2024post}. The replica field-theory treatment of these problems has  features in common with the discussion here. 
(However,  for quantum ground states, the nontrivial effects of the measurements take the form of a boundary condition in a field theory, while in the problems we study here they are a bulk effect.)
Separately, Ref.~\cite{agrawal2022entanglement} 
introduced the notable idea of a ``charge sharpening'' transition for the dynamics of a quantum system in which an experimentalist makes ongoing measurements of the local charge density. 
This transition separates a phase where the experimentalist can rapidly infer the global charge from one where this information is ``hidden'' for a long time \cite{barratt2022field}. 
Despite being quantum and dynamical, this process is closely related to one of the classical problems we will discuss.
Analogies have also been made recently between quantum measurement criticality and monitored classical random walkers \cite{jin2022kardar} or random paths \cite{p2025planted}.

Finally, we  note 
rigorous results \cite{garban2022continuous,abbe2018group,garban2020statistical} on  Bayesian inference problems that generalize the Nishimori case \cite{iba1999nishimori} (see main text).

\smallskip

In this paper we study 
measured classical critical points, viewing them as new problems in \textit{bulk critical phenomena} that are to be understood using the renormalization group.
We set out the formal description using replicas and use this to discuss general features of RG flows (Secs.~\ref{sec:generalities}--\ref{sec:generalitiesweak}),
and give a unified discussion of the various physical interpretations of such ensembles (in terms of partial quenching and real-space RG as well as measurement, Sec.~\ref{sec:otherinterpretations}).
More than this, however, we aim to demonstrate through an extensive set of examples that measured classical systems (conditioned ensembles)  give rise to interesting phase transitions and new bulk renormalization group fixed points (Secs.~\ref{sec:potts2d}--\ref{sec:imagingpolymers}). 
These examples 
 illustrate more general RG mechanisms in measured ensembles.
They suggest (as does  an analogy with disordered systems, where a remarkable variety of phase transitions have been found) that there is a  wide range of  ``Bayesian critical points'' to be investigated.

While our main focus is on  \textit{measured critical equilibrium ensembles}, we also use the results to 
shed light on two related topics.  
We describe a formalism for  \textit{monitored classical stochastic dynamics}, based on the Martin-Siggia-Rose formalism, and apply this to classical particle systems. 
This shows that charge sharpening is a universal classical phenomenon that relies only on properties of classical fluctuating hydrodynamics (and not on any quantum effects).
We also give RG results for charge sharpening.
Second, we  revisit Bayesian inference/error correction problems \cite{dennis2002topological,wang2003confinement,weinstein2024computational}  involving \textit{non-critical} ensembles
(either paramagnetic or classically topologically ordered) in order to classify some closely related universality classes connected with ``Nishimori physics''.
We distinguish ``Nishimori inference'' from ``gauged Nishimori inference'' problems which are physically distinct but share exponents. This classification of error correction problems in terms of replica gauge theory may be more intuitive than the standard approach using  explicit mappings to disordered systems (which conceals the full symmetry).

Our basic framework will be replica field theory (the replica trick allows averaging over measurement randomness).
In the conditioned ensemble problem, the measurement outcomes $M$ play a role that is formally similar to that played by quenched disorder in a disordered system.
However the key difference with quenched disorder is that, instead of being drawn from, say, a trivial uncorrelated distribution, the probability distribution for $M$ is itself generated using the original statistical ensemble. 
At the formal level, taking into account the nontrivial distribution of $M$ leads to an enhancement of replica--permutation symmetry 
\cite{gruzberg2001random,jian2020measurement,nahum2023renormalization,garratt2023measurements,nahum2021measurement,barratt2022field}, from the ``$S_N$ with $N\to 0$'' of standard disordered systems to ``$S_N$ with $N\to 1$''.
Therefore the study of partially quenched classical systems gives an application to replica field theories in a different limit to the one usually studied. 
(Analogously, monitored quantum systems provided new applications of various nonlinear sigma-model theories in an ${N\to 1}$ limit \cite{jian2023measurement,fava2023nonlinear,poboiko2023theory,fava2024monitored,poboiko2024measurement}.)

\smallskip

To conclude this introduction, we now summarize the main classes of problems studied.

We start by discussing {\bf measurement of nontrivial critical states}, including cases where   explicit results can be obtained using perturbative RG.
After a description of replica and RG formalism in Secs.~\ref{sec:generalities}--\ref{sec:generalitiesweak},
Sec.~\ref{sec:potts2d} describes measurement of critical Ising and Potts models in arbitrary dimensionality $d$.
This yields examples of several different kinds of critical points in the  post-measurement ensemble. 
In addition to  RG fixed points at   ``weak'' measurement strength $\lambda$, which   can be accessed with perturbative RG  in $\lambda$, 
we find fixed points at intermediate $\lambda$, 
and  fixed points at infinite $\lambda$ (i.e.\ for perfectly accurate measurement), 
where  measurement  imposes hard constraints on the conditioned ensemble. 
We find that the phase diagram of measured Ising and Potts models is surprisingly rich (open questions remain for the future).

2D ensembles that are described by free fields form a special class of critical states, with some special features (e.g.\ exact marginality).
In Secs.~\ref{sec:freefieldandchargesharpening} and \ref{eq:interactingfixedlinefreefield}
we  discuss measurement of 2D models   involving  flux lines, or dimers, or height fields, or XY spins. 
These various models all share the feature that the pre-measurement ensemble  can be described in the IR by  free-field theory.
(Note also the rigorous results in Ref.~\cite{garban2020statistical}.)
But while the pre-measurement ensemble is free, the post-measurement ensemble is described by an interacting replica theory and can show nontrivial phase transitions and fixed points.

One of the 2D problems we discuss is closely related to  1+1D charge sharpening  \cite{agrawal2022entanglement,barratt2022field}. 
We fit that problem into a broader class of monitored classical stochastic processes,
 giving a simple derivation of the   effective field theory for charge sharpening, 
for a 1D classical particle system,
that relies only on standard fluctuating hydrodynamics.
Many of the techniques discussed in this paper generalize directly to monitoring of classical stochastic processes, as discussed in Secs.~\ref{sec:chargesharpening},~\ref{sec:othermarkov}.
We also give a more detailed discussion of the RG for charge sharpening than previously available,  emphasizing the  differences between the RG flows for charge sharpening 
and for a closely related disordered system \cite{cardy1982random}.

In general, the replica field theory for a given measured system often has a formal similarity to a ``related'' disordered system (see below). One lesson from the examples we study is that, despite this formal similarity, the two kinds of problems often have very different phase diagrams and RG flows. The RG flows for the measured system obey constraints that are special to that context (the strong-measurement regime is also often qualitatively different).

In Sec.~\ref{sec:3dfluxes} we give effective replica field theories for  3D  problems involving measurement of flux lines  or equivalently of free gauge fields. 
We again emphasize the contrast between the measurement problem and the analogous disordered system, which is the  ``vortex glass''   \cite{fisher1989vortex}.  We show that the topology of the RG flows for the measurement problem differs from that of the disordered system in $2+\epsilon$ dimensions. (These results are also  relevant to 2+1D charge sharpening.)

In Sec.~\ref{sec:imagingpolymers},
the imaging of polymers and of percolation configurations
provides examples of measured critical systems in which the interesting observables involve nonlocal connectivity information.

We discuss 
{\bf measurement of non-critical lattice models}, either ``trivial'' paramagnetic states 
or deconfined states of gauge theories, in  Sec.~\ref{sec:paramagneticstates}.
We start this Section with some review:
measurement of the infinite-temperature Ising paramagnet corresponds to a well-studied Bayesian inference problem \cite{sourlas1994spin,iba1999nishimori,nishimori2001statistical,zdeborova2016statistical},
 which maps to the Nishimori line in the random-bond Ising model. Several reconstruction, error-correction or quantum measurement problems have been shown to map to the Nishimori line \cite{dennis2002topological, wang2003confinement, weinstein2024computational,zhu2023nishimori,lee2022decoding,
fan2024diagnostics, hauser2024information}.
We use effective field theory to give a unified discussion of these problems,
making a distinction between some closely related universality classes and, we hope, making the ubiquitous appearance of ``Nishimori'' exponents in various measurement and error correction problems more intuitive.

(We also comment on the differences in the effective field theories for measurement of non-critical states and those for measurement of nontrivial critical states.
One basic difference is that  problems involving measurement of non-critical states, such as the examples which relate to the Nishimori line,
can be mapped to problems in ``conventional'' disordered systems, which is not the case for measurement of generic critical states.)

Moving  away from the discussion of conditioned ensembles in terms of measurement, Sec.~\ref{sec:otherinterpretations} 
briefly discusses the relation to partial quenches and to real space RG (RSRG). We suggest that ``RSRG--breaking'' transitions can occur where a given coarse-graining rule ceases to give sensible results. 

\smallskip

Further discussion and technical details can be found in several appendices.

\tableofcontents

\section{Imperfect measurements: generalities}
\label{sec:generalities}

This section develops some basic formalism
for the measurement problem which is independent of the specific model being studied
(some of the basic structure is a standard consequence of Bayes' theorem \cite{zdeborova2016statistical}), and then
 makes some general points about the symmetries and classification of conditioned ensembles in lattice models.

For a given model and a given measured observable, 
there is  considerable freedom in how the measurement process is defined, but  many of the basic points are independent of these choices. 
For concreteness we start with the example of Gaussian measurements.
That is, for each measured local observable $\mathcal{O}$, the corresponding measurement outcome is a Gaussian random number, centered on the value $\mathcal{O}$, but with a variance $\Delta^2$ which quantifies the precision of the measurement process. Loosely speaking, ${\lambda\equiv1/2\Delta^2}$
quantifies the ``measurement strength''.
(An alternative way to vary the measurement strength would be to make the measurements dilute --- we will also comment on this case.)
Using Gaussian measurements as a representative example, we describe the basic features of the replica formalism
and its symmetries. Then, in the next Section, we describe some general features of RG flows at small measurement strength $\lambda$, before moving on to specific models. 

A description of other physical interpretations of conditioned ensembles, in terms of partial quenches or real-space RG, is deferred to  Sec.~\ref{sec:partialquench}.

\subsection{Gaussian measurements}
\label{sec:gaussianmeasurements}

We start with a lattice model
for degrees of freedom collectively denoted by $S$. We will refer to these as spins,  but the formalism is more general:  $S$ could represent either discrete spins or continuous fields.
We will absorb the factor of $1/k_BT$ into the couplings, so that the partition function is
\be\label{eq:initialBW}
Z = \int_S \exp \lf - \mathcal{H}[S] \ri,
\ee
where $\int_S$ is the sum or the integral over the spins $S$.

We imagine that imperfect measurement is made of a set of real-valued observables 
$\{\measO_x\}_x$, where  $x$ labels spatial positions (e.g.\  sites or bonds of the lattice).
In the most trivial case $\measO_x$ could be the local spin value $S_x$ itself, but more generally it could be an arbitrary local or quasi-local function of the configuration.
Let us assume that the corresponding measurement outcomes $M=\{M_x\}_x$ are  centered around the true values, with a Gaussian error of variance $\Delta^2$. In other words, their conditional probability density is 
\be
P(M| S) = \exp \lf -  \frac{1}{2\Delta^2} \sum_x (\measO_x - M_x)^2  + \ln d \ri,
\ee
where $d=(2\pi\Delta^2)^{-(\text{no.\ measurements})/2}$ is a normalization constant.

The \textit{joint} distribution of $S$ and $M$ may be written
\be\label{eq:jointprobMandS}
P(S,M) = {Z^{-1}} \exp \lf  - \mathcal{H}_{\rm meas}[S,M]
 \ri,
 \ee
where we introduce an effective ``Hamiltonian'' for both spins and measurement outcomes:
\ba\label{eq:Hmeasdefn}
\mathcal{H}_{\rm meas}[S, M] 
& = 
\mathcal{H}[S] + \f{1}{2\Delta^2}\sum_x (\measO_x - M_x)^2 - \ln d.
\end{align}

Starting from the joint distribution (\ref{eq:jointprobMandS}) and integrating out/summing over the spins gives
the marginal probability distribution for the measurement outcomes:
\ba\label{eq:Mmarginal}
P(M) & =  \f{Z(M)}{Z}, 
& Z(M) & = \int_S \exp \lf  - \mathcal{H}_{\rm meas}[S,M]\ri.
\end{align}

Our interest is in what a set of measurement outcomes tells us, typically, about the spins.
Therefore  imagine that we know the measurement outcomes $\{M_x\}_x$, but we are not given further information about the spin configuration $S^\text{ref}$ that they were obtained from.\footnote{We now denote this measured configuration by $S^\text{ref}$, to distinguish it from a sample drawn from the a posteriori measure $P(S|M)$ discussed below. In one terminology, $S^\text{ref}$ is the ``ground truth'' \cite{zdeborova2016statistical}.}
Assuming that we also know the physical Hamiltonian $\mathcal{H}[S]$, 
Bayes' theorem gives the conditional probability distribution of the spins
 as ${P(S|M) = P(S,M)/P(M)}$:
\be\label{eq:PSgivenM}
P(S|M )
=
\f{1}{Z(M)}
 \exp \lf - \mathcal{H}_{\rm meas}[S,M]  \ri.
\ee
As expected, the measurement information deforms the distribution in the direction of the outcomes.
We emphasize that the basic structure of these formulas  is a simple consequence of Bayes' theorem and, modulo the notation and the choice of $P(S|M)$, is common to many inference problems  in statistical physics: see for example the review \cite{zdeborova2016statistical}.

Eq.~\ref{eq:PSgivenM} resembles a Hamiltonian for spins $S$ in a disordered system, with the (spatially varying) values $\{M_x\}_x$ playing the role of disordered couplings. 
One of the distinctions from a standard disordered system is that the outcomes $\{M_x\}_x$ should be drawn from the  distribution $P(M)$, which may encode highly nontrivial correlations arising from the initial Hamiltonian. (But in the special case where  $P(M)$ becomes short-range-correlated, we can make a correspondence with a conventional disordered system with short-range-correlated disorder. We discuss this in Sec.~\ref{sec:paramagneticstates}.)

One kind of question about distributions such as (\ref{eq:PSgivenM}) is  algorithmic:  given a \textit{specific} set of measurement outcomes $M$ for some model, how do we efficiently estimate properties of the spin configuration \cite{zdeborova2016statistical,nishimori2001statistical,gamarnik2022disordered}?
Here, we instead ask about properties of the distribution $P(S|M)$, for \textit{typical} measurement outcomes.
As a result we are ultimately interested in quantities that are averaged over $M$.

As a very simple example, 
we may ask how much the knowledge of the measurements reduces the uncertainty in some observable such as the local spin $S_x$ itself. 
We denote the a posteriori average of the spin
(computed using Eq.~\ref{eq:PSgivenM}, given the measurement outcomes $M$) by   $\<S_x\>_M$.\,\footnote{This is also called the ``minimal mean-squared estimator''  \cite{zdeborova2016statistical}.}
After conditioning on the outcomes $M$, 
the spin has variance
${\< (S_x)^2 \>_M  - \< S_x\>_M^{2}}$,
where the subscript indicates that the averages are  \textit{conditioned} on a particular realization $M$ of measurement outcomes. 
On average,
 the post-measurement uncertainty is 
\be\label{eq:variance}
\mathbb{E}_M \left[ \< (S_x)^2 \>_M  - \< S_x\>_M^{2} \right],
\ee
where $\mathbb{E}_M$ is the average over measurements, with distribution (\ref{eq:Mmarginal}).
Note that the second term in Eq.~\ref{eq:variance} may be interpreted in the spirit of the replica trick: we can think of 
$\mathbb{E}_M \left[  \< S_x\>_M^{2}\right]
= 
\mathbb{E}_M \left[  \< S_x^1 \>_M  
\< S_x^2 \>_M  
\right]$
as a correlation between two distinct samples, $S^1$ and $S^2$, which share the same measurement outcomes.
 Note also a trivial identity: if $\bullet$ represents an observable for the spins (in a single sample),
\be\label{eq:trivialidentity}
\< \bullet\> = \mathbb{E}_M \< \bullet \>_M.
\ee
The left-hand side is the conventional average taken with the Boltzmann distribution $P(S)$.
The equality just says that we are free to average over $S$ in two steps, first averaging over $S$ at fixed $M$, and then averaging over $M$. By contrast, objects such as $\mathbb{E}_M \< \bullet\>_M^{\,\, 2}$ do not reduce to conventional averages.

Eq.~\ref{eq:variance} involves a single site. Long-distance correlation functions are  more interesting, as they can reveal universal behavior about the post-measurement ensemble.
For example, we may ask how much the measurement information reveals about the relative orientation of two distant spins.
These examples again involve  correlations between distinct configurations (``replicas'') that share the same measurement outcomes. 
As we discuss next, these averaged quantities may be addressed using the replica limit applied to a Hamiltonian for $N$ coupled spin configurations:
\be\label{eq:Hreplica}
\mathcal{H}_N [S^1,\ldots, S^N]
= 
\sum_{a=1}^N
\mathcal{H}[S^a] 
+
\f{1}{2 N \Delta^2} \sum_{a<b} \sum_x \lf \mathcal{O}_x^a - \mathcal{O}_x^b  \ri^2,
\ee
up to an additive constant that vanishes in the replica limit.

A formally similar coupling between replicas also arises in structural glasses, where, for example, an infinitesimal coupling can be used to diagnose replica symmetry breaking in mean-field models \cite{monasson1995structural,franz1995recipes,franz1997phase}.

\subsection{Replica formalism}
\label{sec:replicaformalism}

We wish to express correlations between distinct configurations (``replicas'') that share the same measurement outcomes. 
Let ${\bullet\hspace{-1mm}\bullet}$ be a quantity that depends on several replicas, $S^1, \ldots, S^k$ (with ${k\geq 1}$) and potentially also on  the measurement outcomes $M$ themselves.
Then it is straightforward to show (see App.~\ref{app:replicaidentity}) that its average may be written
\be\label{eq:firstreplicaidentity}
\mathbb{E} \left[ {\bullet\hspace{-1mm}\bullet} \right]
=
\lim_{N\rightarrow 1}
\< {\bullet\hspace{-1mm}\bullet} \>_{N},
\ee
where the average on the right-hand side is  computed using  an  effective replica 
partition function in which we integrate over both the measurements and over $N$ replicas, $S^1, \ldots, S^N$, with the effective Hamiltonian
($\mathcal{H}_{\rm meas}$ is defined in Eq.~\ref{eq:Hmeasdefn})
\be
\mathcal{H}_N[S^1,\ldots, S^N, M] = \sum_{a=1}^N \mathcal{H}_{\rm meas}[S^a, M].
\label{eq:HSandM}
\ee
As usual (\ref{eq:firstreplicaidentity}) is well-defined for $N\geq k$, but we hope to be able to analytically continue to $N=1$. 
As in standard applications of the replica trick to disordered systems, we expect this procedure to be well-controlled at least at the level of perturbative RG \cite{cardy1996scaling}.

Since $M$ appears quadratically, we can integrate it out (App.~\ref{app:replicaidentity}).
Abusing notation, we also denote the resulting effective Hamiltonian, given above in Eq.~\ref{eq:Hreplica}, by $\mathcal{H}_N$. 
For many purposes (e.g.\ for computing RG flows)  we may take the limit $N\to 1$ in the coefficients in the Hamiltonian, giving 
\be\label{eq:replicaHsimplified}
\mathcal{H}_N [S^a]
\approx
\sum_{a=1}^N
\mathcal{H}[S^a] 
-
\f{1}{ \Delta^2} \sum_{a<b} \sum_x \mathcal{O}_x^a \mathcal{O}_x^b. 
\ee

The nontrivial term in the replica Hamiltonian 
Eq.~\ref{eq:Hreplica} 
(with the full $N$ dependence) is similar to the one that 
appears when the replica trick is used to describe quenched disorder,\footnote{Using the replica trick to average over quenched disorder, coupling to $\mathcal{O}$, with variance $\Delta_\text{disorder}^2$
gives a replica Hamiltonian with the term 
$-\f{\Delta_\text{disorder}^2}{2}\sum_{a,b}\mathcal{O}^a \mathcal{O}^b$ \cite{cardy1996scaling}.} but with different $N$-dependence for the coefficients. 
For the measurement problem,  the coefficient of the diagonal terms ($\mathcal{O}^a \mathcal{O}^b$ with $a=b$) must vanish when $N\to 1$ to  ensure Eq.~\ref{eq:trivialidentity}.
Many  RG calculations can be done for the measurement problem and the disordered problem in parallel,  taking the replica limit $N\to 1$ for measurements and $N \to 0$ for  disorder, but there are important structural differences between the RG flows in the two cases (see Sec.~\ref{sec:nofeedback}, 
and the examples in Secs.~\ref{sec:isingddims},~\ref{eq:2Dheightlocking},~\ref{sec:3dfluxes}).

Returning to our simple example, the averaged post-measurement uncertainty in  the spin would be written
\ba\notag
\mathbb{E}_M \left[ \< (S_x)^2 \>_M  - \< S_x\>_M^{2} \right]
& = 
\lim_{N\to 1} \left[
\< (S_x^1)^2 \>- 
\< S_x^1 S_x^2\>\right]
\\
& = \lim_{N\to 1}  \f{1}{2} \< (S^1_x - S^2_x)^2 \>,
\label{eq:spinuncertaintyreplica}
\end{align}
where the expectation values on the right-hand sides are evaluated using $\mathcal{H}_N$ (Eq.~\ref{eq:Hreplica}),
and to avoid clutter we have suppressed the subscript ``$N$'' on the replica expectation values.
Note that the post-measurement uncertainty maps to the ``inter-replica'' fluctuations.

So far we have used the replicas to represent independent samples drawn from the a posteriori measure $P(S|M)$.  However it is easy to check that we can also take one of the replicas, say $S^N$, to represent the ``true'' configuration $S^\text{ref}$ on which the measurements were performed \cite{zdeborova2016statistical} (see App.~\ref{app:replicaidentity}).This allows, for example, the computation of the average overlap $S^\text{ref}_x S_x$ between the ``true''  configuration $S^\text{ref}$ and a sample $S$ from the a posteriori distribution $P(S|M)$ (see e.g. Sec.~\ref{sec:paramagneticstates}).

\subsection{More general measurement protocols}
\label{sec:binarymeasurements}

Consider a more general measurement process, 
where the measurement probabilities  $P(M|S)$  depend on $S$ in an arbitrary way.
The effective Hamiltonian for both measurements and spins is then
(it is convenient to use an informal notation\footnote{ 
Strictly speaking $\ln P(M|S)$ may not be well defined (e.g. $P(M|S)$
may vanish for some $M$) but this is not a problem, as the left and right hand sides of Eq.~\ref{eq:Hmeasgeneral} appear only in the exponent. We will use a loose notation that effaces the distinction between continuous and discrete $M$. It is straightforward to replace the formulations with more precise ones in terms of the appropriate probability measures.})
\be\label{eq:Hmeasgeneral}
\mathcal{H}_\text{meas}[S,M]=
\mathcal{H}[S] - \ln P(M|S)
\ee
and the replica Hamiltonian is obtained from (see Eqs.~\ref{eq:ZNintMS}--\ref{eq:ZNintS})
\be
e^{-\mathcal{H}_N[S^1,\ldots,S^N]}
=
\int_M e^{-\sum_{a=1}^N 
\mathcal{H}_\text{meas}[S^a,M]
}.
\ee
We will denote the associated partition function by $Z_N = \exp(-F(N))$, where $F(N)$ is the free energy for $N$ replicas.
Expectation values may again be computed as indicated in the previous section.

The example of binary-valued measurements 
of a binary-valued observable
is discussed in App.~\ref{app:binarymeasurements}.
If the binary  measurements are very imprecise, then the leading coupling in $\mathcal{H}_N$ is again of the form in Eq.~\ref{eq:replicaHsimplified}.
Perfectly precise, but \textit{dilute,} binary measurements give a different term which is nonzero only if $\mathcal{O}_x$  is identical in all replicas. (This case is analogous to the case of ``heralded'' errors in error correction, App.~\ref{app:binarymeasurements}.)

Whether these differences are important will depend on the setting. If our starting point is a critical state, and the measurements are ``weak'' (either in the sense of being imprecise or in the sense of being dilute) then we should decompose the coupling term in $\mathcal{H}_N$ in terms of scaling operators of the (replicated) critical state.
The leading perturbation induced by measurement will then generically be of the form in Eq.~\ref{eq:replicaHsimplified}.
For this reason, the  lessons of the  discussion in the previous section apply more widely than just for Gaussian measurements.
On the other hand if measurements are not weak, the precise measurement protocol may be important (see for example Sec.~\ref{sec:FKmeasurement}).

We may also consider simultaneous measurement of multiple types of observable: i.e.\ several kinds of local operator $\measO_{x,\mu}$, which we label here by an index $\mu$.
In general, a  Gaussian measurement (at position $x$) of the  operators $\mathcal{O}_{x,\mu}$  is characterized not only by the variance of the error for each outcome $M_{x,\mu}$, but also by the covariances of these errors. We will comment on this when we discuss RG flows in Sec.~\ref{eq:multiplemeasuredops}.

\subsection{Symmetry}
\label{sec:symmetry}

Finally, the symmetries of the measurement protocol determine the symmetries of the replicated theory, as discussed next. 
(The following discusion is not essential for understanding most of the concrete examples, but we will appeal to it in Sec.~\ref{sec:paramagneticstates}.)
Very recent work also discusses symmetry in Bayesian inference \cite{semerjian2025some} 
(with different motivations, but also distinguishing the situations we refer to as invariance and covariance).

Assume that there is a symmetry group $G$ that acts on the initial ``physical'' ensemble via  $S\to g S$ for ${g\in G}$, 
leaving the Boltzmann weight in Eq.~\ref{eq:initialBW} invariant.
This notation for the symmetry transformation is schematic: $G$ could be a global symmetry, or a local symmetry that acts independently at each site\footnote{In Sec.~\ref{sec:paramagneticstates} it will be useful to distinguish  ``local symmetries'', which relate distinct physical states, from  ``gauge symmetries'' which relate different redundant parameterizations of the same physical state. However, mathematically the action is similar.}  (or a higher-form symmetry \cite{gaiotto2015generalized});
the action on $S$ does not have to be multiplicative.

At first sight the question is how the measured operators transform under $G$. 
This is not quite the end of the story, because in a general protocol the  measurement ``strength'' could also depend on local observables that transform nontrivially under symmetry. App.~\ref{app:binarymeasurements} discusses the various possibilities for the specific example of  an Ising-like global $\mathbb{Z}_2$ global symmetry.

More formally, for a given symmetry group $G$, the measurement process might have any of three (successively weaker) symmetry properties:

$\bullet$ Invariance: ${P(M|gS)=P(M|S)}$ for all ${g\in G}$.
This implies that measurements can only reveal ``singlet'' information.
In the Ising case, this would correspond to measuring 
a $\mathbb{Z}_2$-even operator such as the local energy $E$.
The invariance implies that the replicated theory has a \textit{separate} $G$ symmetry for each replica, giving a $G^N$ symmetry.

$\bullet$ Covariance: ${P(gM|gS)=P(M|S)}$,
for some symmetry action $M\to gM$ on $M$.
This holds (for example) if we  measure the local order parameter $S$ in the Ising model using 
the  Gaussian protocol discussed in Sec.~\ref{sec:gaussianmeasurements}.
The covariance means that the replicated theory has at least a single $G$ symmetry that acts simultaneously on all replicas. 
Note that, in the Ising example, the term  ${\sum_{a\neq b}S^a S^b}$ which appears in the replica Hamiltonian is invariant under such a  $\mathbb{Z}_2$ symmetry.

$\bullet$ No condition imposed on $P(M|S)$.
An example is measurement of the Ising spin $S$, but with a more general ``detector'' which is more accurate in detecting $+$ spins than $-$ spins, or vice versa  (App.~\ref{app:binarymeasurements}).
In this case the replica Hamiltonian does not have the $G$ symmetry of the previous case for general $N$.
However, the fact that the original ensemble (prior to introducing replicas) possessed $G$ symmetry imposes an important constraint: any 
terms which break $G$ symmetry must involve at least two distinct replicas 
(or have a coefficient that vanishes as ${N\to 1}$; see App.~\ref{app:binarymeasurements}). 
In the Ising example one such term involves products $S^aS^bS^c$ for three distinct replicas.
For some critical points these $G$-breaking terms will be RG irrelevant, 
so that $G$ emerges in the IR, 
taking us back to the previous case.

Interestingly, symmetries can also be \textit{created} by the measurement process.
In the limit of perfect measurement accuracy  ($\lambda=\infty$), the measurements effectively impose a hard constraint on the conditioned ensemble.
Similarly, in the replica theory, all replicas are constrained to agree on the value of the measured operator (see Eq.~\ref{eq:Hreplica}). 
In some cases this constraint --- which might for example take the form of a conservation law for a lattice-flux field ---
can be interpreted as a symmetry or higher-form symmetry of the replica theory 
(examples in Sec.~\ref{sec:nishimoricloserelatives}). 

We note that there is a subtlety here about terminology.
In the language of quantum field theory or quantum statistical mechanics, 
it is natural to consider states that are related by a local change of basis in Hilbert space as equivalent.
This is not, however, natural in classical statistical mechanics, where there is a preferred basis for the transfer matrix, in which classical observables are diagonal.
Similarly, symmetries/conservation laws that are equivalent in the quantum field theory language may have to be distinguished in the classical context.  
We comment on this in App.~\ref{app:symmetrynote}.

\subsection{Replica free energy and physical entropies}
\label{sec:entropies}

The free energy of the replica theory may be related to physical entropies. In cases where the replicated theory is a 2D conformal field theory, this gives an interpretation for the replicated central charge $c(N)$ close to $N=1$. This is similar to the role played by $c(N)$ near $N=0$ for disordered systems \cite{LudwigCardy1987} {(cf.\ also the quantum measurement transition \cite{zabalo2022operator,kumar2024boundary}).

We can choose to look at various closely-related quantities. 
First, knowledge of the measurement outcomes will in general decrease our uncertainty about the spin configuration. This can be quantified by the entropy reduction
\be
\Delta \mathcal{S}_\text{spins} 
= \mathcal{S}_\text{thermo} - 
\mathbb{E}_M \mathcal{S}_\text{spins}^{(M)},
\ee
where $\mathcal{S}_\text{thermo}$ is the standard  thermodynamic entropy (the Shannon entropy of the  Boltzmann distribution for the spins) and $\mathcal{S}_\text{spins}^{(M)}$ is the entropy that remains after  conditioning on a realization $M$ of measurement outcomes.

Second, $\Delta \mathcal{S}_\text{spins}$ is also related to the entropy\footnote{For simplicity we continue to consider Gaussian measurement outcomes: since these are continuous variables, the above is strictly speaking speaking  referred to as the continuous entropy or the differential entropy, rather than the Shannon entropy. Discrete measurements may however be considered similarly and the universal features will be unchanged.} of the probability distribution of measurement outcomes,
\be\label{eq:Smeas}
\mathcal{S}_\text{meas} = - \int_M P(M) \ln P(M),
\ee
via\footnote{Given a joint distribution $P(X_1,X_2)$,
let  $S_1$ be  the entropy of the marginal distribution $P(X_1)$ of $X_1$ (and similarly for $S_2$). Define $S_1^\text{cond}$ as the average (over $X_2)$  of the entropy of the conditional distribution $P(X_1|X_2)$ (and similarly with the variables reversed).
Then 
$S_i -S^\text{cond}_i$ is independent of $i$. We can apply this to $P(S,M)$ to get the result in the text.}  
${\Delta \mathcal{S}_\text{spins}
  =
 \, \mathcal{S}_\text{meas} - \mathcal{S}_\text{triv}}$.
Here $\mathcal{S}_\text{triv}$ is the ``trivial'' contribution to the measurement entropy that would be obtained if the measured observables did not fluctuate: i.e.\ it is the entropy of $V$ Gaussian variables of variance $\Delta^2$, where $V$ is the number of measurements: ${\mathcal{S}_\text{triv}=\f{V}{2}\ln \lf 2\pi e \Delta^2  \ri}$.

Finally, we can define a free energy conditioned on measurements, $\mathcal{F}^{(M)}$,  using the effective partition function (cf.\ Eq.~\ref{eq:Hmeasgeneral})
\be
Z(M) = \sum_S e^{-\mathcal{H}[S]} P(M|S) \equiv e^{-\mathcal{F}^{(M)}}. 
\ee
Again we may relate this to $\mathcal{S}_\text{meas}$,
\be
\mathbb{E}_M \mathcal{F}^{(M)}
=  F + \mathcal{S}_\text{meas},
\ee
where $F$ is the physical free energy.

Letting the free energy for $N$ replicas be ${F(N) = - \ln Z_N}$, where $Z_N$ is the replica partition function defined in App.~\ref{app:replicaidentity}, a straightforward calculation (App.~\ref{app:replicaidentity}) shows that 
\ba
\label{eq:Fprimeidentities}
\mathbb{E}_M \mathcal{F}^{(M)} 
& = F'(1),
\\
\label{eq:Fprimeidentity}
\mathcal{S}_\text{meas} & = F'(1) - F(1),
\\
\Delta\mathcal{S}_\text{spins} & = 
F'(1) - F(1) - \mathcal{S}_\text{triv},
\end{align}
where the prime is the derivative with respect to the number of replicas (note that $\mathcal{S}_\text{triv}$ is a nonuniversal extensive term). 

The scaling behavior of any of these quantities contains information about the IR fate of the monitored system.
If the replicated theory flows to a 2D conformal fixed point that depends on $N$, the central charge $c(N)$ of the replicated theory may be obtained from universal subleading terms in the free energy $F(N)$ on appropriate manifolds.
Eq.~\ref{eq:Fprimeidentities}
shows that it is natural to define
\be\label{eq:ceffectivemeasurement}
c_\text{eff} = c'(1)
\ee
which (together with the central charge $c(1)$ of the unmeasured theory) will determine  universal finite-size corrections to the above entropies.
This effective central charge  can be compared with that   often discussed for disordered systems, defined via the $N\to 0$ replica limit as \cite{LudwigCardy1987}
\be
c_\text{eff} = c'(0).
\ee
The effective central charge has been discussed in the context of the  quantum measurement phase transition in Refs.~\cite{zabalo2022operator,kumar2024boundary}.

If the measurements reveal no information about the system (App.~\ref{app:replicaidentity}) 
then $c_\text{eff}$ is the same as the original central charge $c(1)$, while if measurements perfectly reveal the entire configuration, then the partition sum defining $\mathcal{F}^{(M)}$ becomes trivial and $c_\text{eff}=0$.
However, the examples later on will show that $c_\text{eff}$ 
can be either larger or smaller than the central charge of the unmeasured system.
In Sec~\ref{sec:meas_ceff} we give a protocol for measuring the effective central charge (\ref{eq:ceffectivemeasurement}) in a simulation.

\subsection{Some classes of conditioned ensembles}

The space of possible conditioned ensembles is very large, and we will not attempt an exhaustive classification here. 
However we discuss some types of behavior which we will encounter in this paper.

To begin with, we can classify the problem according to the nature of the state we are measuring. 
Our main focus on this paper will be on measurement of critical states.
We also discuss the measurement of  non-critical  states --- some of those  classical problems are related by various mappings to quantum error correction   \cite{dennis2002topological,wang2003confinement,weinstein2024computational}
and to quantum dynamics problems from the more recent literature  \cite{zhu2023nishimori,lee2022decoding,
fan2024diagnostics, hauser2024information}.

\subsubsection{Measuring critical states}\label{sec:meascritstates}

First consider  critical states, in which, for simplicity, we measure a single kind of operator (everywhere in space). 

One possibility is that these measurements are {\bf relevant} or marginally relevant, and also  completely reveal the large-scale structure of the configuration.

A simple example of this is measurement of the spin itself in the critical Ising model. 
The dimension of this operator is small, so that measurements are highly relevant (see Sec.~\ref{sec:rgsmalllambda}). 
We expect that the flow  of the measurement strength is ``to infinity'' in this example. 
This means that even very imprecise measurement of the microscopic spins gives us precise information about the block spins on large scales.

In the replica language, the flow of the spin measurement strength to infinity means that all the replicas are perfectly locked together in the infra-red, and the inter-replica fluctuations
(of the form ${S^a-S^b}$)
vanish.

Another possibility is that measurements of the critical state are relevant, but the flow leads to a nontrivial fixed point. 
In later sections we give some examples in which this nontrivial fixed point is at small $\lambda$, so that the RG can be treated in a controlled manner.
As examples we will discuss the   Potts model in 2D; Ising and Potts models above two dimensions;
examples involving measured free fields; and some problems involving polymers.

If the scaling dimension of the measured operator is sufficiently large, then measurements of the critical state may instead be {\bf irrelevant} or marginally irrelevant at small measurement strength. 
This means that while weak measurements give  us some information about microscopic spins,  they tell us essentially nothing about the configuration of block spins on large scales.

An example (discussed in Sec.~\ref{sec:potts2d}) is the critical 2D Ising model, 
where we measure the local energy density. The measurement strength is here marginally irrelevant.

However, even if measurements are irrelevant or marginally irrelevant at small strength, there may be a phase transition at a finite strength of measurement.\footnote{It is interesting to ask what general constraints on RG flows exist for conditioned ensemble problems. (The replica theory is not a conventional unitary theory, and so conventional monotonicity results such as Zamolodchikov's $c$-theorem \cite{zamolodchikov1986irreversibility} are not immediately applicable. In quenched random systems 
fixed points are stationary points of $c_\text{eff}$ \cite{JacobsenPicco2000}.)}
We will argue that this is the case for the critical 2D Ising model with bond measurements.
Related things happen for paramagnetic states (Sec.~\ref{sec:initialsummary:noncrit}).
For critical Potts and Ising models in ${d>2}$ we argue that there is a still more complex phase diagram, with multiple fixed points.

Some of these strong-coupling fixed points differ qualitatively from the weak-coupling fixed points mentioned above. For many of the former, the natural fields in a Landau-Ginzburg-like field theory are  the replicated order parameters $S^a$, 
carrying a single replica index. 
At the strong-coupling fixed points it may be natural also to introduce fields $X^{ab}$ carrying multiple replica indices into the effective field theory (Edwards-Anderson-like ``overlap'' order parameters).

It is natural to distinguish another special class of nontrivial fixed points, in which (heuristically) the fixed point is at infinite measurement strength but, because of the nature of the measured operator, measurement only freezes a subset of the inter-replica fluctuations. We will give some examples in Secs.~\ref{sec:FKmeasurement},~\ref{eq:interactingfixedlinefreefield},~\ref{sec:imagingpolymers}.

In special cases measurements can be {\bf exactly marginal}. 
A simple example is measurement of the height gradient $\nabla h$ for a free field $h$, which allows us to continuously tune the stiffness of the ``inter-replica'' modes. 
This example becomes less trivial 
when other operators are measured, as the replica theory becomes interacting.
If cosine interactions between replicas are induced by measurement,  a Kosterlitz-Thouless-like transition can occur. 
This is essentially the field-theory mechanism for the charge-sharpening transition introduced in \cite{agrawal2022entanglement,
barratt2022field}.
In other cases the replica theory can show a nontrival interacting RG fixed line, even though the initial ensemble is free.

We  explore a range of  critical equilibrium lattice models involving measured spins, dimers, flux lines or height fields, for which the \textit{pre}-measurement ensemble reduces  at large scales to free-field theory, in Secs.~\ref{sec:freefieldandchargesharpening},~\ref{eq:interactingfixedlinefreefield}. 
We find that nontrivial  RG fixed points and flows are possible in 2 and $2+\epsilon$ dimensions. Sec.~\ref{sec:3dfluxes} describes analogous  3D ensembles  involving flux lines or gauge fields in Sec.~\ref{sec:3dfluxes}. 
(We touch on other gauge-theory problems in Sec.~\ref{sec:nishimoricloserelatives}.)

In Secs.~\ref{sec:chargesharpening},~\ref{sec:othermarkov},~\ref{sec:chargesharpeninghigherd} we take a detour to give some new results for  {\bf monitored classical dynamical systems}, describing how charge-sharpening can be understood using  classical hydrodynamics via the Martin-Siggia-Rose formalism (Sec.~\ref{sec:othermarkov}).

\subsubsection{Measuring non-critical states}\label{sec:initialsummary:noncrit}

It is also possible to obtain nontrivial transitions even in the case where the measured state is short-range correlated, i.e.\ a paramagnet \cite{iba1999nishimori}.
In some cases these examples are  related, by   ``duality'', to measurement in gauge theories (which may be in a nontrivial deconfined state).
The essential point in the current discussion is that \textit{local} observables have exponentially-decaying correlations.
In these cases weak measurements will not give a nontrivial state, but there may be  transitions at finite measurement strength. 

Cases where the initial ensemble defines a paramagnet have a  special feature:
the measurement outcomes [drawn from $P(M)$] are essentially short-range correlated. 
This means that the effective Hamiltonian ${\mathcal{H}_\text{meas}[S,M]}$ in Eq.~\ref{eq:Hmeasdefn}
can be viewed as a Hamiltonian for spins $S$ in the background of short-range correlated quenched disorder~$M$.

Therefore the fixed points that appear in this context also have an interpretation in terms of systems with quenched disorder.
A simple example is the paramagnetic lattice Ising model with  measurements of $S_x S_y$ on bonds $\<xy\>$.
This is related to the ``Nishimori line'' 
\cite{nishimori1980exact,nishimori1981internal,ozeki1993phase,
le1988location,le1989varepsilon,
gruzberg2001random,
singh1996high,reis1999universality,
honecker2001universality,
hasenbusch2008multicritical,
sourlas1994spin,
nishimori1993optimum,
nishimori1994gauge,
iba1999nishimori,
nishimori2001statistical,
zdeborova2016statistical}
in the phase diagram of an Ising model with random bonds. This connection between the Nishimori line and Bayesian inference is discussed in \cite{iba1999nishimori, zdeborova2016statistical, nishimori2001statistical,tanaka2002statistical}.

Above one dimension, this example has a stable phase at sufficient  measurement strength, in which the replicas are locked. 
Physically, in this phase, measurements are able to reveal information about the relative orientation of arbitrarily distant spins.

Formally, the stability of this phase above 1D is due to the fact that the locking of replicas is the breaking of a discrete symmetry (see Sec.~\ref{sec:paramagnetsymmetry}). Similar examples can be constructed where the physical symmetry is continuous instead of discrete, and where the locked phase is stable only above two dimensions (Sec.~\ref{sec:ctsnishimori}).

There is also a wealth of problems involving measurement of {\bf discrete gauge theories}, some with relevance to error correction \cite{dennis2002topological, wang2003confinement, weinstein2024computational},
that are closely related to the above Nishimori inference problem.
We discuss the classification of these problems in 2D and 3D, and some generalizations where further investigation may be interesting  in Sec.~\ref{sec:nishimoricloserelatives}.
Some of these problems reveal an additional interesting phenomenon, which is the emergence in the 
conditioned ensemble
of symmetries (or higher-form symmetries) that have no precursor in the pre-measurement ensemble.

\section{Measurement of critical states: weak measurement regime} 
\label{sec:generalitiesweak}

In preparation for considering specific models in the following sections, it is useful to consider the general structure of the RG flows for \textit{weak} measurement of a critical state.  Let us look at the replica Hamiltonian~(\ref{eq:replicaHsimplified})
 \be\label{eq:repeatreplicaHforRG}
\mathcal{H}_N
=
\sum_{a=1}^N \mathcal{H}_\alpha
- 
\lambda \sum_{a\neq b} \sum_i \measO^a_i \measO^b_i
\ee
in the case of weak measurements (large $\Delta$). We have defined a ``measurement strength'' 
\be\label{eq:lambdadefn}
\lambda = \f{1}{2 \Delta^2}.
\ee
We assume for now that no other relevant couplings are allowed.
The effect of the measurement is to try to ``lock'' the values of $\measO^a$ in the different replicas $a=1,\ldots, N$ (this is more evident in the rewriting in Eq.~\ref{eq:Hreplica}). This locking may be either relevant or irrelevant.

The replica formalism follows that for  analogous disordered systems (the $N\to0$ limit). In particular, our discussion of the measured Potts model in Sec.~\ref{sec:potts2d} draws on the discussion of the disordered Potts model by Ludwig and Cardy and others in Refs.~\cite{LudwigCardy1987,Ludwig1990,DotsenkoPiccoPujol1995,DotsenkoJacobsenLewisPicco1999}. More generally, however, there are structural differences between disordered systems and monitored clean ones,
both in the weak measurement regime and in the strong measurement regime. 
We discuss some of these in the following two subsections (and in the context of specific models,  e.g.\ in Sec.~\ref{sec:freefieldandchargesharpening}).

\subsection{RG equations at small measurement strength}
\label{sec:rgsmalllambda}

A standard result gives the RG equation, to order $\lambda^2$, in terms of an  operator product expansion (OPE) coefficient  of the replicated theory \cite{LudwigCardy1987,cardy1982random}.\footnote{Defining the perturbing operator ${\mathcal{P} =  \sum_{a\neq b} \ \measO^a \measO^b}$, we write the OPE as
\be
\mathcal{P}\cdot \mathcal{P}
= 
a_N \mathbb{I} 
+ C_N  \mathcal{P} + \ldots,
\ee
where dependence on the spatial coordinate is omitted, and where other terms on the RHS are assumed to be irrelevant. (The constant $a_N$ is not necessarily $1$, i.e.\ we do not assume that $\mathcal{P}$ is canonically normalized.) Then the RG equation to order $\lambda^2$ is 
\be
\partial_\tau \lambda  
= 
(d - 2 x_\mathcal{O}) \lambda + 
\f{\pi^{d/2}}{\Gamma(d/2)}
C_N  \lambda^2 + O(\lambda^3).
\ee 
We will implicitly assume that $\lambda$ is rescaled to absorb the constant ${\pi^{d/2}}/{\Gamma(d/2)}$.}
In turn, this OPE coefficient can be written in terms of the OPE of the original, unreplicated theory. 
Normalizing the physical operator $\mathcal{O}$ so that this OPE takes the form
\be\label{eq:1replicaOPE}
\measO(0) \cdot \measO(r) = \f{1}{r^{2 x_\measO}}
+ \f{C}{r^{x_\measO}} \measO(r/2) + \ldots,
\ee
the RG equation in the limit $N\to 1$ may be written 
(similarly to the $N\to 0$ case, e.g. \cite{shimada2009disordered}):
\be\label{eq:rgeqnsimple}
\partial_\tau \lambda  
= 
(d - 2 x_\mathcal{O}) \lambda + 
(2C^2-4) \lambda^2.
\ee 
We  have rescaled $\lambda$ to eliminate a factor of ${\pi^{d/2}}/{\Gamma(d/2)}$ from the second term.
The RG time $\tau$ is the logarithm of the observation lengthscale.

The first term in Eq.~\ref{eq:rgeqnsimple} is the same as in the case $N=0$, which describes a system with quenched disorder coupling to $\measO$.
The criterion for the measurements to be relevant is therefore  the same as the Harris criterion \cite{Harris1974} for quenched disorder to be relevant, viz. 
\be\label{eq:Harris}
x_\measO < d/2
\qquad
\text{(criterion for relevance)}.
\ee
If measurements are weakly relevant, and \textit{if} the coefficient $2C^2-4$ of the second term is negative, then the RG equation will show a stable nontrivial fixed point at a small value of $\lambda$: we will discuss some examples below.

The standard form for the RG equations in terms of OPE coefficients \cite{cardy1996scaling,Ludwig1990} also gives  scaling dimensions at the new fixed point. 
In particular, consider a scaling operator $\psi$ (with dimension $x_\psi$) that is distinct from $\mathcal{O}$, 
and assume that symmetry enforces $\<\psi\>_M=0$ (this will be the case if there is some symmetry under which $\psi$, but not $\mathcal{O}$, is charged).
Let the OPE of  $\mathcal{O}$ with $\psi$ be $\mathcal{O} \cdot \psi = C_{\mathcal{O}, \psi}^\psi \, \psi + \ldots$. 
In order to determine the probability distribution of the correlator conditioned on measurements, $\<\psi \psi\>_M$, we
need the moments ${\mathbb{E}_M \<\psi \psi\>_M^k}$, which map to 
correlators of the multi-replica operator \cite{Ludwig1990} $\psi^1\cdots\psi^k$, for all $k$.
To lowest order in $\lambda_*\sim (d-2x_\measO)$, these operators have dimension 
\be
x_{\psi^1\cdots \psi^k} = k x_\psi
- \f{k(k-1) (C_{\mathcal{O}, \psi}^\psi)^2(d-2x_\measO)}{2-C^2}+\ldots
\ee
 in the ${N\to 1}$ limit.\footnote{At higher orders, or at the lowest order if $\psi=\mathcal{O}$, it may be necessary to consider operator mixing \cite{Ludwig1990}.} 
This is ``multifractal'' scaling \cite{Ludwig1990}.
In particular,
considering the $k\to 0$ limit for $\mathbb{E}_M\<\psi(0)\psi(r)\>^k$ shows that 
the power-law exponent governing the decay of correlations in a \textit{typical} realization of measurement outcomes $M$,
\ba
\mathbb{E}_M 
\ln \<\psi(0)\psi(r)\>_M 
& \simeq - 2 
x_\psi^\text{typ} \ln r,
\end{align}
with 
\be
x_\psi^\text{typ}
=x_\psi + 
\f{(C_{\measO\psi}^\psi)^2(d-2x_\measO)}{2-C^2},
\ee
is larger than the exponent $x_\psi$ governing the usual average $\<\psi(0)\psi(r)\>$.
We will discuss examples of models with this kind of scaling in Sec.~\ref{sec:pottsweak},~\ref{sec:fluxlinestwopluseps},~\ref{sec:isingddims},~\ref{eq:interactingfixedlinefreefield}.

At the level of the perturbative discussion above,
the replica approach goes through similarly for the case ${N=0}$ (describing quenched disorder) and for ${N=1}$  (describing measurements), 
though with different nontrivial coefficients in the RG equations
which may lead to different flow topologies.

However, this similarity may be misleading, as there are basic structural differences between the $N=1$ case and the $N=0$ case.
These differences are not  apparent in Eq.~\ref{eq:rgeqnsimple},
since there we (a) consider the flow only of the measurement strength $\lambda$, and (b) consider small $\lambda$. 

One key difference is that, in the limit $N\to 1$,  \textit{single-replica} quantities are independent of the value of $\lambda$, since they could equally well be formulated in the non-replicated  system, in which $\lambda$ does not appear (Eq.~\ref{eq:trivialidentity}).
This is not the case for $N=0$.
This independence simplifies the structure of the RG equations for $N=1$: the RG flow of 
``conventional'' couplings  
(i.e.\ those which are already present in the non-replicated theory)
is independent of $\lambda$  when $N=1$. 
This means that RG flows can be very different from those that arise in analogous problems with quenched disorder.
We will give a simple example in Sec.~\ref{sec:nofeedback}.

The regime of strong measurement is also typically very different. Heuristically,  the measured systems lack the strong ``frustration'' possible in generic disordered systems.
Later we will show examples where 
a replica-locked phase is possible for ${N=1}$ but not for ${N=0}$.
(Formally, this has to do with differences between the replica group theory in the ${N\to 1}$ and ${N\to 0}$ limits.) 
Systems of line defects in three dimensions give one example.
Line defects (e.g. superconducting vortices) that are subjected to pinning by disorder
\cite{fisher1989vortex}
show only a trivial weak-disorder phase (there is no ``vortex glass'' at nonzero temperature \cite{bokil1995absence,
kisker1998application,
pfeiffer1999numerical}, Sec.~\ref{sec:fluxlinestwopluseps}).
By contrast, in the analogous measurement problem
there is both a stable weak-measurement phase and a  stable strong-measurement ``replica-locked'' phase (Secs.~\ref{sec:3dfluxes},~\ref{sec:fluxlinestwopluseps}). ``Charge sharpening'' yields another example of  significant differences between the ${N\to 1}$ and ${N\to 0}$ universality classes, as discussed in Sec.~\ref{sec:freefieldandchargesharpening}.

\subsection{Multiple measured operators}
\label{eq:multiplemeasuredops}

For completeness, we state the generalization of the above RG equation to the case where we measure several different types of operators $\measO_\mu$, with ${\mu = 1, \ldots, \mathcal{N}_\text{meas}}$.
We have corresponding measurement outcomes $M_\mu$. 
For a general Gaussian measurement of $\{M_\mu\}_{\mu = 1, \ldots, \mathcal{N}_\text{meas}}$, the measurement outcomes at a single spatial location have a conditional probability distribution of the form
\be
P(M|S) \propto e^{-\f{1}{2} (M_\mu - \measO_\mu)K_{\mu\nu} (M_\mu - \measO_\mu)}
\ee
where we omit site labels.
Here the inverse of the matrix $K$ determines the covariance of the measurement errors. 
This leads to a replica Hamiltonian (cf. Sec.~\ref{sec:replicaformalism})
\be
\mathcal{H}_N = 
\sum_a H[S^a]
- \f{1}{2} \sum_{a\neq b} O^a_\mu K_{\mu\nu} O^b_\nu. 
\ee
Defining $\lambda_{\mu\nu} = \f{1}{2}K_{\mu\nu}$, and taking the single-replica OPE to be
\be
\measO_\mu \measO_\nu = \f{\eta_{\mu\nu}}{r^{x_\mu + x_\nu}} + \sum_\lambda \f{C_{\mu\nu}^\lambda}{r^{x_\mu + x_\nu- x_\lambda}} \measO_\lambda + \ldots,
\ee
 a straightforward calculation of the OPE in the replica theory gives
\ba\label{eq:quadraticRGmoregeneralmaintext}
\notag
\partial_\tau \lambda_{\mu\nu}
=& \, 
(d - x_\mu - x_\nu) \lambda_{\mu\nu}
 +
2 \sum_{\alpha\beta \gamma\delta} C^\mu_{\alpha\beta} C^\nu_{\gamma\delta} \lambda_{\alpha\beta}\lambda_{\gamma\delta}
\\
&
- 4 \sum_{\alpha\beta} \lambda_{\mu\alpha}
\eta_{\alpha\beta} \lambda_{\beta \nu}.
\end{align}
Note that, even if the physical measurements have a diagonal covariance matrix, 
off-diagonal elements of $\lambda$ may  appear under RG, depending on the structure of the OPE.
More importantly, operators that are not measured ``microscopically'' can effectively be measured at larger scales. For example, the RG equation above shows that measuring $\measO$ effectively generates measurements of all the operators that appear in the $\measO \cdot \measO$ OPE.

\subsection{No feedback on conventional couplings}
\label{sec:nofeedback}

Let us illustrate a simple but important feature of the RG flows for measured systems. 
The measurements do not affect single-replica quantities: i.e.\ conventional expectation values are unchanged by the decision to measure!
(See Eq.~\ref{eq:trivialidentity}.)
Therefore, any coupling constants that are already present in the single-replica theory are unaffected by the flow of the measurement rate.

This is shown most clearly in an example. Consider the critical Ising model in $4-\epsilon$ dimensions, in the Landau-Ginzburg formulation, with measurements of the energy~($\phi^2$):
\be
\mathcal{H}_N = \f{1}{2} \sum_a  (\nabla\phi^a)^2
+ 
\f{g}{4!} \sum_a (\phi^a)^4
-
\f{\lambda}{4!} \sum_{a\neq  b}
(\phi^a)^2 (\phi^b)^2.
\ee
Here $g$ is a ``conventional'' coupling that specifies the initial thermodynamic ensemble, whereas $\lambda$ characterizes the measurements.
These couplings' RG equations   are
\ba
\partial_\tau g & = \epsilon g  -3 g^2 -\f{N-1}{3}\lambda^2,
\\
\partial_\tau \lambda & = \epsilon \lambda -2 g\lambda+ \f{N+2}{3} \lambda^2.
\end{align}
Note that when $N\to 1$ the conventional coupling $g$ evolves autonomously, as it should:
\ba\label{eq:4mepsilonIsingmeasurement}
\partial_\tau g & =  \epsilon g -3 g^2,
\\
\partial_\tau \lambda & = (\epsilon  -2 g) \lambda +  \lambda^2.
\label{eq:4mepsilonIsingmeasurement2}
\end{align}
This is in contrast to the $N=0$ case, where the quenched disorder, represented by $\lambda$, would feed back into the flow of $g$ (and also of the mass).
The RG flows described by (\ref{eq:4mepsilonIsingmeasurement},~\ref{eq:4mepsilonIsingmeasurement2}) are qualitatively different (even in four dimensions) from those at ${N=0}$, which are discussed in Ref.~\cite{komargodski2017random}.

At the Ising fixed point, $g_* = \epsilon/3$,
\ba\label{eq:4epsilonIsinglambdaonly}
\partial_\tau \lambda & = \f{\epsilon}{3} \lambda +  \lambda^2,
\end{align}
showing that in $4-\epsilon$ dimensions the critical Ising model  does not have a weak-measurement fixed point. We will discuss the measured Ising model
(which turns out to be interesting)
futher in Sec.~\ref{sec:isingddims}.

In general, for a measured system, we can divide the couplings into  the conventional ``thermodynamic'' couplings $\{g\}$, and the couplings $\{\lambda\}$ which appear only when we use the replica trick (i.e.\ measurement strengths). 
The $\{\lambda\}$ couple to operators 
such as $\sum_{a\neq b} \measO^a\measO^b$
that involve more than one replica and which formally vanish when $N=1$.
As a result the RG equations have the schematic form (suppressing indices on $g$ and $\lambda$ that label different operators)
\ba\label{eq:rgeqnssectors}
\partial_\tau g & = \beta (\{g\}),
& 
\partial_\tau \lambda & = \beta_{\text{meas}} (\{g\}, \{\lambda\}).
\end{align}
If, prior to considering measurements, the  model is at a fixed point $g=g_*$, then in the second equation we can set $g=g_*$, as in Eq.~\ref{eq:4epsilonIsinglambdaonly}.
Note that the full replicated theory may flow to a nontrivial interacting fixed point even if the ``conventional'' sector flows to a free fixed point, for example as a result of being above its upper critical dimension  (analogous phenomena occur for geometrical observables in some lattice models \cite{newman1984q,nahum2013phase,wiese2024two}).

In the next few sections we discuss a variety of models which give rise to interesting fixed points. 
We start with critical Ising and Potts models.

\section{Critical Ising and Potts models}
\label{sec:potts2d}

The $Q$-state Potts model is defined initially from the nearest-neighbor interaction $-J \delta_{\sigma_x,\sigma_y}$ between spins $\sigma_x = 1,2,\ldots,Q$, with the case $Q=2$ being the Ising model:
\be\label{eq:Zpotts}
Z = \sum_{\{\sigma\}} e^{J \sum_{\<x,y\>} \delta_{\sigma_x, \sigma_y}}
\ee
We will start off in 2D, and for concreteness will consider the model on the square lattice.
The model can be extended to any real ${Q \ge 0}$ at the cost of considering models with nonlocal Boltzmann weights
(for example via the Fortuin-Kasteleyn expansion).
We assume throughout that the couplings are ferromagnetic, $J > 0$, and tuned to the critical point.

As in the general setup (Sec.~\ref{sec:meascritstates}) we can then measure arbitrary local observables 
built from the spins $\sigma_x$.
We focus here on measurement of the bond energies, setting ${\cal O} = \delta_{\sigma_x,\sigma_y}$ in the lattice model.
We will find an interesting phase diagram as a function of the measurement strength $\lambda$, with multiple  fixed points.
The first of these fixed points can be understood in detail using perturbative RG in $\lambda$.
We discuss this fixed point  in Sec.~\ref{sec:pottsweak}.
(Later on we will  denote it with the symbol $\fpW$, since it is accessible in perturbation theory at ``weak'' measurement.)
We then turn to the global phase diagram in Sec.~\ref{sec:pottsglobal}, and higher dimensions in Sec.~\ref{sec:isingddims}.

\subsection{Nontrivial perturbative fixed point in 2D}
\label{sec:pottsweak}

In the continuum limit, ${{\cal O}=\varepsilon}$, where $\varepsilon$ denotes the energy operator. This has scaling dimension $\protect{x_{\cal O} = (3-2g)/(2g)}$ with $g=\tfrac{1}{\pi} \arccos(-\sqrt{Q}/2)$ \cite{denNijs1979}, so by Eq.~\ref{eq:Harris} the coupling in the replica theory is 
marginal in the Ising case and
weakly relevant for small positive $Q-2$. 
This fact has been exploited in the analysis of the related disordered system \cite{LudwigCardy1987,Ludwig1990,DotsenkoPiccoPujol1995,DotsenkoJacobsenLewisPicco1999}, and we can carry over many results simply by taking ${N\to 1}$ instead of ${N\to 0}$.

The RG equation for the coupling $\lambda$ 
in Eq.~\ref{eq:repeatreplicaHforRG} has been computed,
for arbitrary $N$,
to  order $\lambda^2$ in \cite{LudwigCardy1987} and  to order $\lambda^3$ in \cite{DotsenkoPiccoPujol1995}. It reads 
\ba\label{eq:rgeqn2dpotts}
 \partial_\tau \lambda = y_\lambda\, \lambda + 4 \pi(N-2) \lambda^2 - 16 \pi^2(N-2) \lambda^3 + \cdots,
\end{align}
% {\red [Remark on signs and conventions (comment out, but let's keep it safe for the record): Agrees with (3.4) in \cite{Ludwig1997}, and with (3a)+(51) in \cite{LudwigCardy1987}, and also with (3.9) in \cite{DotsenkoPiccoPujol1995} upon identifying our $y_\lambda$ with their $-3\epsilon$ (see below their (4.14)). Moreover, our $\lambda^*$ agrees with \cite{DotsenkoPiccoPujol1995} below (4.14).]}
where $y_\lambda = 2-2 x_{\cal O}$,
which will be the small parameter,
is the RG eigenvalue of the perturbation:
\ba
 y_\lambda = \frac{4 (Q - 2)}{3 \pi} - \frac{4 (Q - 2)^2}{9 \pi^2} + \cdots.
\end{align}
We have $y_\lambda = \tfrac25$ for $Q=3$.
We will write $y=y_\lambda$ below.

For the Ising model weak energy measurements are marginally irrelevant, but 
for small $Q-2 > 0$ and $N<2$ there is a flow to the \textit{stable} weak-coupling fixed point at
\ba
 \lambda^* = \frac{y}{4\pi(2-N)} + \frac{y^2}{4\pi(2-N)^2} + \cdots.
 \end{align}
This applies both to the random-bond problem $(N \to 0)$ and to the problem of imperfect measurements $(N \to 1)$.

For the random-bond model it has been shown \cite{CardyJacobsen1997,JacobsenPicco2000} that this fixed point extends beyond the perturbative regime. In fact, it exists for any $Q>2$, even when the transition in the pure model is first order ($Q>4$). 
This cannot be the case in the measurement problem, since by Sec.~\ref{sec:nofeedback} a sector of the unperturbed model ($\lambda=0$) carries over unchanged to the perturbed one at $\lambda^*$. We therefore expect the weak-coupling fixed point to exist only for ${2 < Q \le 4}$.

The central charge of the replicated system has been computed to order $y^3$ in \cite{LudwigCardy1987} and to order $y^4$ in \cite{DotsenkoPiccoPujol1995,DotsenkoJacobsenLewisPicco1999}. 
Denoting by $c(N)$ the value of the central charge at $\lambda^*$, and $N c(1)$ its value at $\lambda=0$, the result reads
\ba
 c(N) = N c(1) - \frac{N(N-1)}{8(N-2)^2} \left( y^3 - \frac{3}{2(N-2)} y^4 \right).
\end{align}
In particular
\ba
 c'(1) - c(1) = -\frac{y^3}{8} - \frac{3 y^4}{16},
\end{align}
which is $-8/625 = -0.0128$ for $Q=3$.

Results for operator scaling dimensions can also be taken over directly from the random-bond case, where  Ludwig \cite{Ludwig1990} has considered the  dimensions $x_\psi^{(k)}$ of the disorder-averaged $k$-th moments, $\overline{\langle \psi(0) \psi(R) \rangle^k}$, for an operator $\psi$.
(The brackets denote the statistical average, while the overline is the disorder average.)
The interpretation in the replica setup is that an operator $\psi$ is inserted into each of $k$ distinct replicas out of the $N$.

In general models with perturbative fixed points, or in the present model if  $\psi$ is taken to be the energy operator \cite{Ludwig1990}, this is a nontrivial problem, because
an operator like $\psi^{a_1}\cdots\psi^{a_k}$ is not necessarily a scaling operator.
There are $N \choose k$ ways of choosing the $\{a_i\}_{i=1}^k$, and a scaling operator at the $g^*$ fixed point corresponds to a suitable linear combination of these choices that mixes them into an irreducible representation (irrep) of the symmetric group $S_N$. In other words, one needs to diagonalise the renormalisation matrix by introducing a new basis of operators. Consequently, the scaling dimension $x_\psi^{(k,\mu)}$ will generally depend not only on the numerical value of $k$, but also on the chosen irrep $\mu$.

However, for spin operators ($\psi = \sigma$), the operator mixing problem is already diagonal, 
not only at lowest order,\footnote{The renormalization of the scaling dimensions of the operators $\psi^{a_1}\cdots\psi^{a_k}$follows from the RG equation for corresponding infinitesimal couplings $g_{a_1,\ldots a_k}$. At lowest nontrivial order in $\lambda_*$ this is of the form
$\partial_\tau g_{a_1,\ldots a_k}= 
\big( y^{(k)}_0 \delta_{a_1,\ldots, a_k}^{a_1',\ldots, a_k'}
+ \lambda_* M_{a_1,\ldots, a_k}^{a_1',\ldots, a_k'}
\big) g_{a_1',\ldots a_k'}$,
where $ M_{a_1,\ldots, a_k}^{a_1',\ldots, a_k'}$ is proportional to the OPE coefficient for obtaining 
$\psi^{a_1}\cdots\psi^{a_k}$
when the operator 
$\psi^{a_1'}\cdots\psi^{a_k'}$
is combined with the perturbing operator 
$\sum_{a\neq b}\epsilon_a \epsilon_b$. However, 
unless $\psi$ is the energy operator itself,
this  OPE coefficient is proportional to 
$\delta_{\{a_1,\ldots,a_k\}}^{\{a_1',\ldots,a_k'\}}$.
In other words, the equation is already diagonal and the operators do not mix at this order.} but  at any order, 
as was noticed  in the explicit RG calculation \cite{Ludwig1990,DotsenkoJacobsenLewisPicco1999}. This is in fact a simple consequence of the Potts symmetry in each replica, i.e.\ the global
 ${S_Q\times \cdots \times S_Q}$ symmetry which  is present  in addition to $S_N$.
The operators $\sigma^{a_1}\cdots\sigma^{a_k}$
with distinct sets $\{a^1,\ldots,a^k\}$ belong to distinct irreps of ${S_Q\times \cdots \times S_Q}$,  so cannot mix.
It follows that the multiscaling exponent $x_\sigma^{(k)}$ depends only on $k$ in this case.

Results for energy operators can be read off from Table 1 in \cite{Ludwig1990}. Let us focus here instead on spin operators. Let $x_\sigma$ denote the scaling dimension in the unreplicated Potts model:
${x_\sigma = \frac{1-4(1-g)^2}{8g}}$ \cite{nienhuis1980magnetic}, so that ${x_\sigma=\frac{2}{15}}$ when ${Q=3}$. 
For $k=1$ one finds \cite{DotsenkoPiccoPujol1995}
\ba
 \label{xsigma1}
 x_\sigma^{(1)} = x_\sigma - \frac{N-1}{8(N-2)^2} \frac{\Gamma^2(-\tfrac23) \Gamma^2(\tfrac16)}{\Gamma^2(-\tfrac13) \Gamma^2(-\tfrac16)} y^3
\end{align}
and since this is a one-replica quantity the correction vanishes
when we set $N=1$
(indeed to all orders in $y$), in agreement with the discussion in Sec.~\ref{sec:nofeedback}.
The exponent for general $k$ was shown in
\cite{Ludwig1990} to take the form, to order $y^2$,
\ba
 \label{spin-exps}
 x_\sigma^{(k)} = k x_\sigma + \frac{k(k-1)}{8(N-2)} y + k(a_2 - b_2 k - c_2 k^2) y^2,
\end{align}
for some coefficients $a_2, b_2, c_2$ which were left
undetermined at the time. They can however be fixed by comparison with the later result
\eqref{xsigma1} and the two-loop results for $x_\sigma^{(2)}$ and $x_\sigma^{(3)}$ reported in
\cite{DotsenkoJacobsenLewisPicco1999}. The result
is
\ba
 a_2 &= \frac{1}{96(N-2)^2} \big( 11 + 12(N-2) \ln 2 - 2 \Xi \big), \\
 b_2 &= \frac{1}{96(N-2)^2} \big( 11 + 12(N-2) \ln 2 - 3 \Xi \big), \\
 c_2 &= \frac{\Xi}{96(N-2)^2},
\end{align}
where we have defined
\ba
 \Xi = -\frac{33 \sqrt{3}-29 \pi}{2\sqrt{3}}.
\end{align}
Let us rewrite \eqref{spin-exps} in the form $x_\sigma^{(k)} = k x_\sigma + \Delta x_\sigma^{(k)}$, where $\Delta x_\sigma^{(k)}$ is the
correction term to order $y^2$. Taking the $N \to 1$
replica limit relevant for the measurement problem, the numerical values are $\Delta x_\sigma^{(2)} \simeq -0.109$ and $\Delta x_\sigma^{(3)} \simeq -0.425$ for $Q=3$.

Ref.~\cite{Ludwig1990} also discusses
the universal scaling function $H(\alpha)$
 describing the  probability distribution of the rescaled correlator $\alpha=-\tfrac12 \ln G / \ln R$, 
where ${G(R)=\langle \sigma(0)\sigma(R)\rangle}$ denotes the spin correlation function before the disorder average. $H(\alpha)$
is essentially the Legendre transform of the function $x_\sigma^{(k)}$. 
In a {\em typical} fixed sample the correlator decays like
\ba
 G(R) \sim R^{-2 x_\text{typ}}, \quad
x_\text{typ} = \left. \tfrac{\rm d}{{\rm d}k} x_\sigma^{(k)} \right|_{k=0}.
\end{align}
For ${Q=3}$ we find $x_\text{typ}\simeq 0.1551$, showing that the typical value of the correlator is parametrically smaller than the mean (whose decay is governed by ${x_\sigma=2/15}$).
The higher derivatives of $x_\sigma^{(k)}$ with respect to $k$ (at ${k \to 0}$), of which the next two can be computed from \eqref{spin-exps}, give access to the higher cumulants of the probability distribution
of $\ln G(R)$.

\subsection{Global phase diagram for 2D Ising and Potts}
\label{sec:pottsglobal}

We now ask what happens at larger measurement strength.
For concreteness, we consider the model on the square lattice.

First consider the Ising model ($Q=2$).
Weak measurement of the energy (small $\lambda$) is marginally irrelevant, as discussed above, so there is a stable weak-measurement phase.
However, it is easy to see that there is \textit{also} a stable phase at strong measurement (large $\lambda$).
In the replica language, this is a phase where
the \textit{relative} spin between distinct replicas,
$S^a_x S^b_x$, acquires an expectation value, spontaneously breaking the symmetry of the replica theory in the pattern\footnote{Because whereas  the replica Hamiltonian has a separate $\mathbb{Z}_2$ symmetry for each replica, the Edwards-Anderson-like order parameter $S^a S^b$
breaks this down to a single $\mathbb{Z}_2$ that acts simultaneously on all replicas.}
\be
(\mathbb{Z}_2\times \cdots \times \mathbb{Z}_2) \rtimes S_N \longrightarrow \mathbb{Z}_2 \times S_N.
\ee
Physically, the signature of this symmetry breaking is long-range order in ${\mathbb{E}_M\<S_x S_y\>_M^2}$.
(The phase is analogous to the strong measurement phase for the  Ising paramagnet, discussed in Sec.~\ref{sec:paramagneticstates}.)
The associated fixed point $\fpL$ is a relatively trivial ``replica-locked'' state in which all replicas are equal, modulo possible  global spin reversals.

Given the existence of stable phases at small and large $\lambda$, the simplest hypothesis is that an unstable critical point $\fpU$ exists at some order-1 value of $\lambda$.
This hypothesis is shown in Fig.~\ref{fig:2DIsingPottsFlows}~(a).

\begin{figure}
\begin{tikzpicture}
\tikzset{flow/.style = {thick, double = black,
      double distance = 0pt}}
\tikzset{fixedp/.style = {minimum width=\minsize,inner sep=\innersep}}
\node at (-2,0) {{\bf Ising $d=2$}};
 \node[fixedp] (o) at (0,0) [draw, circle,fill=\colZ] {$0$};
  \node[fixedp] (u) at (2.5,0) [draw, circle, fill=\colU] {$\mathcal{U}$};
    \node[fixedp] (inf) at (5,0) [draw, circle,fill=\colI] {$\mathcal{L}$};
 \draw [very thick,flow,gray][<-] (o.east) -- (u.west);
  \draw [very thick, flow][->] (u.east) -- (inf.west);
    \draw[thick][->] (0,-0.8) -- (0.9,-0.8);
  \node at (1.15,-0.77) {$\lambda$};
\end{tikzpicture} \\
\vspace{20pt}
\begin{tikzpicture}
\tikzset{flow/.style = {thick, double = black,
      double distance = 0pt}}
\tikzset{fixedp/.style = {minimum width=\minsize,inner sep=\innersep}}
\tikzset{flow/.style = {thick, double = black,
      double distance = 0pt}}
\node at (-2,0) {{\bf Potts $d=2$}};
 \node[fixedp] (o) at (0,0) [draw, circle,fill=\colZ] {$0$};
   \node[fixedp] (w) at (5/3,0) [draw, circle, fill=\colW] {$\mathcal{W}$};
  \node[fixedp] (u) at (10/3,0) [draw, circle, fill=\colU] {$\mathcal{U}$};
        \node[fixedp] (inf) at (5,0) [draw, circle,fill=\colS] {$\mathcal{S}$};
 \draw [very thick,flow,\colW][->] (o.east) -- (w.west);
  \draw [very thick,flow,\colW][<-] (w.east) -- (u.west);
  \draw [very thick, flow,\colS][->] (u.east) -- (inf.west);
  \draw[thick][->] (0,-0.8) -- (0.9,-0.8);
  \node at (1.15,-0.77) {$\lambda$};
  \end{tikzpicture}
\caption{{\bf (a)} Phase diagram and schematic RG flows for the critical 2D critical Ising model, on the lattice, as a function of the strength $\lambda$ of bond-energy measurement. 
There is a stable phase
 ({\protect\tikz[baseline=-2.5pt]
  \protect\draw[ultra thick,gray][-] (0,0) -- (0.4,0);})
  where the measurement strength $\lambda$ flows to $0$,
  and another stable phase
 ({\protect\tikz[baseline=-2.5pt]
  \protect\draw[ultra thick,black][-] (0,0) -- (0.4,0);})
  that flows to the trivial replica-locked fixed point 
 {\protect\tikz[baseline=-2.5pt]
  \protect\node[minimum width=0,inner sep=0.5] (i) at (0,0) [draw, circle, fill=\colI] {$\mathcal{L}$};} at ${\lambda=\infty}$.
 An unstable fixed point 
{\protect\tikz[baseline=-2.5pt]
  \protect\node[minimum width=0,inner sep=0.5] (u) at (0,0) [draw, circle, fill=\colU] {$\mathcal{U}$};}  separates them.
{\bf (b)} Flows for the critical 2D Potts model with ${Q\gtrsim 2}$.
The phase for ${\lambda< \lambda_c}$    ({\protect\tikz[baseline=-2.5pt]
  \protect\draw[ultra thick,\colW][-] (0,0) -- (0.4,0);}) 
  is now nontrivial and  governed by a ``weak measurement''  fixed point     {\protect\tikz[baseline=-2.5pt]
  \protect\node[minimum width=0,inner sep=0.2] (u) at (0,0) [draw, circle, fill=\colW] {\footnotesize $\mathcal{W}$};}. 
  The phase for ${\lambda>\lambda_c}$ also becomes nontrival and is governed by a stable strong-measurement fixed point
{\protect\tikz[baseline=-2.5pt]
  \protect\node[minimum width=0,inner sep=0.5] (u) at (0,0) [draw, circle, fill=\colS] {$\mathcal{S}$};}.
  This fixed point lies (conjecturally) at ${\lambda=\infty}$ 
  but has nontrivial fluctuations.
[Sec.~\ref{sec:FKmeasurement} describes another fixed point for Ising and Potts, {\protect\tikz[baseline=-2.5pt]
  \protect\node[minimum width=0,inner sep=0.5] (u) at (0,0) [draw, circle, fill=white] {\tiny$\mathcal{FK}$};}, but we argue that this is highly fine-tuned and cannot be identified with any of the fixed points above.]}
\label{fig:2DIsingPottsFlows}
\end{figure}
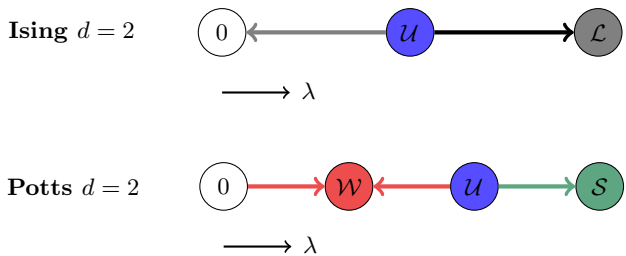

Next consider the Potts model for $Q\gtrsim 2$, assuming that we can continuously vary $Q$.\footnote{The replica partition function has a well-defined continuation to real $Q$, and we assume this is sufficient.
Whether we can define a ``physical'' measurement process (where all  probabilities are positive) for noninteger $Q$ depends on the measurement protocol. It is not possible for the present protocol (App.~\ref{app:FKdetails}). It is possible for (generalizations of) the one in Sec.~\ref{sec:FKmeasurement}, but those cannot access the strong measurement phase discussed here.}
We know that there is then a stable fixed point $\fpW$
at ${\lambda_*}$ of order ${Q-2}$
(discussed in the previous section) which 
governs the weak-measurement phase.
The simplest expectation is that 
the unstable fixed point $\fpU$, which was present in the Ising case, has  a continuation to $Q>2$, 
with the exponents changing continuously in $Q$.

Next consider the fixed point at $\lambda=\infty$.
In the Ising case, this was trivial (e.g.\ the effective central charge $c'(1)$ vanished):
perfect measurement of the energy 
fixes the locations of all domain walls, and once we know the domain wall configurations, the Ising configuration is fixed up to a global sign.
For $Q>2$, however, the fixed point at $\lambda=\infty$ is nontrivial, because fixing the domain wall configuration in general leaves freedom in how the domains are ``colored'' (by assigning spin values).\footnote{In the $\lambda\to\infty$ limit, the replica partition function is a sum over configurations $C$ of ``nets'', representing domain walls, with a weight proportional to $\chi_C(Q)^N$, where where $\chi_C$ is the chromatic polynomial of the graph of domains. (For integer $Q$, this is the number of ways of coloring the domains with $Q$ colors, such that domains separated by a domain wall have a different color.)}
Nevertheless, we conjecture that the nontrivial $\lambda=\infty$ fixed point $\fpS$ remains stable for $Q>2$.

These assumptions give the phase diagrams shown in Fig.~\ref{fig:2DIsingPottsFlows}.
The properties of the fixed points $\fpW$,  $\fpU$, $\fpS$ of course depend on $Q$, but we suppress this label.

Numerics would be required to determine whether this flow topology remains the same all the way up to $Q=4$, but this is possible.

\subsection{Aside: measuring Fortuin-Kasteleyn clusters}
\label{sec:FKmeasurement}

Above we argued that the Potts model exhibits a fixed point $\fpU$ at intermediate measurement strength that is unstable to variation in the strength (see Fig.~\ref{fig:2DIsingPottsFlows}). 
Unlike the \textit{stable} fixed point $\fpW$ which we found for $Q>2$,
the unstable fixed point $\fpU$ is not accessible within the above perturbation theory, so it would require another theoretical approach.

We now briefly discuss a another unstable fixed point, $\fpFK$.
This unstable fixed point appears naturally in a different --- 
highly fine-tuned --- 
measurement protocol that is  related to the Fortuin-Kasteleyn (FK) representation of the Potts model.
At first sight, we might hope that $\fpFK$ could be identified with $\fpU$. 
Instead, we  argue that $\fpFK$  is a much more highly fine-tuned object, with, in principle, an infinite number of relevant perturbations. Therefore, while it arises in the fine-tuned measurement protocol that we describe next, it is not pertinent to the more generic measurement process which we described in the previous section.

Recall the FK representation of the Potts-model partition function that  starts from rewriting (\ref{eq:Zpotts}) as
\ba\label{eq:pottsSW}
Z & = \sum_{\{\sigma\}, \{M\}} 
\prod_{\<x,y\>} 
\lf 
\delta_{M_{x,y},0} + z\delta_{M_{x,y},1} \delta_{\sigma_x, \sigma_y}
\ri,
\end{align}
with ${z=e^J-1}$.
This representation introduces bond  variables $M_{x,y}=0,1$.
We think of $M_{x,y}=0$ as denoting an unoccupied bond, and $M_{x,y}=1$ an occupied bond.
The connected clusters of occupied bonds are the ``FK clusters''.
Note that, once we condition on the FK cluster configuration,  spins in distinct FK clusters become independent, while spins in the same FK cluster are forced to take the same value. (This fact underlies the Swendsen-Wang cluster Monte Carlo algorithm \cite{newman1999monte}.)

We have denoted the bond variables by $M$ because they can be viewed as measurement outcomes: i.e.\ Eq.~\ref{eq:pottsSW} can be viewed as 
a combined measure
$e^{-\mathcal{H}[\sigma]}P(M|\sigma)$ for spins and measurements, for a certain binary measurement process defined by
\be\notag
P(M_{x,y}|\sigma) = ({1 + z  \delta_{\sigma_x, \sigma_y}})^{-1} \lf {\delta_{M_{x,y},0} + z\delta_{M_{x,y},1} \delta_{\sigma_x, \sigma_y}}\ri.
\ee
This measurement process has a simple interpretation. 
Unsatisfied bonds (those with $\delta_{\sigma_x,\sigma_y}=0$) are detected with perfect reliability:
if $\delta_{\sigma_x,\sigma_y}=0$, then $M_{x,y}=0$ with probability 1. However, satisfied bonds  have a nonzero probability $(1+z)^{-1}$ of being misdiagnosed as unsatisfied.

We refer to this as the FK measurement process  since, colloquially, it reveals the geometry of the FK clusters with perfect precision.
Since the spins on different FK clusters are independent, 
the correlator 
${G_M(x,y) = 
(1-Q^{-1})^{-1} \< \delta_{\sigma_x,\sigma_y}-Q^{-1} \>_M}$
conditioned on measurements is easy to write down:
\be
G_M(x,y)
= 
\left\{
\begin{array}{ll}
 1 &\text{if $x,y$ on same cluster,}     \\
 0 &  \text{otherwise.}
\end{array}
\right.
\ee
Therefore, for any $k>0$,
\be
\mathbb{E}_M  G_M(x,y)^k =\mathbb{E}_M G_M(x,y) \equiv G(x,y),
\ee
where $G(x,y)$ is the conventional correlator without any conditioning on measurements.
This fixed point is therefore not ``multifractal'', unlike the stable fixed point for $Q>2$ discussed in the previous section. 

It is also straightforward to write down the replica partition function. Each FK cluster carries a spin value from each replica --- i.e.\ it is characterized by a vector $\vec{\sigma} = (\sigma^1, \ldots, \sigma^N)$ of spin values.
Since $\vec{\sigma}$ takes $Q^N$ values, we effectively have a Potts model with $\mathcal{Q}=Q^N$ states. This may be used to write the effective central charge,
\be
 c_{\rm eff} = \frac{3(1-g^2)}{g^2 \pi} \sqrt{\frac{Q}{4-Q}} \ln Q,
\ee
which satisfies $c_{\rm eff} > c(1)$ for $1 < Q \lesssim 2.989$.

This picture may also be used to assess the stability of the 
present fixed point, $\fpFK$, to slight changes in the measurement protocol. 
We find that it is highly unstable.

%zc = Sqrt{Q}
% for FK point, prob of CORRECTLY measuring F bond is 
% 1 - 1/(1+z) = z/(1+z) ---> Sqrt{Q}/(1+Sqrt{Q}).
%Ising: P(F correct)=0.586
%3statePotts: P(Fcorrect)=0.634
%4statePotts: P(Fcorrect)=2/3
%(0.586-0.5)*2 = 0.172
%(0.634-0.5)*2 = 0.268
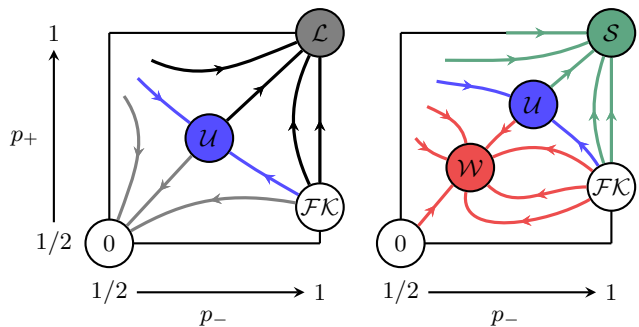
\begin{figure}
\begin{tikzpicture}[
        edge/.style={circle,fill,minimum size=1.5mm,inner sep=0pt},
        every path/.style={draw, thick,decoration={markings,mark = at position 0.5 with {\arrow{>}}},>=stealth},
        every label/.style={font=\footnotesize\sffamily},
        x=0.7cm,y=0.7cm]
%The axes of this phase diagram are p+, p- where p+ is the probability of CORRECTLY measuring a bond with \delta_{sigma sigma}=1 and p- is the probability of CORRECTLY measuring a bond with \delta_{sigma sigma}=0
\tikzset{flow/.style = {thick, double = black,
      double distance = 0pt}}
\tikzset{fixedp/.style = {minimum width=\minsize,inner sep=0}}
\tikzset{flow/.style = {thick, double = black,
      double distance = 0pt}}
      %axes
      \newcommand{\figext}{4}
      % axes
        \draw[thick][-] (0,0) -- (\figext,0);
             \draw[thick][-] (0,0) -- (0,\figext);
              \draw[thick][-] (\figext,0) -- (\figext,\figext);
               \draw[thick][-] (0,\figext) -- (\figext,\figext);
      %fixed points
 \node[fixedp] (o) at (0,0) [draw, circle,fill=\colZ] {$0$};
%   \node[fixedp] (w) at (5/3,0) [draw, circle, fill=\colW] {$\mathcal{W}$};
  \node[fixedp] (u) at ({-0.1+\figext/2},\figext/2) [draw, circle, fill=\colU] {$\mathcal{U}$};
        \node[fixedp] (inf) at (\figext,\figext) [draw, circle,fill=\colI] {$\mathcal{L}$};
         \node[fixedp] (fk) at (\figext,{(0.172)*\figext}) [draw, circle,fill=white] {$\mathcal{FK}$};
  %axislabelling
  \node at (\figext/2,{-\figext/2.8}) {$p_{-}$};
%   \node at (0,-\figext/5) {$1/2$};
%    \node at (\figext,-\figext/5) {$1$};
    \node at ({-\figext/2.5},\figext/2) {$p_{+}$};
       \node (h1) at (-\figext/3.8,0) {$1/2$};
    \node (h2) at (-\figext/3.8,\figext) {$1$};
    \draw [thick][->] (h1) -- (h2);
     \node (v1) at (0,-\figext/4.3) {$1/2$};
    \node (v2) at (\figext,-\figext/4.3) {$1$};
    \draw [thick][->] (v1) -- (v2);
            %flow lines
            \path[postaction={decorate},very thick,gray] (u.south west) -- (o.north east);
            \path[postaction={decorate},very thick] (u.north east) -- (inf.south west);
            \path[postaction={decorate},very thick] (fk.north) -- (inf.south);
            \path[postaction={decorate},very thick,\colU] (fk) to[out=150,in=-35] (u);
             \path[postaction={decorate},very thick,gray] (fk) to[out=170,in=30] (o);
               \path[postaction={decorate},very thick] (fk) to[out=115,in=-115] (inf);
                \path[postaction={decorate},very thick,\colU] ({0.135*\figext},{0.785*\figext}) to[out={-45},in=142] (u);
                   \path[postaction={decorate},very thick] ({0.2*\figext},{0.87*\figext}) to[out={-25},in=202] (inf);
                   \path[postaction={decorate},very thick,gray] ({0.07*\figext},{0.7*\figext}) to[out={-55},in=70] (o);
  \end{tikzpicture}
  \hspace{5pt}
  %%%%%
  %%%%%
  %%%%%
  \begin{tikzpicture}[
        edge/.style={circle,fill,minimum size=1.5mm,inner sep=0pt},
        every path/.style={draw, thick,decoration={markings,mark = at position 0.5 with {\arrow{>}}},>=stealth},
        every label/.style={font=\footnotesize\sffamily},
        x=0.7cm,y=0.7cm]
%The axes of this phase diagram are p+, p- where p+ is the probability of CORRECTLY measuring a bond with \delta_{sigma sigma}=1 and p- is the probability of CORRECTLY measuring a bond with \delta_{sigma sigma}=0
\tikzset{flow/.style = {thick, double = black,
      double distance = 0pt}}
\tikzset{fixedp/.style = {minimum width=\minsize,inner sep=0}}
\tikzset{flow/.style = {thick, double = black,
      double distance = 0pt}}
      %axes
      \newcommand{\figext}{4}
      % axes
        \draw[thick][-] (0,0) -- (\figext,0);
             \draw[thick][-] (0,0) -- (0,\figext);
              \draw[thick][-] (\figext,0) -- (\figext,\figext);
               \draw[thick][-] (0,\figext) -- (\figext,\figext);
      %fixed points
 \node[fixedp] (o) at (0,0) [draw, circle,fill=\colZ] {$0$};
%   \node[fixedp] (w) at (5/3,0) [draw, circle, fill=\colW] {$\mathcal{W}$};
  \node[fixedp] (u) at ({0.63*\figext},{0.66*\figext}) [draw, circle, fill=\colU] {$\mathcal{U}$};
    \node[fixedp] (w) at ({0.33*\figext},{0.36*\figext}) [draw, circle, fill=\colW] {$\mathcal{W}$};
        \node[fixedp] (inf) at (\figext,\figext) [draw, circle,fill=\colS] {$\mathcal{S}$};
         \node[fixedp] (fk) at (\figext,{(0.268)*\figext}) [draw, circle,fill=white] {$\mathcal{FK}$};
  %axislabelling
  \node at (\figext/2,{-\figext/2.8}) {$p_{-}$};
%   \node at (0,-\figext/5) {$1/2$};
%    \node at (\figext,-\figext/5) {$1$};
%    \node at ({-\figext/2.5},\figext/2) {$p_{+}$};
%       \node (h1) at (-\figext/3.8,0) {$1/2$};
%    \node (h2) at (-\figext/3.8,\figext) {$1$};
%    \draw [thick][->] (h1) -- (h2);
     \node (v1) at (0,-\figext/4.3) {$1/2$};
    \node (v2) at (\figext,-\figext/4.3) {$1$};
    \draw [thick][->] (v1) -- (v2);
            %flow lines
            \path[postaction={decorate},very thick,\colW] (o.north east) -- (w.south west);
            \path[postaction={decorate},very thick,\colS] (u.north east) -- (inf.south west);
            \path[postaction={decorate},very thick,\colS] (fk.north) -- (inf.south);
            \path[postaction={decorate},very thick,\colU] (fk) to[out=122,in=-40] (u);
            % fk --> w
             \path[postaction={decorate},very thick,gray,\colW] (fk) to[out=180,in=-45] (w);
                 \path[postaction={decorate},very thick,gray,\colW] (fk) to[out=210,in=-100] (w);
                 \path[postaction={decorate},very thick,gray,\colW] (fk) to[out=140,in=25] (w);
               \path[postaction={decorate},very thick,\colS] (fk) to[out=108,in=-108] (inf);
                \path[postaction={decorate},very thick,\colU] ({0.17*\figext},{0.77*\figext}) to[out={-10},in=160] (u);
                         \path[postaction={decorate},very thick,\colW] ({0.125*\figext},{0.65*\figext}) to[out={-25},in=100] (w);
                   \path[postaction={decorate},very thick,\colS] ({0.21*\figext},{0.87*\figext}) to[out={0},in=202] (inf);
                   \path[postaction={decorate},very thick,\colW] ({0.07*\figext},{0.5*\figext}) to[out={-50},in=160] (w);
                      \path[postaction={decorate},very thick,\colW] (u) to[out=-138,in=50] (w);
                      \path[postaction={decorate},very thick,\colS] ({0.5*\figext},\figext) -- (inf);
  \end{tikzpicture}
  \caption{Conjectured topology of RG flows for the critical Ising model (left) and the critical 3-state Potts model (right), in 2D, superimposed on a schematic phase diagram in the plane $(p_-, p_+)$ for the generalized  measurement protocol described at the end of Sec.~\ref{sec:FKmeasurement}. Here $p_+$ is the probability that a satisfied bond ($\delta_{\sigma,\sigma'}=1$) is correctly measured to be satisfied, and $p_-$
  is the probability that an unsatisfied bond ($\delta_{\sigma,\sigma'}=0$) is correctly measured to be unsatisfied. Note that in each case there are two stable phases.}
  \label{fig:2dflowdiagramspotts}
\end{figure}

[In more detail:
a generic protocol involving measurement of the energy yields the replica symmetry 
\be
G_{Q,N}\equiv (S_Q \times S_Q \times \cdots \times S_Q )\rtimes S_N,
\ee
which includes a separate Potts $S_Q$ symmetry for each replica, together with replica exchanges. 
The fine-tuned process in which FK clusters are measured perfectly has a much larger $S_{\mathcal{Q}}$ symmetry, allowing permutations of all the $\mathcal{Q} = Q^N$ possible states of the vector $\vec{\sigma}$ defined just above.
When we move away from this fine-tuned limit, $S_{\mathcal{Q}}$ is broken down to $G_{Q,N}$, allowing additional relevant perturbations in the action. In fact, in the replica limit, a formally infinite number of relevant perturbations is allowed by symmetry, all with the dimension $x = 2g-(g-1)^2/(2g)$ (Sec.~\ref{sec:pottsweak}).
This mechanism is loosely analogous to the one \cite{nahum2021measurement} that destabilizes the large Hilbert-space dimension limit \cite{skinner2019measurement,jian2020measurement} in which the quantum measurement phase transition  maps to classical percolation. Further details in App.~\ref{app:FKdetails}.]

We may consider a generalized measurement protocol in which the error probability differs for satisfied and unsatisfied bonds, giving a two-dimensional phase diagram.
The instability of $\fpFK$ leads to the conjectural  phase diagram topologies shown in Fig.~\ref{fig:2dflowdiagramspotts}. 

For those flow lines for which can compute the effective central charge in the UV and the IR (involving $\fpZ$, $\fpW$, $\fpFK$), we find that
$c^\text{UV}_\text{eff}>c^\text{IR}_\text{eff}$.\footnote{$c_\text{eff}^{\mathcal{FK}}$ starts off at $Q=2$ being  larger than the central charge of the unmeasured model (consistent with the flow line $\fpFK\to \fpZ$ in the Ising case),
but has become smaller than the central charge of the pure model by $Q=3$.}

\subsection{Critical Ising model in $d > 2$ dimensions}
\label{sec:isingddims}

We have argued that, in two dimensions, measurement of bond energies in the critical Ising model 
leads to relatively trivial weak and strong measurement phases, separated by a phase transition.
Now we consider ${d>2}$, where the picture turns out to be surprisingly  intricate.

For the purposes of the following discussion, we will assume that the phase diagram can be continued to non-integer dimensions.
Continuations to non-integer dimensions are ubiquitous in statistical physics, but in general have not yet been rigorously defined. 
Nevertheless, in the present case they will help us to formulate conjectures that are ultimately to be tested for integer dimensions.
The continuous-$d$ formulation of the measured Ising model is also interesting because it allows simplifications not only near two dimensions but also  near four and six dimensions.

First consider ${2+\epsilon}$ dimensions.
Ref.~\cite{komargodski2017random} proposed  an expansion of the RG equations for the random-bond Ising model in ${\epsilon}$ that is easily adapted to measurements: 
\ba\label{eq:ising2pluseps}
\partial_\tau \lambda & = y_\lambda \lambda - 4\pi \lambda^2 + \ldots, 
&
y_\lambda & \simeq 0.4 \epsilon + \ldots
\end{align}
In Eq.~\ref{eq:ising2pluseps} the   scaling dimension of the Ising energy operator is taken \cite{komargodski2017random} from 
conformal bootstrap \cite{el2014conformal}.
The initial formulation of the bootstrap in noninteger dimensions is no longer believed to be exact
(since unitarity bounds cannot be assumed for noninteger $d$). However, the resulting errors are believed to be numerically small \cite{rychkov2024new}; in any case,  similar results for $y_\lambda$ can be obtained from the  resummed expansion around four dimensions~\cite{el2014conformal,le1987accurate}.

Eq.~\ref{eq:ising2pluseps} shows that when $d$ is increased above 2, a new stable weak-coupling fixed point appears 
(similarly to what we saw when $Q$ was increased above 2 in the 2D Potts model).
This leads to the flow topology  in Fig.~\ref{fig:3DIsingFlows}, where the stable weak-coupling fixed point is denoted  $\fpW$. By continuity, the unstable fixed point $\fpU$ persists to $2+\epsilon$ dimensions.

\begin{figure}
\vspace{20pt}
\begin{tikzpicture}
\tikzset{flow/.style = {thick, double = black,
      double distance = 0pt}}
\tikzset{fixedp/.style = {minimum width=\minsize,inner sep=\innersep}}
\tikzset{flow/.style = {thick, double = black,
      double distance = 0pt}}
\node at (-2,0) {{\bf Ising $d=3$}};
 \node[fixedp] (o) at (0,0) [draw, circle,fill=\colZ] {$0$};
   \node[fixedp] (w) at (5/3,0) [draw, circle, fill=\colW] {$\mathcal{W}$};
  \node[fixedp] (u) at (10/3,0) [draw, circle, fill=\colU] {$\mathcal{U}$};
        \node[fixedp] (inf) at (5,0) [draw, circle,fill=\colI] {$\mathcal{L}$};
 \draw [very thick,flow,\colW][->] (o.east) -- (w.west);
  \draw [very thick,flow,\colW][<-] (w.east) -- (u.west);
  \draw [very thick, flow,black][->] (u.east) -- (inf.west);
  \draw[thick][->] (0,-0.8) -- (0.9,-0.8);
  \node at (1.15,-0.77) {$\lambda$};
  \end{tikzpicture}
\caption{Conjectural phase diagram and schematic RG flows for the critical 3D  Ising model, on the lattice, as a function of  strength $\lambda$ of bond-energy measurement. 
The topology of the flows is similar to the example of the 2D critical Potts model for $Q\gtrsim 2$ shown in Fig.~\ref{fig:2DIsingPottsFlows}~{\bf (b)}, with a nontrivial ``multifractal'' weak measurement phase.  However the strong measurement phase is now trivially locked.}
  \label{fig:3DIsingFlows}
\end{figure}
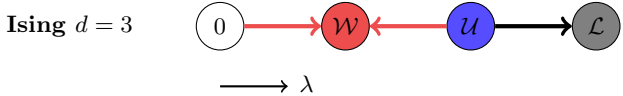

What happens as we continue to increase the dimension?
At first glance, a natural possibility might seem to be that 
the fixed points $\fpW$ and $\fpU$ meet with each other and annihilate at some critical value of the dimensionality.
If this was to occur below four dimensions, 
it would leave \textit{no} nontrivial fixed points.
As a result, there would be uninterrupted RG flow all the way from ${\lambda=0^+}$ to ${\lambda=\infty}$,
meaning that even very weak measurement would be sufficient to access the strong-measurement phase.

We argue  below that this is not possible: it is impossible for the strong-measurement phase to extend to arbitrarily small $\lambda$.
(The argument uses the FK representation.)
Therefore we conclude that $\fpW$ and $\fpU$ should not annihilate with each other,
at least not before some other fixed point appears on the scene.
The next dimensionality where we expect a new fixed point to appear is ${d=4}$,
as we discuss in a moment.

Therefore we conjecture that the flow topology in 
Fig.~\ref{fig:3DIsingFlows}
also holds in three dimensions. 

Now consider the vicinity of ${d=4}$. The RG equation for $\lambda$ has already been given in Eq.~\ref{eq:4mepsilonIsingmeasurement2}, where now ${\varepsilon=4-d}$:
\ba
\partial_\tau \lambda & = \f{\varepsilon}{3} \lambda +  \lambda^2 & & (\varepsilon>0),
\\
\partial_\tau \lambda & = \varepsilon \lambda +  \lambda^2 & & (\varepsilon<0).
\end{align}
(The nonanalyticity arises because the conventional quartic coupling $g$ in Eq.~\ref{eq:4mepsilonIsingmeasurement2} vanishes, at the Ising critical point, above 4D.)
We see that there is no weak-coupling fixed point just below four dimensions, but there is an unstable weak-coupling fixed point,
at ${\lambda_*\simeq |\varepsilon|}$,
 \textit{above} four dimensions. 
(We comment in passing that the flows in and below 4D are very different from the ones that occur in the random-bond problem, i.e.\ at $N=0$. The latter are discussed in Ref.~\cite{komargodski2017random}.)

\begin{figure}
    \centering
\includegraphics[width=0.45\columnwidth]{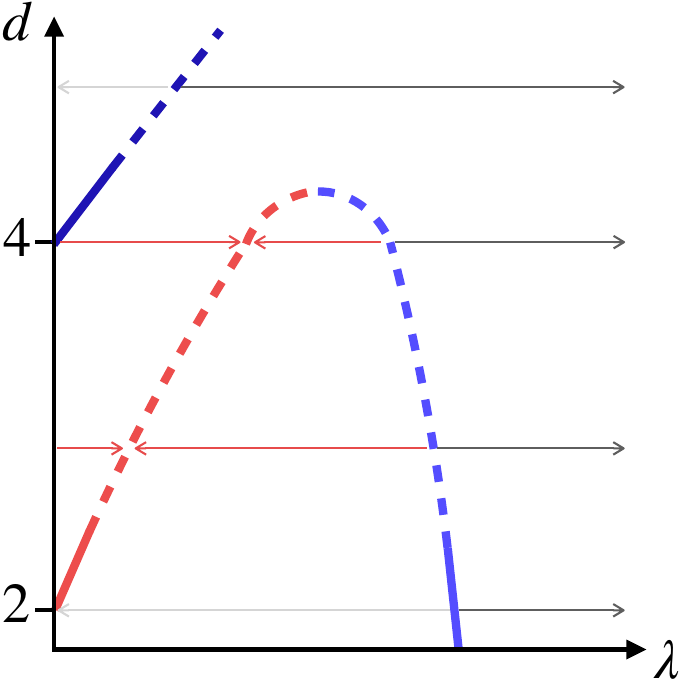}
\includegraphics[width=0.45\columnwidth]{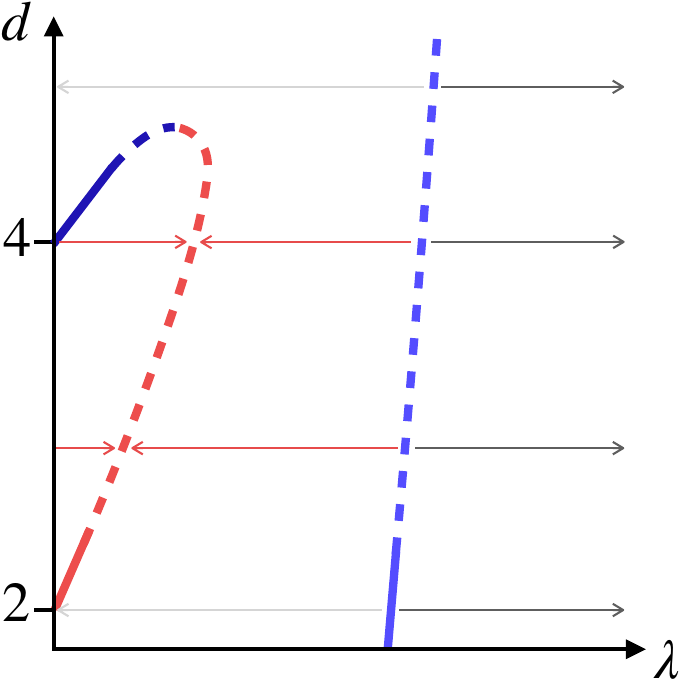}
    \caption{Two possibilities for the topology of the RG fixed points and flows of the critical Ising model with bond energy measurements as a function of dimension $d$ and measurement strength $\lambda$.
    Solid lines are results near ${d=2}$ and ${d=4}$. Dashed lines are speculative extensions to a wider range of $d$.}
\label{fig:possibleisingflowsgenerald}
\end{figure}

An interesting question (for the future) is 
how the fixed points we have found in $2+\epsilon$ and $4+|\varepsilon|$ connect up. 
Fig.~\ref{fig:possibleisingflowsgenerald} shows  two possibilities ---
 not the only ones --- for the global topology.
In the scenarios shown in Fig.~\ref{fig:possibleisingflowsgenerald},
two of the fixed points annihilate at some ${d_*\geq 4}$, so that
there is (if $d_*$ is strictly greater than 4) 
a range of $d$, viz. ${4<d<d_*}$, with a surprisingly complex phase diagram as a function of measurement strength, with three distinct stable phases.

Simplifications also arise near six dimensions, where it is natural to formulate an effective field theory by promoting the overlap operator parameter $S^a S^b$ to an independent field $X^{ab}$ (for $a\neq b$):
\ba\notag
\mathcal{H} =\,&  \f{1}{2}\sum_{a} (\nabla S^a)^2 + 
 \f{1}{2}\sum_{a\neq b} (\nabla X^{ab})^2 +
 \f{m^2}{2} \sum_{a\neq b} (X^{ab})^2
 \\
& - g_1 \sum_{a\neq b} X^{ab} S^a S^b
- g_2 \sum_{\substack{a,b,c  \\  \text{distinct}}}
X^{ab}X^{bc}X^{ca}.
\label{eq:criticalisingsixminuseps}
\end{align}
The leading interaction terms allowed by $S_N\ltimes \mathbb{Z}_2^N$ symmetry are marginal in six dimensions. 
There is no mass term for $S$ because we are assuming that the initial measured Ising model is at its critical point.
In high enough dimensions we expect two stable phases, for positive and negative $m^2$ (with the latter corresponding to large $\lambda$ in the microscopic model).
The interacting theory below six dimensions will be discussed elsewhere \cite{ANWieseUnpublished}. 
The same Hamiltonian, but for a smaller number of replicas, appears in the context of spin glasses \cite{chen1977mean,le1988location,le1989varepsilon}.

The discussion above assumed that the phase diagram of the hypercubic lattice model could be continued into noninteger $d$. 
Somewhat analogous continuations can be studied directly in integer dimensions if we make the ``interactions'' long range.
In the presence of measurements there are various ways of doing this, since we can control both the interactions of the original ensemble, and the nonlocality of the measurement process.

Finally, let us substantiate the claim made above, that the strong-measurement phase cannot extend to arbitrarily small $\lambda$.

First we consider the modified measurement protocol in Sec.~\ref{sec:FKmeasurement}, in which FK clusters are measured with perfect precision. 
We saw that this protocol does not lie in the strong-measurement phase (instead it lies at a critical point).
Therefore, \textit{weak} measurement of FK clusters will certainly not lie in the strong-measurement phase either.
But, on symmetry grounds, \textit{weak} measurement of FK clusters maps to the problem considered in this section, in which the replicated Ising CFT is perturbed by small $\lambda$, together with irrelevant terms that are also small.
Therefore we infer that the continuum theory with small $\lambda$ cannot lie in the strong measurement phase either. If $d\leq 4$, so that $\lambda$ is relevant or marginally relevant, then the theory with small $\lambda$ must flow to some nontrivial fixed point.

\section{Measuring 2D free fields \& flux lines, \& charge sharpening in {1+1D}}
\label{sec:freefieldandchargesharpening}

In this Section we will discuss ``Kosterlitz-Thouless-like'' phase transitions that can occur as a result of measurement in  microscopic models whose IR description is free field theory.
The models we discuss fall into two  classes:
one where the classical configurations obey a conservation law,
such as a system of flux lines, or fully-packed dimers, and one where there is a  U(1) symmetry action on classical configurations, such as the classical XY model.  
Though these are physically different, the IR  description is the same in both cases.\footnote{Formally this is because the emergence in the IR of the enlarged $\mathrm{U}(1)\times\mathrm{U}(1)$ symmetry of the compactified free boson
erases the symmetry difference between the microscopic models (see below and App.~\ref{app:symmetrynote}).}

There is also a close relationship between these isotropic 2D problems for ``static'' equilibrium statistical mechanics, 
and  anisotropic 1+1D problems describing the dynamics of charged particles in one spatial dimension. 
The latter have a ``charge sharpening'' transition that was introduced in Ref.~\cite{agrawal2022entanglement} 
and related to field theory in Ref.~\cite{barratt2022field}.
 
After this work was completed, we also learned of rigorous mathematical results for Bayesian reconstruction of free-field configurations \cite{garban2020statistical}. This paper demonstrates the existence of two distinct phases for a  certain choice of microscopic model. (That work does not address universal properties of the phase transition).

In Sec.~\ref{eq:2Dheightlocking} we  begin with a class of static equilibrium problems. We first describe the relevant microscopic models and then the  RG flows of the resulting  field theory. 
These RG flows  require only a reinterpretation of existing results, because the replica field theory was studied in celebrated work on the  random-field XY model \cite{cardy1982random,toner1990super}
(described by $N\to 0$ instead of $N\to 1$).
Further, as a result of the decoupling \cite{barratt2022field} of the replica-symmetric mode,
the relevant sector of the replica field theory matches that for charge sharpening.
The basic features of the resulting ``Kosterlitz-Thouless-like'' transition were anticipated in Ref.~\cite{barratt2022field} even in the absence of explicit RG equations,
so the RG discussion in  
Sec.~\ref{eq:2Dheightlocking}  just adds more detail.
However, explicit consideration of the RG equations is worthwhile for several reasons.
It is interesting to note that the RG flows for the  measurement problem are qualitatively different to those for the disordered system
(the limit of strong measurement is also very different from the limit of strong disorder).
The RG equations also show that while the transitions are, loosely speaking, Kosterlitz Thouless--like, 
the detailed RG flows differ from those of the true KT problem. 
The discussion in Sec.~\ref{eq:2Dheightlocking} will also be useful preparation for the later  Sec.~\ref{sec:3dfluxes}, where we will give effective field theories and RG results for  higher-dimensional static and dynamic problems.

In the second part of the present Section (Sec.~\ref{sec:chargesharpening}) we discuss the dynamical problem of  charge-sharpening  directly.
Though charge sharpening was introduced in the context of quantum dynamics, 
the basic phenomenon is classical and can arise in a purely classical stochastic model
(indeed an effective classical model appears after  quantum-circuit averaging  in \cite{barratt2022field})
or even in a deterministic but chaotic classical system. We give a  derivation of the continuum theory for charge sharpening that relies only on  classical fluctuating  hydrodynamics.
This demonstrates the universality of the  phenomenon,
and is arguably much simpler  than the derivation  of Ref.~\cite{barratt2022field} for a circuit using particle-vortex duality.

In Sec.~\ref{sec:othermarkov}
we briefly note that  a much larger class of monitoring problems for stochastic hydrodynamic problems and classical dynamical critical phenomena can be formalized in a similar manner to Sec.~\ref{sec:chargesharpening}, by applying the replica trick to the Martin-Siggia-Rose formalism.

While the 2D measurement protocols  discussed in this section do lead to continuous phase transitions, 
the RG fixed points governing these transitions are Gaussian 
(similarly to the KT transition). Later, in Sec.~\ref{eq:interactingfixedlinefreefield}, we will show that alternative measurement protocols for free fields can give  {non-Gaussian} RG fixed points, with nontrivial  ``interactions''  due  to measurements.
(In fact some readers may prefer to start there, since the discussion of the RG equations is simpler.)

In the following Section (Sec.~\ref{sec:3dfluxes}), which is closely related to this one, 
we will analyze measurement of a family of 3D systems involving flux lines.
We formulate an effective field theory for these problems,
as well as a $2+\epsilon$ expansion.
The results are also applicable to higher-dimensional charge sharpening.
(For example, they show that in higher dimensions that transition is not related to  standard XY ordering transitions, ruling out a previous speculation.)
We again make the contrast between the measured system and  the analogous disordered system, 
which is the ``vortex glass''  \cite{fisher1989vortex}
(the effective field theory in Sec.~\ref{sec:3dfluxes} could also be applied there).

\subsection{Height fields, phase fields and flux lines in 2D}
\label{eq:2Dheightlocking}

% conventions (checked):
% S = (K/2) (d h)^2
% allowed non-random cosine term is cos(4 h) 
% <hh> = -(log r)/(2pi K)
% <(h-h)^2> = (log r)/(pi K)
% disorder coupling to e^{ih} is marginal at K=1/(4pi).

Here we will discuss 2D systems that are described in the deep-IR by free field theory,
\be\label{eq:freefieldonly}
\f{K}{2}\int \dd^2 x \, (\nabla h)^2
+ \text{irrelevant terms},
\ee
and for which the measurements involve the operators 
\be\label{eq:gradientscosines}
\partial_x h, \quad \partial_y h,  \quad 
\cos( h), \quad \sin( h).
\ee
This is natural in many settings, for example in the context of models of flux lines or dimers in the plane, or the 2D XY model (see below).

The discussion of RG flows closely parallels that for the analogous disordered systems,
so in this section we need only 
 review the RG equations for general $N$,  and  note the significant qualitative differences between the flows at ${N=0}$ and ${N=1}$.
The RG equations for Eq.~\ref{eq:coHamiltonian} below were obtained by Cardy and Ostlund \cite{cardy1982random}
(and have been applied to a range of problems, including crystal surface roughening \cite{toner1990super} and 2D vortex glasses
\cite{fisher1989vortex,hwa1994vortex,zeng1999thermodynamics,zeng2000universal}).

Note that one reason for considering all 
four observables in Eq.~\ref{eq:gradientscosines}
is that measuring 
$(\cos h, \sin h)$
is equivalent, after coarse-graining, to measuring all four observables.
This is clear in the replica approach \cite{cardy1982random}. 
More physically, if we acquire information about the phase $h$ at nearby points, then we also acquire information about its gradient.

Let us review types  of model described by Eq.~\ref{eq:freefieldonly}.

$\bullet$ 
Fully-packed dimer configurations on the square lattice map to configurations of a height field $h$. At the microscopic level $h$ is  discrete, but the coarse-grained $h$ is real-valued
\cite{kondev1994conformal,kondev1995geometrical,kondev1996kac,zeng1997zero,fradkin2004bipartite,papanikolaou2007quantum,tang2011properties,wilkins2023derivation}. The  dimer occupation number ${d_\ell=0,1}$ on  a link $\ell$
may be written as a linear combination of the continuum operators in Eq.~\ref{eq:gradientscosines}, with coefficients which depend on the sublattice \cite{zeng1997zero,fradkin2004bipartite,papanikolaou2007quantum,tang2011properties,wilkins2023derivation}; these formulas  are reviewed in App.~\ref{app:dimer}.
In the case where all dimer configurations are equally weighted then ${K=\f{1}{4\pi}}$, 
but  $K$ can be varied with interactions \cite{alet2005intdimers,alet2006intdimers}.\footnote{The action in the mid-IR also contains a term $\propto \cos(4 h)$, but this is irrelevant for $K<2/\pi$. 
If the dimers are fully packed, 
the functional integral does \textit{not} allow vortices in $h$ (App.~\ref{app:dimer}).}

$\bullet$ The example above falls into a larger class of models involving  configurations of discrete flux lines in the plane \cite{fisher1989vortex,zeng1999thermodynamics}.
For example, a well-known mapping relates dimer configurations 
to configurations of vertically directed flux line configurations: see e.g. Fig.~1 in \cite{zeng1999thermodynamics}.
In a larger class of models, 
microscopic fluxes, defined say on the links of some lattice, 
correspond to a discretized version of 
the continuum flux 
\be
{J_\mu = \epsilon_{\mu \nu} \nabla_\nu h}
\ee
that may be defined in terms of $h$.
(More precisely, in general they correspond to ${J_\mu = \epsilon_{\mu \nu} \nabla_\nu h + J_\mu^{(0)}}$ for 
some fixed background flux configuration 
$J_\mu^{(0)}$.)
The picture in terms of measuring flux lines makes clear why there is a close connection to the dynamical problem of  charge sharpening \cite{agrawal2022entanglement} (Sec.~\ref{sec:chargesharpening}).  In the latter case the relevant ``lines'' are worldlines of particles undergoing stochastic dynamics in 1+1D.

$\bullet$ Eq.~\ref{eq:freefieldonly} also describes the XY model for a spin ${\vec S=(\cos h, \sin h)}$, if vortices can be neglected
--- either because they are RG irrelevant\footnote{For vortices to be irrelevant we require $K>2/\pi$. As in the disordered system \cite{cardy1982random}, this threshold can effectively be reduced by measuring a higher harmonic, $(\cos q h, \sin q h)$ for $q>1$, instead of $\vec S$ itself.} or because they have been suppressed energetically.

Microscopically, the flux line models and the XY model differ at the level of symmetry and therefore in the  irrelevant terms in (\ref{eq:freefieldonly}).
Loosely speaking, the two kinds of models are ``dual'' to each other. 
But in the IR they have the same symmetry --- due to symmetry emergence under RG ---  
and the same free-field description, 
so they are equivalent for the purposes of the following discussion.
The analogous distinction between 
a model and its dual will become more important when we move to higher dimensions.

[In more detail:
in quantum field theory language, the 2D free field possesses a $\mathrm{U}(1)\times\mathrm{U}(1)$ symmetry.
If this free field was describing a one-dimensional quantum Hamiltonian, then these two U(1) symmetries would be exchanged by a simple change of basis, so we would regard them as equivalent.
But in a classical model there is a preferred  basis  of classical observables (i.e.\ observables that we can measure),
so the two symmetries are not equivalent.
(See App.~\ref{app:symmetrynote} for other  examples.) 
Here the observables are functions of $h$.
In terms of $h$, the first $\mathrm{U}(1)$ imposes a \textit{constraint} on configurations,  viz.\ the conservation law $\nabla \cdot J=0$,\footnote{Equivalent to absence of vortices in $h$}
while the second U(1) imposes \textit{invariance} under continuous shifts of $h$. Microscopically, the flux-line models preserve only the first U(1),
while the XY model preserves only the second. Both models acquire the full 
$\mathrm{U}(1)\times\mathrm{U}(1)$ in the IR, for appropriate microscopic couplings.]

The XY model with random field disorder coupling to $\vec S$ was analyzed in Ref.~\cite{cardy1982random}, and it was pointed out that disorder coupling to $\nabla h$ will also be generated under the RG. 
This is also true for the measurement problem:
measuring the local order parameter $\vec S$ will effectively generate measurements of the other operators in (\ref{eq:gradientscosines}).
Physically, this is just because information about $h$ at nearby points also gives us information about its gradient.

Letting the measurement strength for the cosine and sine in Eq.~\ref{eq:gradientscosines} be  $\lambda$ (cf. Eq.~\ref{eq:lambdadefn}) and that for the two gradients be $\kappa$, we obtain the replica Hamiltonian  density
\ba\label{eq:coHamiltonian}
\mathcal{H}_N =  & \, 
\f{1}{2} \sum_{ab} K_{ab}  (\nabla h^a) (\nabla h^b)
- {\lambda} \sum_{a\neq b} \cos (h^a - h^b),
%+ \lambda (N-1).
\end{align}
where the 
diagonal and off-diagonal elements of the 
symmetric matrix $K_{ab}$ are, in the limit $N\to 1$,
\ba\label{eq:K11K12values}
K_{11} &= K_{22} = \ldots = K,
& 
\,\,K_{12} &  = K_{21} =  \ldots = - 2\kappa.
\end{align}
Note that $K_{11}$ becomes independent of the measurement strength $\kappa$, a special feature of the replica Hamiltonian for the measurement problem.\footnote{For general $N$, the bare values of the couplings in the replica Hamiltonian derived from the measurement problem are
$K_{11} = K + 2 \lf\f{N-1}{N} \ri \kappa$ and 
$K_{12} = - \f{2\kappa}{N}$, and $\lambda/N$ in place of $\lambda$ in Eq.~\ref{eq:coHamiltonian}. 
For comparison, in the case of the replica Hamiltonian for the disordered system, ${K_{11} = K - \Delta_\text{dis}^2}$ and ${K_{12} = -\Delta_\text{dis}^2}$ (for any $N$), where $\Delta_\text{dis}^2$ is the variance of the random fields  coupling to $\partial_x h$ and $\partial_y h$.}

Although our focus is on the measurement problem,
it is interesting also to review the RG flows for the disordered system ($N=0$)~\cite{cardy1982random}, 
which are strikingly different.
This is because the   ${N\to 0}$ limit is more singular than the ${N\to 1}$ limit. For ${N>0}$ the natural parameterization of the couplings uses stiffnesses\footnote{${K_\text{symm}\propto K_{11}+(N-1)K_{12}}$ and 
${K_\text{asymm}\propto K_{11}-K_{12}}$.} $K_\text{symm}$ and $K_\text{asymm}$ for the replica-symmetric mode $\sum_a h^a$ and the ${N-1}$ remaining modes, respectively \cite{cardy1982random}.
Since the former decouples from the interactions, and does not renormalize, the RG equations reduce entirely to RG equations in the $(K_\text{asymm},\lambda)$ plane 
(and KT-like behavior is possible in this plane \cite{barratt2022field}).
For $N\to 0$ this parameterization is not valid, because ${K_\text{symm}\to K_\text{asymm}}$ in this limit.
Using a complete parameterization shows that the RG equations no longer reduce to equations for two variables \cite{cardy1982random}.

More concretely, the RG equations to  $O(\lambda^2)$ are \cite{cardy1982random,
goldschmidt1982field,
toner1990super,
hwa1994vortex,
carpentier1997glass,
bernard1997perturbed,
ristivojevic2012super,
perret2012super,
le2013exact,
wiese2022theory}
\ba\notag
\dot K_{11} & = (N-1) \lambda^2\,,\,
&
& \dot \lambda  = y_\lambda \lambda + A(y_\lambda) (N-2) \lambda^2,
\\
\dot K_{12} & = - \lambda^2\,, \,
&
& y_\lambda   \equiv 2- \left[ {2\pi (K_{11}-K_{12})}\right]^{-1}.
\notag
\end{align}
(They will be written in more physical variables  below.)  We will not need the value of the positive quantity $A(y_\lambda)$  in this section, but  for the $2+\epsilon$ expansion in Sec.~\ref{sec:3dfluxes} we will need the fact that
\ba
A = 2\sqrt{\pi} + O(y_\lambda)
\end{align}
when $\lambda$ is close to marginality \cite{goldschmidt1982field, carpentier1997glass, ristivojevic2012super, wiese2022theory}.

\vspace{1mm}

{\bf $\bullet$ Flows for  disordered system.} The physical parameterization is ${K_{11} = K - \Delta_\text{dis}^2}$, ${K_{12}=-\Delta_\text{dis}^2}$, where $K$ is the stiffness of the clean model and $\Delta_\text{dis}^2$ is the variance of the random vector field coupling to $\vec \nabla h$. Setting $N\to 0$ gives \cite{cardy1982random}
\ba
\f{\dd K}{\dd \tau} & = 0,
& 
\f{\dd \lambda}{\dd\tau} & = \lf 
2 - \f{1}{2\pi K}
\ri \lambda -2 A \lambda^2,
&
\f{\dd \Delta_\text{dis}^2 }{\dd \tau} &  = \lambda^2.
\end{align}
Remarkably, none of these equations feeds back into the preceding one. 
However,  $K$ does not form a decoupled sector, as it appears in the beta function for~$\lambda$.

Consider first the $(K,\lambda)$ plane.
At high temperature, i.e.\ for ${K<\f{1}{4\pi}}$, the ${\lambda=0}$ axis is stable.
This allows a Gaussian phase. 
But at lower temperature the 
${\lambda=0}$ axis is unstable. 
The low-temperature regime  is governed instead by a stable fixed line at nonzero $\lambda$, which meets the axis at the point $K=\f{1}{4\pi}$.\footnote{The Gaussian high temperature phase is really governed by a fixed plane, since both $K$ and $\Delta_\text{dis}^2$ are exactly marginal \cite{cardy1982random,toner1990super}.}

This fixed line is interesting because, although $K$ and $\lambda$ tend to fixed values, $\Delta_\text{dis}^2$ flows to infinity.
Studying correlators  \cite{toner1990super} shows that 
typical height differences ${|h(0)-h(x)|}$ grow parametrically faster than in the case of a free field. This is a ``glassy phase'', in which the field makes large excursions because of pinning by disorder.

\vspace{0.5mm}

{\bf $\bullet$ Flows for measurement problem.}
Now consider the measurement problem ($N\to1$). Expressing the RG equations  in terms of the  stiffness and the measurement strengths via Eq.~\ref{eq:K11K12values},
\ba
\f{\dd K}{\dd \tau} & \hspace{-0.3mm} =\hspace{-0.3mm}  0,
& 
\f{\dd \lambda}{\dd\tau} & \hspace{-0.3mm} = \hspace{-0.3mm} \lf \hspace{-0.3mm} 
2 - \f{1}{2\pi (K+2\kappa)} \hspace{-0.3mm} 
\ri \lambda - A \lambda^2,
&
\f{\dd \kappa}{\dd \tau} & \hspace{-0.3mm}  = \hspace{-0.3mm} \f{\lambda^2}{2} .
\end{align}
The part of the Gaussian  plane (the ${\lambda=0}$ plane) with ${K+ 2\kappa<\f{1}{4\pi}}$ is stable.
This gives a Gaussian ``weak measurement'' phase with two\footnote{Formally this Gaussian phase generalizes to one with more exactly marginal parameters, since we can (for example) choose to  measure $\partial_x h$ and $\partial_y h$ with different strengths. The resulting theory cannot be made isotropic in space:  we can rescale coordinates to make either the replica symmetric or the replica asymmetric modes isotropic, but not both.} exactly marginal parameters, $K$ and $\kappa$: for example\footnote{Higher correlators are given by Wick's theorem in the replica theory, but are equivalent to the statement that the random variable  $h(0)-h(x)$
is a sum of two independent Gaussian mean-zero pieces:
first $\<h(0)-h(x)\>_M$, which is fixed by $M$, and has variance given by  (\ref{eq:hmhsecond});
and second the residual fluctuation
${(h(0)-h(x))}-{\<h(0)-h(x)\>_M}$ 
whose variance is equal to  the difference of (\ref{eq:hmhfirst}) and (\ref{eq:hmhsecond}).}
\ba\label{eq:hmhfirst}
\big\langle \lf h(0)-h(x)\ri^2 \big\rangle & \sim \f{1}{\pi K} \ln |x|,
\\
\mathbb{E}_M \< h(0)-h(x) \>_M^{\, 2} & \sim \f{1}{\pi K}\f{2\kappa}{K+2\kappa} \ln |x|.
\label{eq:hmhsecond}
\end{align}
(Continuously varying exponents in the charge-sharpening problem are discussed in \cite{barratt2022field}.)

However, unlike the disordered system,  
once $\lambda$ becomes relevant 
in the measurement problem,
it inevitably flows out of the perturbative regime.
We expect that this leads to a trivial locked phase, in which inter-replica fluctuations --- i.e. fluctuations of ${(h^a-h^b)}$ --- become massive \cite{barratt2022field}.
For example, this is the prediction that we arrive at by  taking the RG equations at face value even for non-small $\lambda$. 
When $\lambda$ becomes non-negligible and $\kappa$ becomes large, 
we can expand around the minimum of the cosines $\cos(h^a-h^b)$, 
and we find that the inter-replica
fluctuations are massive. 

This analysis of the locked state reveals another significant difference with the disordered systems problem.
Expanding the potential to quadratic order around a replica-locked state ${h^1 = h^2 = \ldots = h^N}$ shows that the mass eigenvalue for the replica-asymmetric modes is proportional to $N$. 
So a ``trivial'' replica-locked state with massive inter-replica fluctuations is consistent for $N=1$, but not for $N=0$. 
In the disordered system, the ``glassy'' phase takes the place of the locked phase.

Summarizing, in the measurement problem we have a transition between a Gaussian weak-measurement phase and a trivial replica-locked phase. Slightly disappointingly, this example did not give an interacting RG fixed point. However, later we will see that other closely-related 2D theories do lead to new interacting fixed points/lines (Sec.~\ref{eq:interactingfixedlinefreefield}).

\subsection{Charge sharpening for diffusing particles}
\label{sec:chargesharpening}

We now discuss a dynamical problem that is closely related to the static problems in Sec.~\ref{eq:2Dheightlocking}, giving  a new derivation of the effective field theory for charge sharpening that starts from classical continuum hydrodynamics (see the top of Sec.~\ref{sec:freefieldandchargesharpening} for further background).

Consider a one-dimensional system of particles undergoing classical stochastic motion, either on the lattice or in the continuum. 
A very broad range of such systems are described at large scales by a noisy diffusion equation \cite{hohenberg1977theory,bertini2015macroscopic}. 
To begin with, consider the fluctuations, on top of a homogeneous background density   $n_0$, of a continuum density  ${n(x,t)=n_0+\rho(x,t)}$:
\ba\label{eq:noisydiffusion}
\partial_t \rho(x,t) & = D \partial_x^2 \rho(x,t) + 
\partial_x \eta(x,t),
\\  \label{eq:noisydiffusion2}
\< \eta(x,t)\eta(x',t')\> 
& = \kappa \delta(x-x')\delta(t-t').
\end{align}
The diffusion constant $D$ and noise strength $\kappa$ are in general density-dependent, but we can take them to be constant (i.e.\ evaluate them at the constant density $n_0$) since
higher-order terms in $\rho$ are irrelevant in the RG sense.

We now imagine 
that the density is made up of discrete particles, 
and we are making local density measurements in such a system. 
The discreteness of the microscopic particles is important:
the IR description encoded in Eq.~\ref{eq:noisydiffusion} remains valid,  
but discreteness leads to a slightly more nontrivial relation between the 
microscopic density and the   coarse-grained field, 
of the kind which is familiar from bosonization \cite{haldane1981effective}.

Let us first set up the field theory in which these operators will be defined.
Eq.~\ref{eq:noisydiffusion} maps to a field theory via the Martin-Siggia-Rose formalism \cite{cardy1996scaling}. It is convenient to work not with $\rho$, but with the height field or ``counting field'' $h(x,t)$, satisfying $h(0,t)=0$ and
\ba\label{eq:countingfield}
\rho(x,t) & =\f{1}{2\pi} \partial_x h(x,t),
\end{align}
so that (\ref{eq:noisydiffusion}) is 
\be\label{eq:countingfield2}
\partial_t h  = D \partial_x^2 h + 2\pi \eta.
\ee
Then the Martin-Siggia-Rose trick (Sec.~\ref{sec:othermarkov})
gives the effective ``Hamiltonian'' (see App.~\ref{app:diffusion})
\be\label{eq:MSRaction}
\mathcal{H} = 
\f{1}{8\pi^2 \kappa} \int \dd x \dd t \left[
(\partial_t h)^2 + D^2 (\partial_x^2 h)^2
\right]
+ \mathcal{H}_B,
\ee
where $\mathcal{H}_B$ is a boundary term 
(App.~\ref{app:diffusion}) 
 that is important in defining the ``direction of time'' but which
can be neglected for the purposes of studying bulk RG flows.
This ``Hamiltonian'' yields a functional integral $\int_h e^{-\mathcal{H}}$  that integrates over spacetime histories of the stochastic process with the correct probabilities; therefore the formalism in Sec.~\ref{sec:generalities}  goes through essentially unchanged. See  Sec.~\ref{sec:othermarkov} for more detail, where we describe the analogous formalism for stochastic field theories in a more general setting.

The present ``Hamiltonian'' is anisotropic in spacetime, but since it plays the same role in defining the probability measure for the field as does the Hamiltonian for the equilibrium problems in previous sections, the measured operators appear in the replica theory in precisely the same way (Sec.~\ref{sec:generalities}).

The relation between the microscopic density and the coarse-grained counting field has been understood in the context of bosonization \cite{haldane1981effective} 
(we give a brief recap in App.~\ref{app:diffusion}). 
For a broad class of models (either on a 1D lattice or in the spatial continuum, but evolving in continuous time)
\be\label{eq:nmicromaintext}
n_\text{micro} \simeq n_0 + \f{1}{2\pi} \partial_x  h + B \cos  \big(  h + 2\pi n_0 x \big),
\ee
where $B$ is a nonuniversal constant, and we have dropped less relevant terms.
Taking $n_\text{micro}$ to be the measured operator, and using the general expression in Eq.~\ref{eq:repeatreplicaHforRG} (where the limit $N\to 1$ for the coefficient has been taken),
the perturbation to the replica ``Hamiltonian'' density
due to measurement is 
 ${\delta\mathcal{H} = -\lambda \int \sum_{a\neq b} n_\text{micro}^a(x) n_\text{micro}^b(x)}$.
Expanding this expression out using (\ref{eq:nmicromaintext}), terms  that oscillate in $x$ can be neglected
(on large enough scales)
since they lead to irrelevant corrections.
If the microscopic model is in continuous space, this leads to
\be
\delta \mathcal{H} \to -\f{\lambda}{4\pi^2}  \sum_{a\neq b} (\partial_x h^a) (\partial_x h^b)
- \f{\lambda B^2}{2} \sum_{a\neq b} \cos (h^a - h^b).
\ee
If the model is a lattice model, half-filling is a slightly special case because additional terms survive phase cancellation:
\ba\notag
\delta \mathcal{H} 
\to &  -\f{\lambda}{4\pi^2}  \sum_{a\neq b} (\partial_x h^a) (\partial_x h^b)
\\
& - {\lambda B^2} \sum_{a\neq b} \left[ \cos (h^a-h^b)
+
\cos (h^a + h^b) \right].
\end{align}
However, the difference will not matter for the universal behavior, because the term $\cos(h^a+h^b)$ 
is irrelevant.

The resulting replica action
\ba\label{eq:MSRactionReplica}\notag
\mathcal{H}  
=  & \,
\f{1}{8\pi^2 \kappa}\sum_a
 \left[
(\partial_t h^a)^2 + D^2 (\partial_x^2 h^a)^2
\right]\\
&
- \f{\lambda}{4\pi^2} \sum_{a\neq b} (\partial_x h^a) (\partial_x h^b) 
- \f{\lambda B^2}{2}\sum_{a\neq b}
\cos(h^a - h^b)
\end{align}
is  equivalent 
to that first obtained in a discrete circuit model  by Barratt et al.~\cite{barratt2022field},
up to nonuniversal differences.
(For example, in contrast to the Lagrangian in Ref.~\cite{barratt2022field},
Eq.~\ref{eq:MSRactionReplica} does not contain the term
$\sum_{a\neq b} \partial_t h^a \partial_t h^b$,
but this is a nonuniversal difference, as this term will be generated under RG when Eq.~\ref{eq:MSRactionReplica} is coarse-grained.)
The present hydrodynamic derivation makes evident the universality of the charge-sharpening phenomenon 
for classical fluctuating hydrodynamics
and is a starting point for generalizations in~Sec.~\ref{sec:othermarkov}.

As discussed in Ref.~\cite{barratt2022field}, the replica-symmetric mode decouples (and has diffusive scaling). 
For the replica-asymmetric modes the term with four spatial derivatives is irrelevant at large scales (for small $\lambda$).
The action in this sector shows (roughly speaking) Kosterlitz-Thouless-like flows, 
leading to the charge-sharpening transition \cite{barratt2022field}.
We have encountered these flows in Sec.~\ref{eq:2Dheightlocking}.\footnote{The version of the charge sharpening problem with ``forced'' measurements is described by an $N\to 0$ limit, and may show the Cardy-Ostlund universal behavior, with a glassy phase.}
The structure of the  flows differs in detail from those of the true Kosterlitz-Thouless problem:  unlike in  Kosterlitz-Thouless, there is no symmetry $\lambda\to - \lambda$, so that, for example, a term of order $\lambda^2$ appears  in the beta function for~$\lambda$.

\subsection{Other stochastic field theories}
\label{sec:othermarkov}

The inference problem in the previous Section concerned a particular problem in fluctuating hydrodynamics.  We note briefly that the formalism extends to more general stochastic field theories, which could provide interesting examples for future study.

For concreteness, we assume the stochastic dynamics is given by an Itô differential equation of the schematic form
\ba
\partial_t  S = f[S]+ g[S] \eta,
\end{align}
where $\eta$ is white noise of strength $\kappa=1$ (compare Eq.~\ref{eq:noisydiffusion2}).
Depending on the interpretation of $S$ this could represent, for example, one of the Hohenberg--Halperin dynamical universality classes \cite{halperin2019theory}, or a nonequilibrium stochastic process such as directed percolation \cite{janssen2005field} or surface growth.

The Martin-Siggia-Rose approach writes the correctly-weighted
function integral over field configurations  using an (imaginary)  auxiliary field $\widetilde S$ to enforce the equation of motion  \cite{cardy1996scaling}
\be
Z = \int_{S, \widetilde S, \eta} \, e^{-
\int \dd^d x \dd t \lf 
\f{1}{2}\eta^2 
+ \widetilde S \dot S
- \widetilde S f[S]-\widetilde S g[S]\eta
\ri
},
\ee
or, after integrating out $\eta$,
$Z=\int_{S,\widetilde S}e^{-\mathcal{H}}$
with 
\be
\mathcal{H}[S,\widetilde S]=\widetilde S \lf \dot S - f[S] \ri + \f{1}{2} \widetilde S^2 g(S)^2.
\ee
The theory in the previous Section was particularly simple because  $\widetilde S$ could also be integrated out,  leaving an effective  ``Hamiltonian''  for the physical field $S$ alone
(denoted by $h$ above).

The replica treatment goes through as in the case of a spatial ensemble. 
The measured operator is a function $\mathcal{O}[S]$ of the physical field $S$ (but not of the response field $\widetilde S$), so by (\ref{eq:Hreplica})
\be
\mathcal{H}_N = 
\sum_a\mathcal{H}[S^a,\widetilde S^a]
+ \f{\lambda}{N} \sum_{a< b} \lf
\mathcal{O}[S^a] - \mathcal{O}[S^b]
\ri^2.
\ee
It will be interesting in the future to study the RG for a  wider range of such theories.

\section{Measuring dimers and flux lines in 3D}
\label{sec:3dfluxes}

We return to equilibrium statistical mechanics models, and consider 3D generalizations of the 2D problems in Sec.~\ref{eq:2Dheightlocking}.

As mentioned in Sec.~\ref{eq:2Dheightlocking}, we are able to think of the 2D free field theory as describing ensembles of 
discrete integer-valued (divergence-free) currents in the plane --- or, equivalently, ensembles of oriented ``flux lines'' with short range interactions.
We now consider imaging flux lines in  3D, either on the lattice or in the continuum.

A physical example of such an ensemble is the ensemble of flux lines (vortex lines)
in a type II superconductor above $H_{c1}$ \cite{fisher1989vortex}.
Another example is spin ice, which maps to an Ising-like antiferromagnet on the pyrochlore lattice and whose configurations may be mapped to flux loop configurations  (see Refs.~\cite{chalker2017spin,henley2010coulomb} for reviews).
A final concrete lattice model is the fully-packed dimer model on the cubic lattice \cite{huse2003coulomb,alet2006unconventional,powell20082,charrier2008gauge,chen2009coulomb,sreejith2019emergent}.

For these measurement problems, an analogous disordered system (${N\to 0}$ instead of ${N\to 1}$) is the ``vortex glass'' problem for disordered superconductors introduced by Fisher \cite{fisher1989vortex}.
The gauge-Higgs model below should
 apply to both problems. 
(Indeed this is simply the natural effective field theory to write down for vortices, 
given the comments in Ref.~\cite{fisher1989vortex} about an off-diagonal condensate.)
Despite the unification via the $N$-dependent  effective Hamiltonian, we will show
(using a $2+\epsilon$ calculation and results directly in 3D)
that  the RG flows for the disordered system and for the measurement problem are very different, as they were also in the 2D case (Sec.~\ref{eq:2Dheightlocking}).

\subsection{Effective field theory}

\label{sec:3DdimerEFT}

An example of a 3D lattice model in the relevant class, which is simpler than the dimer model, is a model of  fluxes $J_{ij}\in \mathbb{Z}$ that live on the links $(i,j)$ of the cubic lattice ($J_{ij}=-J_{ij}$).
The divergence-free constraint ${(\nabla \cdot J)_i \equiv \sum_j J_{ij}=0}$ is imposed, and we  take the energy to be ${K_\text{latt} \sum_{\<ij\>}J_{ij}^2}$. For small enough $K_\text{latt}$, this model is in a stable power-law-correlated phase known as the ``Coulomb'' phase, whose IR description is  free U(1) gauge theory \cite{youngblood1981polarization,
henley2010coulomb,
huse2003coulomb,henley2005power}:
\ba\label{eq:coulomb}
\mathcal{H} & = 
\f{K}{2} \int \dd^3 x \, 
(\nabla\times A)^2,
& 
J = (\nabla\times A).
\end{align}
Here ${J=(J_x,J_y, J_z)}$ is the coarse-grained current, and $K$ is a renormalized stiffness.\footnote{If $K_\text{latt}$ becomes too large, the model can leave the Coulomb phase (and enter a trivial short-range correlated phase) via a Higgs transition in which a matter field condenses (the dimer model has more intricate analogs \cite{powell20082,charrier2008gauge,chen2009coulomb,sreejith2019emergent}). This matter field is omitted in Eq.~\ref{eq:coulomb} because it is massive in the Coulomb phase.}
The Coulomb phase has a  dual description  as the ordered phase of an XY model \cite{wen2004quantum}, but the gauge description is more useful for us at the moment.

Now consider measuring the lattice currents $J_{ij}$. We claim that the phase diagram is captured by a  replica Hamiltonian density for a U(1)$^N$ gauge theory
 for the replicated gauge field,
 with $N(N-1)/2$ matter fields $z_{ab}$, each coupling to two gauge fields:
\ba\label{eq:replicahiggs}
\notag
\mathcal{H}_N  & =
\sum_{a}  \f{K}{2}   (\nabla\times A^a)^2
- \sum_{a\neq b}  \f{K'}{2}  (\nabla\times A^a)(\nabla\times A^b)
\\ 
& + 
\f{1}{2} \sum_{a<b} \left|
\lf \nabla - i A^a + iA^b \ri  z_{ab} 
\right|^2
+ \f{m^2}{2} \left| z_{ab} \right|^2 + V.
\end{align}
$V$ is a  potential for $z_{ab}$ (see  below).
For  positive $N$ the symmetric combination of the gauge fields, ${A^\text{symm}=\sum_a A^a}$, decouples from the other fields, leaving a U(1)$^{N-1}$ gauge theory.

This theory reproduces the expected phases if we assume that weak measurement corresponds to positive $m^2$ (and small $K'$) and strong measurement corresponds to negative $m^2$ (and large $K'$):

$\bullet$ When $m^2$ is large and positive we can integrate out the matter fields, giving a finite renormalization of $K'$. 
The resulting free theory matches what we obtain (by the general formalism, Sec.~\ref{sec:replicaformalism}) if we make weak measurements of $J$ in the continuum theory (\ref{eq:coulomb}).
This is the weak measurement phase. 
It can be viewed as a line of fixed points, parameterized by the dimensionless ratio ${r=K'/K}$. Moving along this line varies the relative strength of the Coulombic \cite{chalker2017spin,henley2010coulomb} intra and inter-replica correlations:
\ba \label{eq:Coulombcorrelators}
 {\< J_\mu(x) J_\nu (y)\>}& = \f{3 x_\mu x_\nu - \delta_{\mu\nu} x^2}{4\pi K x^5},
 \\ \notag
  {\mathbb{E}_M \< J_\mu(x) \>_M \<J_\nu (y)\>_M} & =\f{3 x_\mu x_\nu - \delta_{\mu\nu} x^2}{4\pi K x^5}  \f{r}{1+r} .
  \end{align}

$\bullet$ When $m^2$ is large and negative we condense the matter fields, giving a Higgs mass for the relative fluctuations of the gauge fields.
Crucially, one may check that the squared-mass for these fluctuations (the eigenvalue of the mass matrix) is proportional to $N$. 
Since this remains finite at $N=1$,
there is a stable phase in which the only
non-massive field is $A^\text{symm}$.
This is the expected strong measurement phase, where there are no nontrivial current fluctuations left after we condition on the measurement.\footnote{That is, the fluctuation
on top of the conditioned average,
 ${J-\<J\>_M}$, has exponentially decaying correlations, in contrast to the weak measurement phase, where (\ref{eq:Coulombcorrelators}) shows that the two-point correlator of  ${J-\<J\>_M}$ 
 is a power law with prefactor proportional to $1/(1+r)$. }

We might hope to study the transition between weak and strong measurement phases, at renormalized mass-squared ${m^2=0}$,
using  RG for  Eq.~\ref{eq:replicahiggs}.
The leading term in the potential $V$ is cubic, $z_{ab}z_{bc} z_{ca}$ (we let $z_{ba}=z_{ab}^*$), which means that the  gauge-matter couplings and the cubic term do not have the same upper critical dimensionality (4 and 6 respectively), making an $\epsilon$ expansion challenging.
It might be possible to apply perturbative or nonperturbative \cite{dupuis2021nonperturbative} RG directly in 3D.  (In the next Section we discuss an alternative approach in $2+\epsilon$ dimensions.)
The gauge theory is interesting for arbitrary $N$ as a simple ``quiver'' gauge theory, where each matter field carries positive charge under one of the gauge groups and negative charge under another.

To end this Section, let us briefly contrast the measurement problem
with the quenched disorder (vortex glass) problem.
We expect that they are  described by the ${N\to 1}$ and ${N\to 0}$ limits of the gauge-Higgs model in
Eq.~\ref{eq:replicahiggs}.
 
For ${N\to 1}$, there is a stable phase in which the inter-replica gauge fields are Higgsed.
However there is no such stable phase for ${N\to 0}$, because as noted above the Higgs mass for these fluctuations is proportional to $N$.  
A priori this phase could be replaced by a nontrivial stable glass phase, but in fact numerics \cite{bokil1995absence,kisker1998application,
pfeiffer1999numerical} suggest that the quenched disorder problem has a unique trivial phase for nonzero temperature (this is also consistent with the $2+\epsilon$ expansion below).

Therefore the phase diagram structure  conjectured for pinned vortices in Ref.~\cite{fisher1989vortex}, 
although not applicable in the original setting, turns out to be applicable to a different problem, namely  measurement of flux lines in a system without quenched disorder.

\subsection{Flux lines in $2+\epsilon$ dimensions}
\label{sec:fluxlinestwopluseps}

There are fundamental differences between disorder problems and measurement problems
(Sec.~\ref{sec:generalitiesweak})
which are easily  missed if we think of them simply as different limits for $N$.
This can be illustrated with a very simple  example. Consider starting with an XY model in its \textit{ordered} phase in 3D.

First imagine coupling very weak quenched disorder to the XY order parameter. This has a radical effect: by the Imry-Ma argument, the ordered phase is completely destroyed.
Vortices will proliferate, taking us  into   a trivial phase that is  smoothly connected to the limit where every spin is pinned by its local field.
(We can also consider a  modification of the XY model, with the constraint that  vortices are not allowed; in this case weak disorder takes us into a nontrivial power-law correlated phase, the Bragg glass~\cite{nattermann1990scaling,
korshunov1993replica,
giamarchi1994elastic,
giamarchi1995elastic,
fisher1997stability,
nattermann2000vortex}.)

Compare this to the problem of imperfect (weak) measurement of the clean system.
At first glance the replica Hamiltonian looks similar, but this is of course misleading. The measurements cannot  have any effect on the long-range order. Measurements are strongly relevant, and fully reveal the coarse-grained configuration. Formally, the inter-replica fluctuations of the Goldstone mode become trivially massive (vortices are not induced). 

This physics has very little to do with that of the disorder problem.  
Below we discuss another setting where the topology of the flows differs between the measurement and quenched disorder problems.

In Sec.~\ref{eq:2Dheightlocking} we considered measurement of  $(\cos h, \sin h)$ and $(\partial_x h, \partial_y h)$, for a 2D free field $h$, and we contrasted the behavior of the ${N\to 1}$ replica theory  with that of the ${N\to 0}$ theory obtained by  coupling disorder to these operators.
We now briefly consider ${2+\epsilon}$ expansions related to these problems \cite{ANunpublished}.

Interestingly, the 2D classical ensemble for $h$ can be obtained  as a ${d\to 2}$  limit of either of   two  classes of theory,
which coincide  in 2D, but are \textit{distinct} in general $d$.

$\bullet$ {\bf Measuring phase field in $d$ dimensions.}
The first class involves a vortex-free phase field $h$ with measurements, or disorder, coupling to ${(\cos h, \sin h)}$.
The regime we will talk about is where the initial ensemble is at large stiffness, so  this is the simple problem discussed above, where we start with a long-range-ordered XY model. 
As noted above, in the case of measurements this just induces a a trivial  mass for
inter-replica Goldstone fluctuations.
However it gives a nontrivial state for quenched disorder~\cite{nattermann1990scaling,
korshunov1993replica,
giamarchi1994elastic,
giamarchi1995elastic,
fisher1997stability,
nattermann2000vortex}. 

One can consider the ${2+\epsilon}$ expansion for these problems based on (\ref{eq:coHamiltonian}) \cite{goldschmidt1985xy,ANunpublished}.
Working close to the point where $\lambda$ is marginal,\footnote{\label{footnote:epsilonscaling} We expand in $\epsilon$ in the regime where  $\lambda=O(\sqrt{\epsilon})$ and 
${(K_{11}-K_{12})=\f{1}{4\pi}+O(\sqrt{\epsilon})}$.
To lowest order in $\epsilon$, the only change to the Cardy-Ostlund RG equations is that the stiffnesses acquire an RG eigenvalue $\epsilon$.}
one finds that the flows lead out of the perturbative regime, to large values of the stiffnesses. 
In the case of measurement we interpret this as flow to the trivial, replica-locked phase. 
In the case of quenched disorder
a trivial locked phase is not consistent\footnote{By logic similar to that discussed for the Higgs mass in Sec.~\ref{sec:3dfluxes}.} and we interpret it as flow to the Bragg glass. 

$\bullet$ {\bf Measurement of flux lines.} The second class of problems, which is of more interest for us here, involves ensembles of discrete flux lines.
We have encountered these problems in 2D (Sec.~\ref{eq:2Dheightlocking}) and 3D (Sec.~\ref{sec:3dfluxes}).
Microscopically we may have, for example,  integer-valued fluxes   
defined on the links of a lattice, satisfying the lattice version of $\nabla \cdot J=0$. 

We have already discussed one way to formulate a continuum description, which is to resolve the divergence-free constraint, writing
\ba\notag
J_\mu = \epsilon_{\mu\nu} \nabla_\nu h
\quad \text{in 2D},
\qquad
J_\mu = \epsilon_{\mu\nu\lambda} \nabla_\nu A_\lambda
\quad \text{in 3D}
\end{align}
(or their lattice analogs), and so on. These fields are massless in the absence of measurement.
Replica field theories in the presence of measurement were discussed in  Secs.~\ref{eq:2Dheightlocking} and~\ref{sec:3dfluxes}.
At weak measurement, the $N$ replicated fields ($\{h^a\}_{a=1}^N$ or $\{A^a_\mu\}_{a=1}^N$) all remain massless in the IR.
At strong measurement, 
replicas are ``locked'' together, and only a single, replica-symmetric mode remains massless.
The locking is effected by ${\cos(h^a-h^b)}$ terms in the 2D case \cite{barratt2022field,cardy1982random} 
and by condensation of Higgs fields $z^{ab}$ in the 3D case. 

In this language, the effective field theories for flux lines look quite different in 2D and 3D
(with a scalar field in 2D and a gauge field in 3D). 

To unify them, we can pass to the dual language \cite{wen2004quantum}.
Physically, this means that we think of the microscopic flux lines as \textit{worldlines} for a complex scalar field ${\psi\sim\exp i \theta}$.
More formally, we interpret $J_\mu$ is the U(1) conserved current for this scalar. 

In outline, the weak-measurement phase is one where 
there is separate long-range\footnote{In 2D there is only quasi-long-range order.} order ${\<\psi^a \> \neq 0}$ in each replica,
while in the strong-measurement phase 
there is long-range order only in a composite field made from all the replicas:  ${\< \psi^1\cdots \psi^N\>\neq 0}$. 

[In more detail: a heuristic way to understand this is by noting that long-range order in $\psi$ corresponds to a proliferation of worldlines.
In the weak-measurement phase, where replicas are not locked, worldlines proliferate in each of the $N$ replicas.
In the strong-measurement phase, 
locking means that the only kind of line that can proliferate is a multi-strand made up of $N$ worldlines, one from each replica. The proliferation of such multi-strands is long-range order for ${\psi^1\cdots\psi^N}$.]

In the weak-measurement phase, long-range order for each of the $\psi^a$ means that there are $N$ free Goldstone modes $\theta^a$, 
which are dual to the free fields $h^a$ (in 2D) or $A^a_\mu$ (in 3D).
In the strong-measurement phase there is only a single Goldstone mode ${\sum_a \theta^a}$.
In 2D this  is dual to the replica-symmetric height field ${\sum_a h^a}$ that is not locked by the cosine term, and in 3D it is dual to the replica-symmetric gauge field ${\sum_a A^a_\mu}$ that does not get Higgsed.

We can be more concrete in 2D. 
Dualizing the replica field theory (\ref{eq:coHamiltonian}) gives the formal Hamiltonian
\be\label{eq:HCOdual}
\mathcal{H}_\text{dual} = 
\f{1}{2} (K^{-1})_{ab} (\nabla \theta^a) (\nabla \theta^b) -2  \lambda \sum_{a\neq b} V_{ab}.
\ee
Here $V_{ab}=V_{ba}^*$ is an operator that inserts a relative vortex \cite{barratt2022field}, i.e.\ a vortex in replica $a$ and an antivortex in replica $b$. That is, $\lambda$ is interpreted as a fugacity for such vortices. 

In higher dimensions vortices are no longer point objects and cannot be associated with a local operator. 
Nevertheless, can we use (\ref{eq:HCOdual}) as the starting point for a $2+\epsilon$ expansion?
Ref.~\cite{nelson1977dynamics} used the Kosterlitz-Thouless RG equations as the starting point for a $2+\epsilon$ expansion for the XY model, which was extended  to the $\mathrm{O}(n)$ model in Ref.~\cite{cardy1980n}.
Such approaches are based on the assumption  that  
(a) the RG fixed point of interest has a meaningful continuation in $d$, and (b) the RG equations can be expanded in $\epsilon$.
It is not clear whether these assumptions are generally valid.
However, this criticism can be made about $\epsilon$ expansions more generally. It is interesting to explore a $2+\epsilon$ expansion of the present RG equations too \cite{ANunpublished}.

The assumption that the RG equations can be power-expanded in $\epsilon$ fixes the leading terms (cf. footnote~\ref{footnote:epsilonscaling}):
\ba\label{eq:COepsilonRGeqsgen}
\dot K_{11}\hspace{-0.2mm} & = -\epsilon K_{11} \hspace{-0.2mm}+ \hspace{-0.2mm}(\hspace{-0.2mm}N\hspace{-0.2mm}-\hspace{-0.2mm}1\hspace{-0.2mm}) \lambda^2,
&
& \dot \lambda  = y_\lambda \lambda \hspace{-0.2mm}+\hspace{-0.2mm} A (\hspace{-0.2mm}N\hspace{-0.2mm}-\hspace{-0.2mm}2\hspace{-0.2mm}) \lambda^2,
\\
\dot K_{12}\hspace{-0.2mm} & = - \epsilon K_{12}\hspace{-0.2mm} - \hspace{-0.2mm}\lambda^2, 
&
& y_\lambda   \equiv 2\hspace{-0.2mm}-\hspace{-0.2mm} \left[ {2\pi (\hspace{-0.2mm}K_{11}\hspace{-0.2mm}-\hspace{-0.2mm}K_{12}\hspace{-0.2mm})}\right]^{-1}
\notag
\end{align}
(notation in Sec.~\ref{eq:2Dheightlocking}).

These equations differ from the  ${2+\epsilon}$ expansion for 
the field $h$, with a cosine term, by the sign of the $\epsilon$ term.
This sign change is related to the fact that  duality exchanges $K$ and $K^{-1}$~(\ref{eq:HCOdual}).

For the flux-line measurement problem (${N\to 1}$), the replica-symmetric stiffness $K$ decouples and,  setting ${K_\text{asymm} = K_{11}-K_{12} =K + 2\kappa}$,
\ba\label{eq:lambdaeq2plusepsilon}
\f{\dd \lambda}{\dd\tau} & = \lf 
2 - \f{1}{2\pi K_\text{asymm}}
\ri \lambda - A \lambda^2,
\\
\f{\dd K_\text{asymm}}{\dd \tau} &  = - \epsilon K_\text{asymm} +  \lambda^2 .
\end{align}
This gives a fixed point at $K_\text{asymm}\simeq \f{1}{4\pi} + \f{A\sqrt{\epsilon}}{16\pi^{3/2}}$, $\lambda \simeq \f{\sqrt{\epsilon}}{2\sqrt{\pi}}$,
with one unstable direction, and  corresponding correlation length exponent $\nu$ given by  (recall $A=2\sqrt{\pi}$)
\be
\nu^{-1} 
\simeq \f{\sqrt{A^2+64 \pi}-A}{4\sqrt{\pi}}\sqrt{\epsilon}
\simeq 1.56 \sqrt{\epsilon}.
\ee

The theory above may also apply to charge sharpening in $1+\epsilon$ spatial dimensions, see Sec.~\ref{sec:chargesharpeninghigherd}.

Note that the universality class of the above transition differs from that of the XY nodel. For the Kosterlitz-Thouless RG equations of the XY model
(extended to $2+\epsilon$ dimensions \cite{nelson1977dynamics})
the $O(\lambda^2)$ term in Eq.~\ref{eq:lambdaeq2plusepsilon} is not present, leading to a smaller correlation length exponent,  $\nu^{-1}\simeq 2 \sqrt{\epsilon}$.

Finally, let us comment on the quenched disorder problem ($N\to 0$), which corresponds here to pinning of vortex lines \cite{fisher1989vortex}.
As for the Bragg-glass problem mentioned above (a random phase field $h$ coupled to quenched disorder), in $2+\epsilon$ the flows exit the perturbative regime where the expansion is formally valid.
However, whereas for the Bragg glass the
stiffness of $h$ and the strength of the disorder coupling to $\nabla h$ flow to larger values, 
 the opposite happens in  the ``vortex glass'':
the stiffness for $J_\mu$ and the disorder coupling to $J_\mu$ flow to smaller values.

This difference arises from the fact that, while these two kinds of disordered system are dual to each other in 2D \cite{fisher1989vortex}, they are no longer dual to each other in ${2+\epsilon}$.
This is a heuristic explanation for the numerically-demonstrated lack \cite{bokil1995absence,kisker1998application,pfeiffer1999numerical} of a stable vortex glass phase in three dimensions.

\subsection{Charge sharpening in $1+\epsilon$ and 2 dimensions}
\label{sec:chargesharpeninghigherd}

The field theories proposed above can also be adapted to charge sharpening in higher dimensions.

First, adapting the discussion in Sec.~\ref{sec:chargesharpening},
we may write ${\rho\propto (\partial_x A_y - \partial_y A_x)}$ for a gauge field $A$ (in the Weyl gauge $A_t=0$).
We may then expect an  anisotropic version of Eq.~\ref{eq:replicahiggs} 
in which the replica--symmetric $A^\text{symm}$ mode has $z=2$ (diffusive) scaling. 

Since $A^\text{symm}$ naively decouples, at first sight we might expect the same critical behavior to apply for the problems in the previous section and for charge sharpening. However, as noted in \cite{barratt2022field} (see also \cite{ha2024measurement}), 
in the anisotropic case one must consider additional symmetry-allowed terms coupling the sectors, in particular $\rho \mathcal{E}$, where $\rho\sim (\nabla \times A_\text{symm})_t$ is diffusive density and ${\mathcal{E}\sim \sum_{a\neq b}|z_{ab}|^2}$ is most relevant operator of ${z=1}$ sector. 
This is irrelevant only if ${\nu > 2/d}$, where $d$ is the \textit{spatial} dimensionality in ${d+1}$ dimensions.

However, Ref.~\cite{barratt2022field} also pointed out that in a lattice model with particle-hole symmetry the condition for irrelevance is relaxed
(the particle-hole symmetry $\rho\to - \rho$ forbids $\rho\mathcal{E}$ but allows $\rho^2 \mathcal{E}$).
By a heuristic argument we find the weaker condition ${\nu > 4/(3d)}$ for irrelevance in this case.\footnote{This differs from  the $\nu>1/d$ stated in
\cite{barratt2022field} so further examination is warranted.}
In 2+1D, the leading-order estimate from our $\epsilon$-expansion in Sec.~\ref{sec:fluxlinestwopluseps} is not far from the threshold for this weaker condition. 
Therefore it could be that the two sectors decouple even in the dynamical problem. This requires further investigation.

In \cite{barratt2022field} it was suggested that the charge-sharpening transition in higher $d$ may have XY exponents.
The results of the previous sections show that the transition is 
not an XY transition, even for $d=1+\epsilon$.
As a result, the exact value of  $\nu$ is not known in $2+1$D.
While the dual formulation (\ref{eq:HCOdual}) is expressed in terms of angular degrees of freedom $\theta^a$,  the fact that the relevant vortices in that theory 
(which become vortex lines in 2+1D) 
are  vortices in the relative phase of two angles means that we are very far from a Landau-theory-like regime described by coupled XY models for the corresponding ``order parameters'' ${\psi^a \sim e^{i\theta^a}}$.\footnote{For example: if we neglect fluctuations of the replica-symmetric mode, we may parameterize the theory by ${N-1}$ independent complex unit vectors ${\psi^a  = \exp(i\theta^a)}$
(Sec.~\ref{sec:fluxlinestwopluseps})
for ${a=1,\ldots,{N-1}}$, 
so that ${\psi_N = (\psi^1\cdots \psi^{N-1})^*}$.
In principle one might try to formulate a soft-spin Landau theory for ${\psi^1,\ldots, \psi^{N-1}}$. However the full symmetry of the theory  (which is hidden by the representation in terms of $\psi^1,\ldots, \psi^{N-1}$) shows that $\psi^1$ and ${(\psi^1\cdots \psi^{N-1})}$ have the same scaling dimension.
This is not possible at a ``perturbative'' fixed point at small values of the  Landau  couplings, 
where the scaling dimension of 
${\psi^1\cdots \psi^{N-1}}$ is close to $(N-1)$ times the scaling dimension of~$\psi^1$.}

\section{A nontrivial RG fixed line from measuring free fields}
\label{eq:interactingfixedlinefreefield}

\begin{figure}
    \centering
\includegraphics[width=0.95\columnwidth]{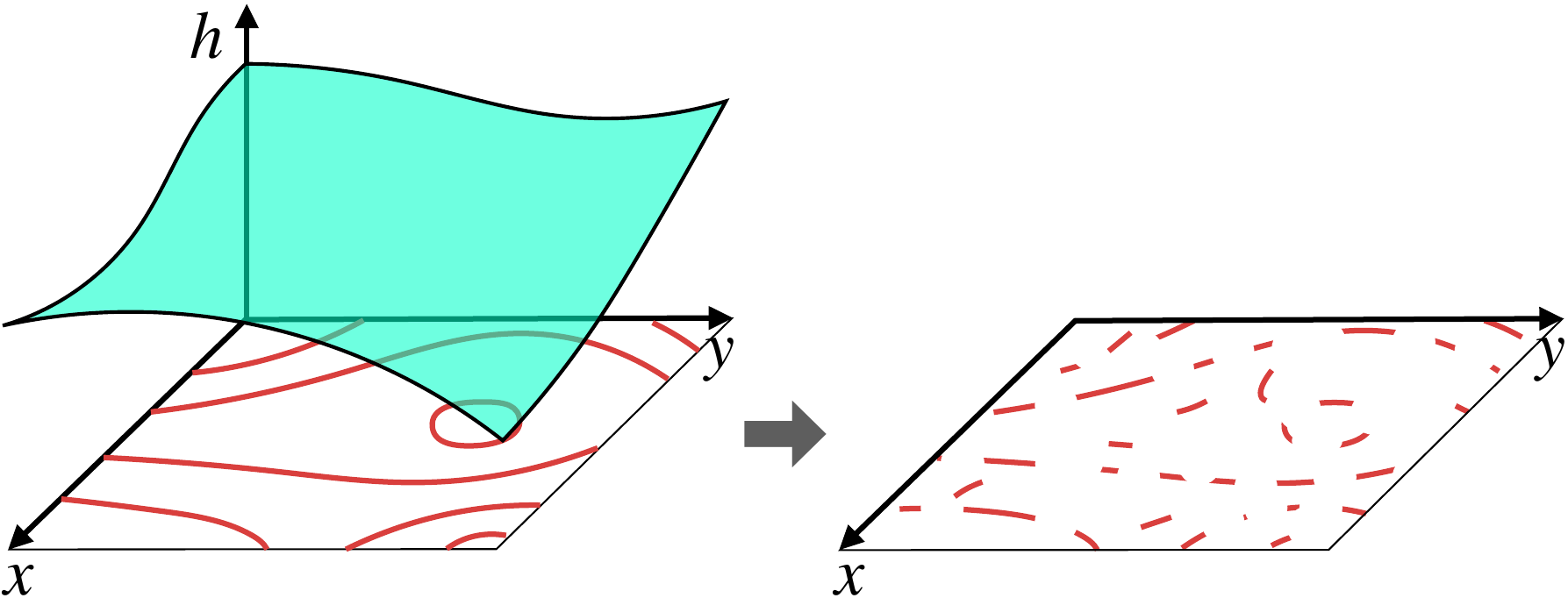}
    \caption{Imperfect measurement of contour lines of a height field (not intended to  accurately represent protocol in text).}
\label{fig:contourmeasurement}
\end{figure}

In this Section we discuss protocols that reduce to measurement of a free field --- so that the pre-measurement ensemble is effectively  
\be\label{eq:Hfree}
\mathcal{H} = \f{K}{2} \int (\nabla h)^2
+ \text{irrelevant terms},
\ee
--- but for which, unlike the measurement protocol in Secs.~\ref{eq:2Dheightlocking},
the post-measurement ensemble is described by a nontrivial conformal field theory in the infrared. 
The microscopic problems considered here reduce to  measurement of the operator
\be
\measO = \cos h
\ee
in (\ref{eq:Hfree}).
We show that, depending on the stiffness $K$, 
weak measurement of $\measO$ can induce flow to a nontrivial line of interacting fixed points in the  replica theory. 
In addition there can be nontrivial fixed points at $\lambda=\infty$.

This problem with $\measO=\cos h$ can be obtained from various 
very different physical models.
These may behave differently at  strong measurement rate but are equivalent   at weak measurement rate as a result of the field theory description.

One possibility is to interpret  ${\vec S = (\cos h, \sin h)}$ as an XY spin.
Then we are measuring a single component of $\vec S$.  For the free-field description to be appropriate,  we must be able to neglect vortices: this is true at arbitrarily large scales if $K>2/\pi$ \cite{cardy1996scaling}.\footnote{This means that in the XY model, with measurement of $\cos h$, a part of the nontrivial fixed line discussed below will be destabilized by vortices. In order to see the full fixed line
we can modify the problem  by instead measuring a higher harmonic of the XY field. Equivalently, we take the XY angular variable to be $H = h/q$. Vortices are then irrelevant over the larger range $K>2/(\pi q^2)$.}

We focus on a more entertaining  interpretation which involves thinking of  $h(x,y) \in \mathbb{R}$ as a \textit{height field}, instead of an angular degree of freedom. 
For concreteness, we imagine a regularization of the theory in which the height field $h(x,y)$ is defined in the spatial continuum, say with a momentum cutoff, rather than on a lattice. 
See the cartoon in Fig.~\ref{fig:contourmeasurement}.

Now imagine that we are given a noisy version of a \textit{contour map} of $h$, 
indicating the lines where $h \in \alpha \mathbb{Z}$.
Here $\alpha$ is a constant that determines the spacing of the level lines: 
we choose to normalize $h$ so that ${\alpha=2\pi}$.
The level lines are  assumed to be unlabelled and unoriented (i.e.\ we are not told  the corresponding value of $h$, 
or the  direction in which the level line should  be traversed in order for $h$ to increase\footnote{If we change those assumptions, 
then the measurements reveal more information, and the universal description will change. In particular, knowledge of \textit{labelled} level lines is clearly enough to reconstruct the full field configuration on large scales. We expect that measuring  labelled level lines is equivalent at large scales to measuring $h$, so that the replica Hamiltonian then has a term $\sum_{ab}(h_a-h_b)^2$, which gives a mass to replica-asymmetric modes. Oriented (but unlabelled) level lines essentially take us back to the setting in Sec.~\ref{eq:2Dheightlocking}.}).
What information does the noisy contour map give us about the configuration?

There are many ways of defining the measurement process. 
(For example, one possibility 
would be to take the contour map to be made up of  discrete pixels, each either black/white, with some probability of error.)
But in the limit where the measurement is weak, i.e.\ the contour map is very noisy, we can decompose the measured operator into scaling operators, and retain only the most relevant. By symmetry, this is $\measO = \cos h$.\footnote{For example, one way to define the measurement process would be via a measurement of a regularized version of the Dirac comb operator $\sum_{n\in \mathbb{Z}} \delta(h - 2\pi n)$, which is peaked on the level lines. When we expand this operator in scaling operators, the most relevant term is $\cos h$, leading to the replica description above.}

Both for the XY model with measurement of a single spin component, and for the height field with noisy contour measurement,  we obtain the usual  form of the replica Hamiltonian at weak measurement strength
\be\label{eq:Hcontour}
\mathcal{H}_N =   \f{K}{2}  \int  \sum_a (\nabla h^a )^2
 -  \lambda  \int  \sum_{a\neq b}   \measO^a \measO^b ,
\ee
where now we take ${\measO^a \propto \cos h^a}$ to be conventionally normalized (its two-point function is a power law with unit coefficient).
This description is universal for weak measurement strength, but it is important to note that   distinct microscopic models may have  different fates at strong measurement. We will describe  the strong measurement limit for a particular model below.

Note that this replica Hamiltonian retains a \textit{separate} $h^a\rightarrow -h^a$ symmetry in every replica $a$. 
This prevents the generation, under the RG, of an off-diagonal stiffness term $\sum_{a\neq b}(\nabla h^a)(\nabla h^b)$. 
Importantly, the stiffness $K$ cannot flow either,
by the general property of the $N\to 1$ limit discussed in  Sec.~\ref{sec:nofeedback}.
As a result, the RG flows in this problem are very simple: we need consider only the RG flow of $\lambda$.

The operator $\measO$ does not appear in its OPE with itself, so the general formula (\ref{eq:rgeqnsimple}) gives 
\be\label{eq:flowscontourmeasurement}
\partial_\tau \lambda = \lf 2- \f{1}{2\pi K} \ri \lambda - 4 \lambda^2
\ee
 (where $x_\measO = 1/4\pi K$ is the scaling dimension of the measured operator).
 Therefore the flow diagram is as in Fig.~\ref{fig:flowscontourmeasurement}, with a \textit{stable} interacting fixed line appearing for $K>1/4\pi$. 
 This fixed point is under analytical control if ${K-1/4\pi}$ is small, since ${\lambda_* =(K-\f{1}{4\pi})/2K}$
is then small.  
In the future, it would be interesting to compute correlation functions on the new fixed line.

\begin{figure}
    \centering
\includegraphics[width=0.65\columnwidth]{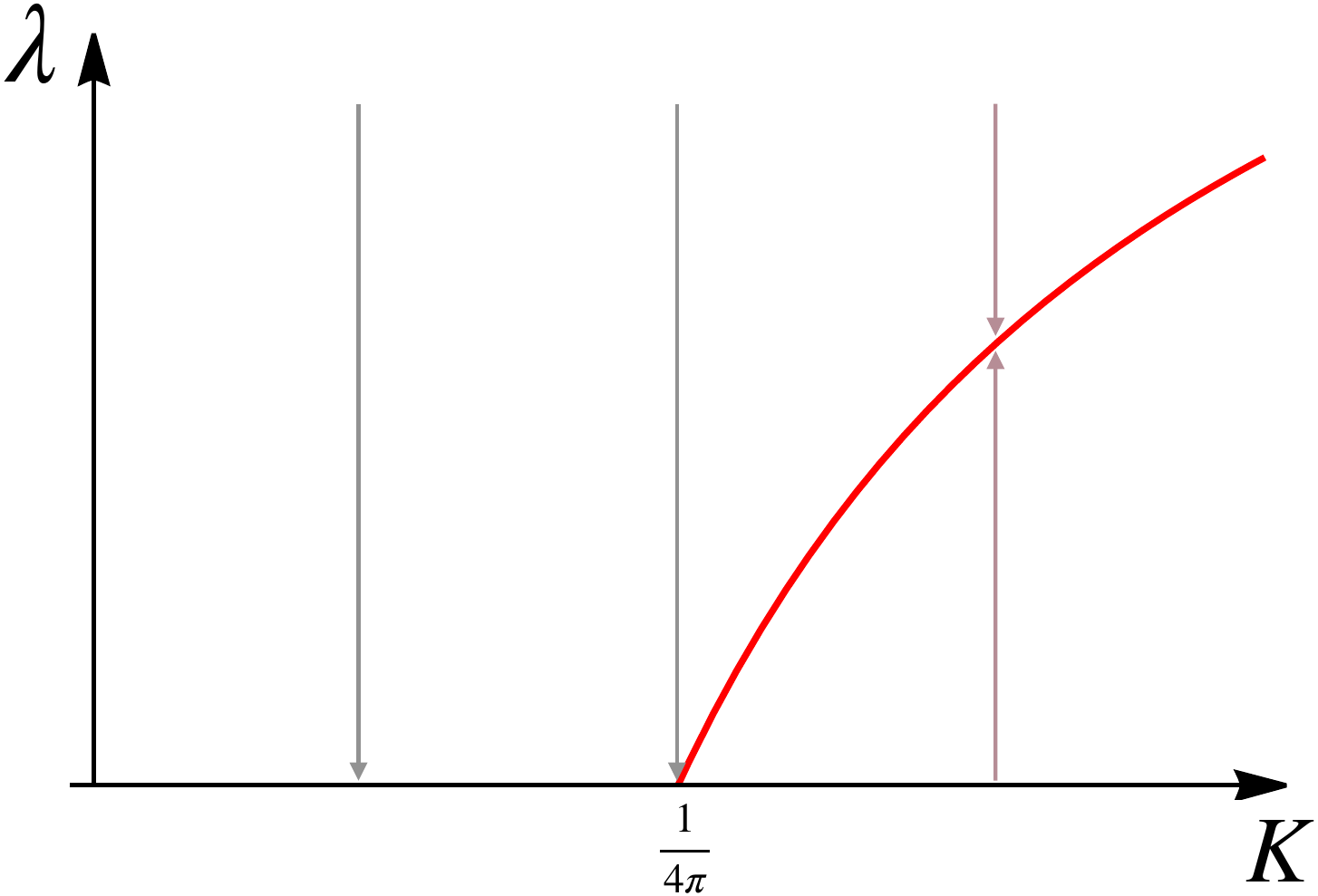}
    \caption{Schematic RG flows (\ref{eq:flowscontourmeasurement}) for $\mathcal{H}_N$ in  Eq.~\ref{eq:Hcontour}.}
\label{fig:flowscontourmeasurement}
\end{figure}

The phase diagram at strong measurement rate may depend on more details of the microscopic realization.
Here we mention a lattice example
having a nontrivial fixed point at $\lambda=\infty$ for which the replica partition function  can be computed exactly using results from loop models.

This is a standard model for \textit{discrete} heights $h\in 2\pi \mathbb{Z}$ on the sites of the triangular lattice, in which adjacent heights differ by at most 1 unit, with an energy cost for height difference.
We can uniquely assign heights ${h\in 2\pi (\mathbb{Z}+1/2)}$ to  level lines that live on the links of the dual honeycomb lattice and form closed loops. 
After a trivial shift of $h$ by $\pi$  this is an example of the kind described above
(we discuss the model in more detail in App.~\ref{sec:looporientationexample}).

An effective field theory for it is a sine-Gordon model with an \textit{irrelevant} cosine potential: in the IR it flows to (\ref{eq:Hfree}) at the specific value $K=\f{1}{8\pi}$. This is in the regime where weak measurement is irrelevant. However, very strong measurement of the level lines gives a nontrivial replica theory. All replicas are forced to agree on the positions of the loops, but  there is still freedom in the ``orientations'' of the loops which determine which side of the loop has a larger value of $h$. This gives an effective loop model. 

We argue (App.~\ref{sec:looporientationexample}) that the effective central charge at $\lambda=\infty$ is 
\be
c_\text{eff} = \f{12 \ln 2}{\pi^2}
\simeq 1 - 0.157.
\ee
Conditioning on the measurements has a subleading (though universal) effect on the expectation value $\<(h(x) - h(0))^2\>_M$. However, it has a strong effect on correlators of vertex operators 
${\< e^{i\alpha h(0)}
e^{-i\alpha h(x)}\>_M}$
conditioned on measurement. 
The mean value of the conditioned correlator is of course the conventional correlator $\< e^{i\alpha h(0)}
e^{-i\alpha h(x)}\>$.
But this is dominated by rare measurement outcomes, so that the value for a typical measurement sample is smaller:
\ba
\< e^{i\alpha h(0)}
e^{-i\alpha h(x)}\>
& = |x|^{-2 x_\text{av}(\alpha)},
\\
\left|\< e^{i\alpha h(0)}
e^{-i\alpha h(x)}\>_M
\right|_\text{typ}
& \sim
|x|^{-2 x_\text{typ}(\alpha)},
\label{eq:typicalcorrelatorlevellinestrongmeasurement}
\end{align}
with 
\ba
x_\text{av}(\alpha) & = 2 \alpha^2,
&
x_\text{typ}(\alpha) &  = -\f{1}{\pi^2}\ln \cos |2\pi \alpha|.
\end{align}
We leave the characterization of the large-$\lambda$ regime in more general models to the future.

\section{Nishimori line and its cousins: measuring non-critical states}
\label{sec:paramagneticstates}

We now consider an important special class of measurement problems, in which the initial state has short-range correlations. 
To be more precise, we assume here that local observables are short-range correlated (which could allow for classical topological order).

Inference phase transitions are still possible in this setting. 
A paradigmatic example corresponds to
measuring $S_x S_y$ on nearest-neighbor bonds $\<xy\>$
in the \textit{infinite-temperature} Ising model \cite{nishimori2001statistical,iba1999nishimori}, 
as illustrated very schematically in 
Fig.~\ref{fig:N3types2D},~Left.
(In the 2D case, this can  be thought of as an inference problem for images made up of uncorrelated bits.)
In this problem there is a \textit{weak} measurement phase in which we have essentially no information about the relative sign  $S_x S_y$ for distant site pairs $x$, $y$, 
and a \textit{strong} measurement phase where we do have information about long-distance relative orientations.
By a change of variable (reviewed below), this Bayesian reconstruction problem can be mapped \cite{iba1999nishimori}  onto the Nishimori line \cite{nishimori1980exact,nishimori1981internal,ozeki1993phase,
le1988location,le1989varepsilon,
gruzberg2001random,
singh1996high,reis1999universality,
honecker2001universality,
hasenbusch2008multicritical,
sourlas1994spin,
nishimori1993optimum,
nishimori1994gauge,
iba1999nishimori,
nishimori2001statistical,
zdeborova2016statistical} in the phase diagram of the random-bond  Ising model.

Our aims in this Section are modest (and we will spend some time on review).
First, we aim to clarify some differences between measurement problems involving critical states (discussed in previous sections) and those  involving noncritical states. Second, we will suggest some generalizations of Nishimori-like problems.

Thirdly, we aim to  clarify the relation between the Nishimori inference universality class and two other --- very closely related --- universality classes. 
We will refer to these as ``gauged Nishimori inference'' problems.
These problems share Nishimori exponents but are really physically distinct.
They are illustrated in Fig.~\ref{fig:N3types2D} and will be explained in Sec.~\ref{sec:nishimoricloserelatives}.
An important example arises from a simple model of error correction in the toric code state,
which was  shown to have an inference phase transition with Nishimori exponents in Refs.~\cite{dennis2002topological,wang2003confinement}.

Further discussion of these problems and higher-dimensional analogs is in Sec.~\ref{sec:nishimoricloserelatives}.
A wide range of problems involving quantum measurement  \cite{zhu2023nishimori,lee2022decoding,
fan2024diagnostics, hauser2024information} or classical inference \cite{weinstein2024computational} have been related to the Nishimori line in the recent literature. 
We hope that the unified discussion in Secs.~\ref{sec:isingparasubsec},~\ref{sec:nishimoricloserelatives} will help make the ubiquity of these exponents more intuitive.

Before discussing concrete models let us note some general ways in which the problems in this Section (arising from measurement of non-critical states)
differ from those in previous Sections (arising from measurement of critical states).

First, the present problems have a special relation to disordered systems.
Since the initial state is short-range correlated, the measurement outcomes $M$ are (loosely speaking) also short-range correlated.
As a result, the Hamiltonian $\mathcal{H}_\text{meas}[S,M]$ (Eq.~\ref{eq:Hmeasgeneral}), which determines the a posteriori probability distribution for the spins (conditioned on the measurements $M$)
can  be 
\textit{reinterpreted}
 as a Hamiltonian in which $M$ represents random local couplings, i.e.\ short-range correlated disorder. In this reinterpretation, we ``forget'' that $M$ arose from measurement.
 
This reinterpretation means that the critical points encountered in measuring paramagnetic states can be mapped to critical points in disordered systems.\footnote{A change of variable is necessary to make these disordered systems look more natural   \cite{zdeborova2016statistical}, see below.}
The disordered systems arising from these mappings are however of a special (fine-tuned) kind. 
The difference is  encoded in the replica symmetry. 
For a generic disordered system,
 the replica trick leads to $S_n$ symmetry with the limit ${n\to 0}$.
The disordered systems  arising from the above mapping have an enhanced symmetry, $S_{N}$ with ${N=n+1}$ and ${N\to 1}$ \cite{gruzberg2001random}. This enlarged symmetry  is  enforced by the underlying 
measurement interpretation.

Second, the critical points in this Section have a different field-theory structure.
The basic field $S_a$, with a single replica index, is not critical in these examples\footnote{The spin $S$ in the inference problem should not be confused with the spin in the random-bond Ising model on the Nishimori line, which we denote $\widetilde S$. The two fields are related by a simple change of variable, reviewed below.} (unlike those in previous Sections) and  instead 
only an  ``overlap field'' of the schematic form ${X^{ab}\sim S^a S^b}$ carrying more replica indices becomes critical.  
The resulting Landau-Ginsburg theories therefore have a similar structure to those for spin glasses.

More formally, the measurement problem for the infinite-temperature Ising paramagnet 
possesses a \textit{local} $\mathbb{Z}_2$ symmetry. 
Similar local symmetries are characteristic of many other paramagnet-measurement problems (and distinguish them from problems in which critical states are measured). 
As clarified below, the local symmetries are microscopically exact in the simplest cases (e.g.\ measurement of the Ising paramagnet at infinite temperature) 
and are emergent in the IR in more general cases (e.g.\ measurement of the Ising paramagnet at some finite temperature above its critical point).

At a heuristic level, the present critical points are rather different from the perturbatively-accessible ``weak measurement'' critical points discussed  in, for example, Secs.~\ref{sec:pottsweak},~\ref{eq:interactingfixedlinefreefield}. 
In the latter, we started with a replicated critical field $S^a$, 
and 
(in 2D)  the weak measurements
(which suppress  inter-replica fluctuations) 
led to a \textit{reduction} of the effective central charge $c_\text{eff}$ compared to that of the unmeasured theory.
In the present Section, we start with paramagnetic states, with  $c=0$. 
Adding measurement is then (loosely speaking!) akin to reducing the temperature for inter-replica fluctuations, leading to a condensation of $X^{ab}\sim S^aS^b$ at some finite measurement strength. 
At the critical point, $c_\text{eff}$ is positive, and therefore is \textit{larger} than that of the unmeasured theory.

\subsection{Measuring the Ising paramagnet: \\ ``Nishimori Inference'' universality class}
\label{sec:isingparasubsec}

\begin{figure}
    \centering
\includegraphics[width=0.96\columnwidth]{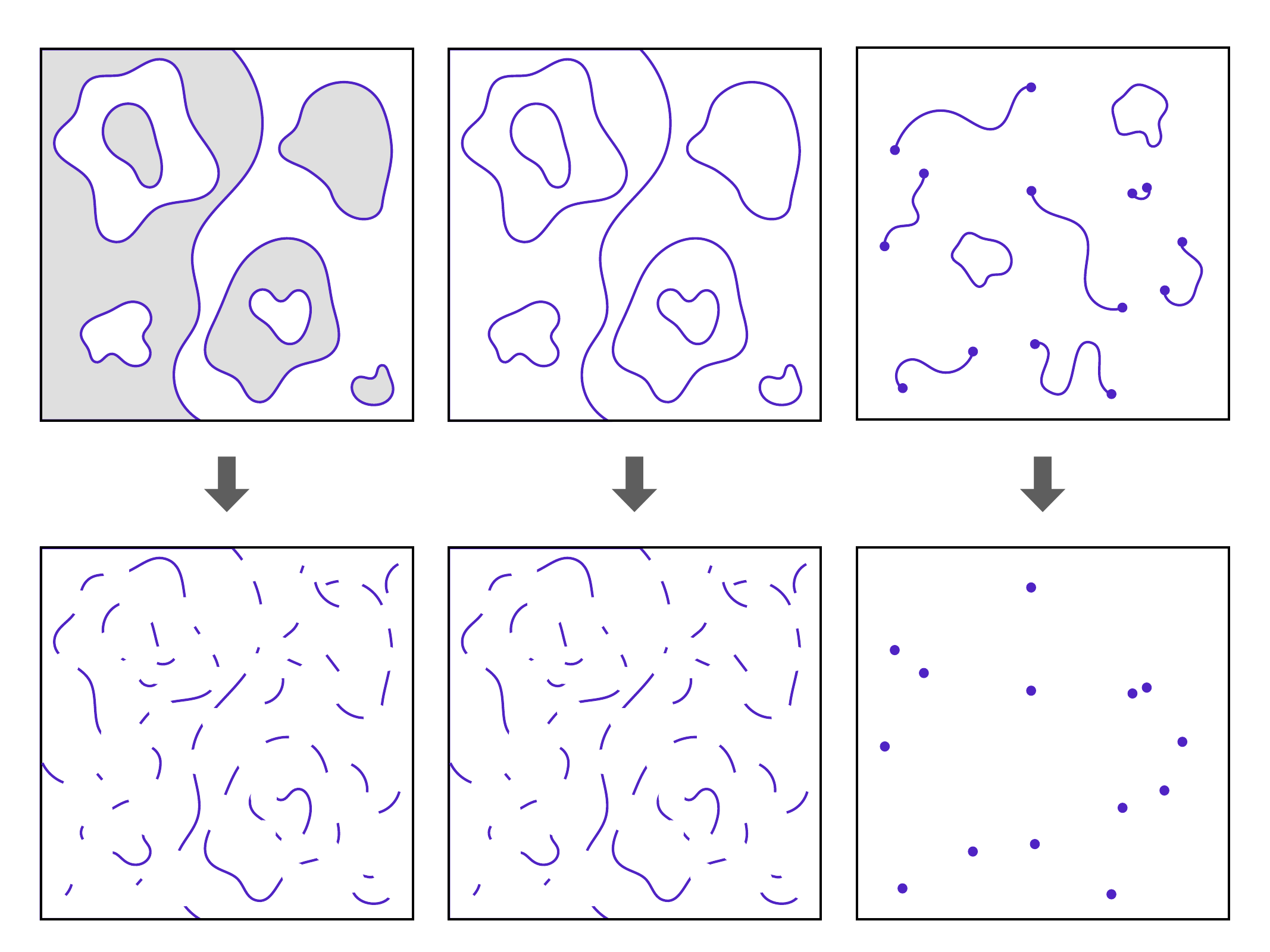}
    \caption{Schematic: we discuss three types of measurement problem that all show transitions with ``Nishimori'' exponents, but which 
    are physically distinct. 
    Each class exists in any dimension: here we show the 2D case.
    {\bf Left:} starting with an Ising spin configuration (white $=$ spin up, gray $=$ spin down), we make a noisy measurement of the domain walls (i.e. of $S_i S_j$ on bonds), indicated schematically by the lower-left image.
    {\bf Centre:} 
    The physical configuration is now not a configuration of spins but a configuration of closed strings. 
    (This can be viewed as a configuration in a gauge theory, see text.)
    We make a noisy measurement of the local string density.
    {\bf Right:}
    The physical configuration is now a configuration of potentially open strings. 
    Instead of measuring the string density, 
    we make a noiseless measurement of the locations of string endpoints. }
\label{fig:N3types2D}
\end{figure}

First we review the Bayesian inference problem \cite{iba1999nishimori} that permits a mapping  to the  Nishimori line \cite{nishimori1980exact,nishimori1981internal,ozeki1993phase,
le1988location,le1989varepsilon,
gruzberg2001random,
singh1996high,reis1999universality,
honecker2001universality,
hasenbusch2008multicritical,
sourlas1994spin,
nishimori1993optimum,
nishimori1994gauge,
iba1999nishimori,
nishimori2001statistical,
zdeborova2016statistical} in the random-bond Ising model.

To begin with, consider a system of Ising spins at infinite temperature (so that the initial reduced Hamiltonian is $\mathcal{H}=0$)
for which we measure $S_xS_y$ on every bond \cite{iba1999nishimori}.
For simplicity we will continue to consider Gaussian measurements, with variance ${\Delta^2 = (2\lambda)^{-1}}$,
but noisy binary measurements may be treated similarly.
The replica Hamiltonian is (dropping terms of order ${N-1}$) 
\ba\label{eq:nishimorilandau}
\mathcal{H}_N & = \lambda \sum_{\<xy\>} \sum_{a\neq b} (S_x^a S_y^a )(S_x^b S_y^b )
\\
& = \lambda \sum_{\<xy\>} \sum_{a\neq b} X_x^{ab} X_y^{ab} .
\end{align}
As usual, we have ${N\to 1}$ replicas, and   $X$ is defined by   ${X^{ab}=S^aS^b}$ for $a\neq b$ (and we can take the diagonal elements of $X$ to vanish).

The above rewriting suggests (correctly) that as long as we are above 1D  there will be some $\lambda_c$, such that  
for  ${\lambda<\lambda_c}$ all correlations decay exponentially, while for $\lambda>\lambda_c$ the overlap field $X$ is ordered
in the pattern ${\<X^{ab}\>= C \chi^a \chi^b}$ 
(for some $C>0$ and some signs $\chi^a=\pm 1$ that select one of the symmetry-related ground states), so that
\be
\lim_{|x-y|\to\infty}\mathbb{E}_M\<S_x S_y\>^2 
=\lim_{|x-y|\to\infty}
\< X^{12}_x X^{12}_y\> 
= C^2.
\ee
In this strong-measurement phase, the local relative measurements together give us useful information about the relative orientation of distant spins. 
Note that $\mathbb{E}_M \<S_x S_y\>_M=0$ regardless of $\lambda$, since as usual this ``single-replica'' quantity reduces to the conventional correlator in the initial Gibbs state defined by ${\mathcal{H}=0}$: $\<S_xS_y\>=0$.\footnote{In this example the inference problem is nontrivial even though the initial ensemble is paramagnetic. A different way in which paramagnetic models can give rise to nontrivial correlators is through nonlocal geometrical observables: for example the infinite-temperature Ising model on a 2D (triangular) lattice may be mapped to critical site percolation (Sec.~\ref{sec:percolation}).}

A Landau theory for the transition may be formulated by promoting $X$ to an independent fluctuating field in the functional integral \cite{chen1977mean,le1988location,
le1989varepsilon},\footnote{In \cite{chen1977mean,le1988location,
le1989varepsilon} the Landau theory is written in terms of an $n\times n$ matrix $Q$ and an $n$-component vector $M$, since this is natural in the disordered systems context 
(Sec.~\ref{sec:rbimmapping}). These are essentially sub-blocks of the $N\times N$ matrix $X^{ab}$ in Eq.~\ref{eq:NishimoriLG} (with $N=n+1$).}
\be\label{eq:NishimoriLG}
\mathcal{H} = \f{1}{2} \sum_{a\neq b} \left[ (\nabla X^{ab})^2 + 
{m^2} (X^{ab})^2  \right] + 
g \hspace{-1.2mm} \sum_{a\neq b \neq c \neq a}
\hspace{-1.2mm}
X^{ab} X^{bc} X^{ca},
\ee
where the terms respect the symmetry discussed immediately below,
and in the $N\to 1$ limit this theory shows a fixed point in $6-\epsilon$ dimensions. 

\subsubsection{Symmetry and emergent symmetry in the paramagnet measurement problem}
\label{sec:paramagnetsymmetry}

Let us take a brief detour to see how the symmetries of the above problem fit into the general framework in Sec.~\ref{sec:symmetry}.

First consider the initial ensemble (the infinite-temperature Ising model), before we introduce replicas.
Since this Ising model has reduced Hamiltonian  $\mathcal{H}=0$, it has not only a $\mathbb{Z}_2$ global symmetry, but in fact a \textit{local} $\mathbb{Z}_2$ symmetry for every site: 
any spin can be flipped, without changing the value of $\mathcal{H}$.
We denote this collection of local symmetries by $\mathbb{Z}_2^\text{local}$.
We do not refer to it as a gauge symmetry, because flipping a spin gives a physically distinct state.

Note that the global Ising symmetry is\footnote{More precisely, for any fixed lattice, $\mathbb{Z}_2^\text{global}$ is a subgroup of $\mathbb{Z}_2^\text{local}$.} a subgroup of $\mathbb{Z}_2^\text{local}$.
The local and global symmetries play different roles, because they have a different effect on the measurements $M_{x,y}$.
The measurements are  in general \textit{covariant} under $\mathbb{Z}_2^\text{local}$, but are \textit{invariant} under the $\mathbb{Z}_2^\text{global}$ subgroup (Sec.~\ref{sec:symmetry}).\footnote{In other words, 
$\mathbb{Z}_2^\text{local}$ has a combined action on the spins and the measurements 
via ${S_x \rightarrow \chi_x S_x}$, ${M_{x,y}\rightarrow \chi_x\chi_y M_{x,y}}$ (for local symmetry transformation parameters $\chi_x=\pm 1$) which leaves $P(M|S)$ unchanged.  For the global symmetry, $\chi_x = \chi$ is independent of position, so that $M$ does not transform.}

As a result, when we introduce replicas, the replica Hamiltonian is symmetric under $\mathbb{Z}_2^\text{local}$ transformations  only if they are applied to all replicas simultaneously, 
but it is symmetric under  global $\mathbb{Z}_2$ transformations for each replica separately, i.e.\ under $\mathbb{Z}_2^N$ transformations.
Replica-uniform $\mathbb{Z}_2$ transformations 
(in the diagonal subgroup of $\mathbb{Z}_2^N$) are already contained in $\mathbb{Z}_2^\text{local}$, so the global symmetry that acts nontrivially is the remaining $\mathbb{Z}_2^{N-1}$, so the  full symmetry of the replica theory for Nishimori inference is
\be\label{eq:NishInfSymm}
{\mathbb{Z}_2^\text{local}\times \mathbb{Z}_2^{N-1}}
\ee
together with replica permutations.

The $\mathbb{Z}_2^\text{local}$ transformations act trivially on the order parameter $X^{ab}_x$.
The global symmetry $\mathbb{Z}_2^{N-1}$ acts nontrivially on $X^{ab}_x$.
This symmetry of the replica Hamiltonian   ensures, for example, 
 that in each term of (\ref{eq:nishimorilandau}), each replica index appears an even number of times. 
In the strong measurement phase, the global  $\mathbb{Z}_2^{N-1}$ symmetry  is  broken completely by the expectation value for $X$.

So far we have considered the case where the initial Hamiltonian is at infinite temperature ($\mathcal{H}=0$). 
We would of course expect essentially the same physics if the initial Hamiltonian is at finite temperature, within the paramagnetic phase, i.e.\ if 
${\mathcal{H} = J \sum_{\<x,y\>} S_x S_y}$ with sufficiently large $J$.
In this case, part of the symmetry in (\ref{eq:NishInfSymm}) is emergent.

The introduction of a nonzero $J$ preserves the $\mathbb{Z}_2^\text{global}$ symmetry of the initial ensemble, but it explicitly breaks the $\mathbb{Z}_2^\text{local}$ symmetry. 
Accordingly the replica Hamiltonian
retains $(\mathbb{Z}_2^\text{global})^N$ symmetry, but 
  its $\mathbb{Z}_2^\text{local}$ symmetry is explicitly broken by the $J$ term:
\ba\label{eq:nishimorilandaumodified}
\mathcal{H}_N & = 
J \sum_{\<xy\>} \sum_a S^a S^b + 
 \lambda \sum_{\<xy\>} \sum_{a\neq b} X_x^{ab} X_y^{ab} .
\end{align}
We can also write a Landau theory, with the same symmetries, in which $S$ and $X$ are treated as independent fluctuating fields, coupled by a term $X^{ab}S^a S^b$   (see Sec.~\ref{sec:isingddims}).

However, this reduction in microscopic symmetry does not change the universality class, because $\mathbb{Z}_2^\text{local}$ is recovered in the IR in a very simple way so long as the initial ensemble is paramagnetic ($J>J_c$).
In this case, $S$ is a massive field, 
and may be integrated out.
In the  vicinity of the phase transition at $\lambda_c$ induced by measurement,\footnote{The value of $\lambda_c$ will depend in general on $J$.} 
we are left in  the IR  with a theory that only includes the field $X$. 
Since $X$ is invariant under $\mathbb{Z}_2^\text{local}$, this means that 
$\mathbb{Z}_2^\text{local}$ is recovered as an emergent symmetry of the replica theory in the IR.

This emergent symmetry may be understood more heuristically without replicas. We start with an Ising model in the high-temperature phase. In a naive real-space RG picture, this flows to the infinite-temperature fixed point with $J=0$, where ${\mathbb{Z}_2^\text{local}}$ is recovered.

{ 

\subsubsection{Mapping to a random-bond Ising model (review)}
\label{sec:rbimmapping}

We now review the connection between the Bayesian inference problem and a disordered Ising model \cite{sourlas1994spin,iba1999nishimori}. 
This mapping to a disordered system will not be needed for our subsequent discussion of measurement problems, but we include it for completeness since it underlies the basic terminology.
The phrase ``Nishimori line'' refers initially not to a Bayesian inference problem, but rather to a line in the phase diagram of the random-bond Ising model.  The connection between this line and Bayesian inference is discussed  in Refs.~\cite{sourlas1994spin,iba1999nishimori,nishimori2001statistical,zdeborova2016statistical}. The enlarged replica symmetry on the Nishimori line is discussed in Refs.~\cite{le1988location,le1989varepsilon,gruzberg2001random}

Let us first describe the mapping without using replicas.
Recall the structure of the inference problem. 
There is the initial or ``true'' configuration, which we denote here by $S^\text{ref}$ (for ``reference configuration''; in the terminology of \cite{zdeborova2016statistical} this is the ``ground truth'').
Measurements, with Gaussian errors $\epsilon_{xy}$ of variance $\Delta^2$, are made on the bonds of this configuration, giving measurement outcomes
\be\label{eq:MintermsofS}
M_{x,y} = S^\text{ref}_x S^\text{ref}_y + \epsilon_{x,y}.
\ee
These measurement outcomes are then fed into Bayes' theorem to give the a posteriori (conditional) distribution $P(S|M)$.
This probability distribution has the form of a Boltzmann weight  with effective Hamiltonian $\mathcal{H}_\text{meas}[S,M]$ (Sec.~\ref{sec:binarymeasurements}),
\be
\mathcal{H}_\text{meas} [S,M] = \lambda \sum_{\<x,y\>} \lf S_x S_y - M_{x,y} \ri^2 .
\ee
We now reinterpret this as the Hamiltonian for spins $S$  with quenched random couplings set by $M$.
For this purpose we can expand out the brackets and drop terms that are independent of $S$. We also recall that instances of $M$ are generated as in Eq.~\ref{eq:MintermsofS}. We denote the resulting effective Hamiltonian   by $\mathcal{H}_\text{quenched}$ (and suppress the arguments):
\be
\mathcal{H}_\text{quenched} = 
2 \lambda \sum_{\<x,y\>} \lf S^\text{ref}_x S^\text{ref}_y + \epsilon_{x,y} \ri S_x S_y.
\ee
This can be understood as a disordered system 
in which the quenched disorder is determined both by  uncorrelated binary variables $S^\text{ref}_x$ on sites and by  Gaussian variables $\epsilon_{x,y}$ on bonds.

This is not yet the Nishimori problem. To obtain that, we make a change of variable, defining
\be
\widetilde S_x = S^\text{ref}_x S_x
\ee
to give
\be\label{eq:HrbimNishimori}
\mathcal{H}_\text{quenched} = 
2 \lambda \sum_{\<x,y\>} (1+ \widetilde \epsilon_{x,y})
 \widetilde S_x \widetilde S_y.
\ee
This is now a disordered system with only Gaussian bond randomness, and with a non-random ferromagnetic coupling.
(Here $\widetilde \epsilon_{x,y}= \epsilon_{x,y} S^\text{ref}_x S^\text{ref}_y$ has the same statistics as~$\epsilon_{x,y}$.)

The variance  of the bond randomness and strength of the ferromagnetic coupling are not independent (recall $\lambda = 1/2\Delta^2$). This is a ``line'' in a two-dimensional phase diagram where the ferromagnetic coupling strength and the disorder variance are independent axes. 
The part of the line with ${\lambda<\lambda_c}$ lies in the paramagnetic phase for $\widetilde S$, and the part of the line with ${\lambda>\lambda_c}$ lies in the ferromagnetic phase (see the discussion below Eq.~\ref{eq:correlatoridentities}).
In 2D, the Nishimori critical point 
(given by Eq.~\ref{eq:HrbimNishimori} at   ${\lambda=\lambda_c}$)
is a multicritical point on the paramagnet--ferromagnet phase boundary. In higher dimensions, the paramagnetic, ferromagnetic and spin glass phases can all meet at this point \cite{le1988location,gruzberg2001random,nishimori2001statistical,nishimori1986geometry}.

In the original measurement interpretation, $\widetilde S$ is the overlap between the (unknown) measured configuration, 
and a sample drawn from the conditional distribution.
When we reinterpret (\ref{eq:HrbimNishimori}) as the Hamiltonian for a disordered system, $\widetilde S$ 
is instead regarded as a physical spin variable.

The change of variable goes through similarly in the replica theory \cite{le1988location}. 
As usual  the measurement problem gives
\be\label{eq:nishimorireplicasymmetric}
\mathcal{H}_N = \lambda \sum_{\<x,y\>} \sum_{\substack{a\neq b \\ 1}}^N
(S_x^a S_y^a)(S_x^b S_y^b).
\ee
Recall that, in the correlation functions of the replica theory, $S^\text{ref}$ may be identified with one of the replicas, let's say $S^N$. 
We define $\widetilde S^a = S^a S^N$ for $a=1,\ldots, {N-1}$, giving \cite{le1988location,gruzberg2001random}
\be
\mathcal{H}_N = 
2 \lambda \sum_{\<x,y\>}  \sum_{a=1}^n
\widetilde S_x^a \widetilde S_y^a
+
\lambda \sum_{\<x,y\>}  \sum_{\substack{a\neq b \\ 1}}^n
(\widetilde S_x^a \widetilde S_y^a)(\widetilde S_x^b \widetilde S_y^b),
\ee
where we have used ${n=N-1}$. The field $S^N$ no longer appears in  $\mathcal{H}_N$, so may be trivially summed over.

Since the replica limit is $n\to 0$, 
this is just the replica formulation of the random-bond Ising model  in Eq.~\ref{eq:HrbimNishimori}.
This replica Hamiltonian has an $S_n$ permutational symmetry and a single global $\mathbb{Z}_2$, giving ${\mathbb{Z}_2\times S_n}$.
But as we have just seen it is thermodynamically equivalent to 
(\ref{eq:nishimorireplicasymmetric}), 
which has a much larger symmetry, discussed in Sec.~\ref{sec:paramagnetsymmetry}.
The global symmetry is ${(\mathbb{Z}_2)^N\rtimes S_N}$, 
and in  addition the replica-uniform subgroup of the $(\mathbb{Z}_2)^N$ symmetry is promoted to a local symmetry.

As is well-known, the random-bond Ising model on the Nishimori line obeys various exact identities between correlators \cite{nishimori1981internal}. In the replica approach, these  follow from the enlarged replica symmetry \cite{gruzberg2001random, le1989varepsilon,zdeborova2016statistical}. For example,\footnote{More generally, 
$\overline{\langle\widetilde S_x \widetilde S_y\rangle^{2k-1}} = 
\overline{\langle\widetilde S_x \widetilde S_y\rangle^{2k}}
=
\mathbb{E}_M \< S_x S_y\>^{2k}
$ for any positive integer $k$ \cite{le1989varepsilon,zdeborova2016statistical}.}
\ba\label{eq:correlatoridentities}
\overline{\langle\widetilde S_x \widetilde S_y\rangle} = 
\overline{\langle\widetilde S_x \widetilde S_y\rangle^2}
=
\mathbb{E}_M \< S_x S_y\>^2,
\end{align}
where the first two quantities have a meaning in the disordered systems problem, and  the overline is the average over the Gaussian bond disorder in Eq.~\ref{eq:HrbimNishimori}, and the final quantity has a meaning in the related inference problem.
To see Eq.~\ref{eq:correlatoridentities}, note that the first of the above correlators can be written (in the replica approach) using the operator $\widetilde S^1 = S^N S^1$ at each site. 
The second can be written using $\widetilde S^1 \widetilde S^2 = (S^N S^1)(S^N S^2) = S^1 S^2$ at each site, giving the same result.
Note that the ``strong measurement'' phase of the inference problem maps to a ferromagnetic phase for $\widetilde S$, since, by Eq.~\ref{eq:correlatoridentities}, the long-range order of $\mathbb{E}_M \< S_x S_y\>^2$ implies long range order in $\overline{\langle\widetilde S_x \widetilde S_y\rangle}$.

There is a terminological subtlety about whether we decide to say that the inference problem at $\lambda_c$ and the related random bond Ising problem at the Nishimori critical point are ``in the same universality class'',
or whether we only say that there is a mapping between these two universality classes. 
We will avoid this terminological decision because in the rest of this section we will only be concerned with the inference problems, and will forget   about the associated disordered systems.

We will refer to the universality class of the Ising-paramagnet inference problem as the ``Nishimori inference'' (NI) universality class.

\subsection{``Gauged Nishimori inference'' problems in  error correction and  $\mathbb{Z}_2$ gauge theory}
\label{sec:nishimoricloserelatives}

Next we discuss two universality classes of measurement transition
 that we argue should be distinguished from that of 
 the ``Nishimori inference'' problem, 
 but which are very closely related to it and share the same exponents (in any $d$). 
We will refer to these as ``gauged Nishimori inference'' problems. 

It has long been known that a standard error correction problem involving matching of pointlike defects in 2D
 can be mapped to the Nishimori line in the 2D random-bond Ising model
\cite{dennis2002topological, wang2003confinement}.
We will formalize this and related problems in terms of measurement of  
classical $\mathbb{Z}_2$ gauge-Higgs theory 
(note also Ref.~\cite{weinstein2024computational}) 
together with replicas.

Fig.~\ref{fig:N3types2D} summarizes the three closely related  problems
 in 2D which all have Nishimori exponents.
The relations are quite simple from a heuristic point of view: 
let us first give a schematic overview.

\smallskip 

{\bf Inference in Ising paramagnet (recap).} Recall that the Nishimori inference problem effectively involves measuring the density of \textit{domain walls} in an Ising model, with some measurement strength $\lambda$. 
In $d$ dimensions, these domain walls are closed $(d-1)$-dimensional hypersurfaces
(loops in 2D, membranes in 3D). See Fig.~\ref{fig:N3types2D},~Left.
The most basic object for which inference can be performed is the relative orientation $S_x S_y$ of distant spins.

\smallskip

{\bf Inference in deconfined state.} An ``almost'' equivalent inference problem arises in other models involving  closed 
$(d-1)$-dimensional hypersurfaces 
which are \textit{not} domain walls for anything
--- 
see Fig.~\ref{fig:N3types2D},~Center.
Loosely speaking, such a state, in which closed hypersurfaces proliferate, is an example of a ``deconfined'' state in the gauge-Higgs theory 
 (details in Sec.~\ref{sec:measuringZ2deconfined}). 
 Physically, a problem of essentially this kind   could arise from an imperfect measurement of a 2D topological quantum wavefunction in the computational basis.
A closely related problem describes \textit{spacetime} 
histories of errors in  the 1D repetition code \cite{dennis2002topological}, where (heuristically)  the loops represent worldlines of point defects, about which partial information is obtained by measurement.

Inference is no longer about local observables:  since the hypersurfaces are not domain walls, ``$S_x S_y$'' has no meaning. 
The role of the latter is instead played by a topological\footnote{The value of $V_P$ is unchanged under deformations of $P$ that preserve its endpoints \cite{nussinov2009symmetry,kapustin2017higher,gaiotto2015generalized}.} line observable $V_{P}$,  labelled by a path $P$   between $x$ and $y$,
  which counts the parity of the number of hypersurfaces intersecting $P$.
  (In the gauge theory language this is just a  Wilson line.)
 The expectation values  of this operator are analogous to those of $S_xS_y$ in the Ising paramagnet inference problem. 
 
Within the deconfined phase the ``single-replica'' expectation value $\<V_P\>$ decays to zero, but we may consider 
\be\label{eq:VPexp}
\lim_{|x-y|\to \infty} \mathbb{E}_M \<V_P\>_{\hspace{-0.5mm} M}^{\,2}, 
\ee
where $P$ is, say, a straight path connecting $x$ and $y$.
This is nonzero at large enough measurement strength $\lambda$, 
indicating a phase where  measurements allow nontrivial inference about the sign of $V_P$ for long paths.

Above we considered  hypersurfaces that were strictly closed.
In the classical gauge-Higgs theory, this is the extreme limit  $K\to\infty$ for the gauge coupling (Sec.~\ref{sec:measuringZ2deconfined}), where gauge fluxes are absent.
In 2D, going away from this limit trivializes the phase diagram, 
but in 3D, the transition survives for large enough finite $K$ (Sec.~\ref{sec:measuringZ2deconfined}). 
In this regime the membranes are not strictly closed, but the holes in them are ``small'' and disappear under RG (so that in the IR there is again a mapping to the Ising paramagnet).
We must however generalize the topological operator $V_P$ to a ``dressed'' version \cite{serna2024worldsheet}.
See Fig.~\ref{fig:N3types3D} for a schematic of these membranes.

\begin{figure}
    \centering
\includegraphics[width=0.8\columnwidth]{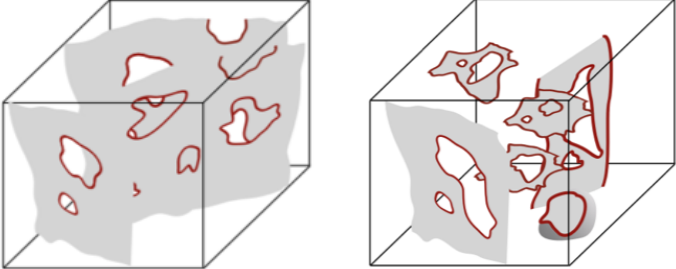}
    \caption{The 3D generalizations of the problems in Fig.~\ref{fig:N3types2D} can be viewed in terms of ensembles of membranes.
    Cases with closed or ``almost closed'' membranes correspond to deconfined states of the gauge-Higgs model (left) and cases where membrane boundary proliferates correspond to confined states (right).
    In the case where membranes are strictly closed they are dual to Ising domain walls.
    (Figure adapted from Ref.~\cite{serna2024worldsheet}.)}
\label{fig:N3types3D}
\end{figure}

\smallskip

{\bf Inference in confined state.} Above the configurations were made up of closed loops in 2D  (or closed/almost-closed membranes in 3D),
making the relation to the inference problem in the Ising paramagnet straightforward.

However there can still be a Nishimori--like transition  even in the case where string endpoints (or membrane boundaries in 3D, see Fig.~\ref{fig:N3types3D})proliferate, so long as they are accurately measured.
This  was shown in 2D  by  mappings to the random-bond Ising model in
\cite{dennis2002topological,weinstein2024computational}.

In the gauge-Higgs language, the proliferation of string endpoints (or membrane boundaries in 3D) is a proliferation of gauge fluxes, meaning that the \textit{physical} gauge field is in a \textit{confined} state. 
However, what is important for the preservation of the Nishimori exponents is that the \textit{inter}-replica gauge fluctuations are deconfined. 
This is the case if the fluxes are measured perfectly (in 2D) or with sufficient accuracy (in 3D).

In the setting of a single round of measurement in the toric code \cite{dennis2002topological},
 the strings correspond to a set of lattice links (a 1-chain)
 where errors have occurred,
 and the string endpoints are revealed by measurement of check operators.
 In the context of classical statistical mechanics,
 Ref.~\cite{weinstein2024computational} described a  slightly more general inference problem, phrased explicitly in terms of 2D gauge theory.
(We will clarify the universality class of this more general transition, which was left open in \cite{weinstein2024computational}.)

We formulate the above transitions in terms of a replicated gauge theory. 
For measurement of both deconfined and confined states, the ``replica-asymmetric'' sector of the critical field theory is the same, and hosts a Higgs transition with Nishimori exponents.

The 2D transitions in Refs.~\cite{dennis2002topological, wang2003confinement,weinstein2024computational}
exist only for perfect measurement accuracy of the string endpoints, i.e.\ of the error locations.
(Note that,  unlike the transition in the infinite-temperature Ising model, the transitions are driven not by a measurement strength but by a ``temperature'' in the initial ensemble.)
In 3D the analogous transition survives for finite measurement strength, so does not require fine-tuning.

In three dimensions, discrete gauge theories 
can give rise to nontrivial phase transitions even when the microscopic model obeys no symmetries (or constraints) whatsoever.
This is because the deconfined phase is absolutely stable in the RG sense.
(The deconfined phase can  be characterized in terms of spontaneously broken \textit{emergent} one-form symmetries \cite{nussinov2009symmetry,
kapustin2017higher,
gaiotto2015generalized,
hastings2005quasiadiabatic,
wen2019emergent,somoza2021self,pace2023exact}.)
The 3D measurement problem in Sec.~\ref{sec:measuremembraneboundary} shows that nontrivial \textit{measurement} phase transitions can arise even when  the microscopic  model   obeys no symmetries or constraints, and the state being measured is thermodynamically trivial. In this case deconfinement and  one-form symmetries are instead emergent properties of the inter-replica fluctuations.

\smallskip

{\bf Other inference transitions in $\mathbb{Z}_2$ gauge theory.} 
In 3D the phase diagrams of discrete gauge-Higgs theories can be 
 quite rich \cite{fradkin1979phase}. 
Refs.~\cite{serna2024worldsheet,somoza2021self} discuss some algorithmic problems involving defects in 3D gauge theory
which could be explored further using the present tools.
Even restricting to cases where the physical (``pre-measurement'') gauge field is non-critical,
the full phase diagram of the replicated gauge-Higgs model is nontrivial, as we discuss briefly in
Sec.~\ref{sec:replicatedGTmoregeneral}, and could be explored further. (One axis of this phase diagram is related to a standard error-correction process for the toric code in 2+1D spacetime \cite{dennis2002topological,wang2003confinement}.)

 \subsubsection{$\mathbb{Z}_2$ gauge-Higgs model --- review}
 \label{sec:Z2gtreview}
 
 For completeness we give a rapid review of the model discussed in the next couple of subsections. For a standard review see \cite{kogut1979introduction}. For a more detailed recap of the geometrical picture in 3D see e.g.\ the early sections of \cite{serna2024worldsheet,somoza2021self} as well as the pioneering paper \cite{huse1991sponge}.
 
 In Sec.~\ref{sec:isingparasubsec} above we discussed measurement of bond energies in an Ising paramagnet,
 \be
  \mathcal{H} = - J \sum_{\<x,y\>} S_x S_y.
 \ee
(We focussed on the paramagnetic phase, 
especially the simple case ${J=0}$.)
Next, we will consider the case where the spins are gauged. 
 In other words, we take the initial ensemble to be defined, on a hypercubic lattice in $d$ dimensions, by
 \be\label{eq:LGT}
 \mathcal{H} = - J \sum_{\<x,y\>} \sigma_{xy} S_x S_y 
 - K \sum_{\square} \sigma\sigma\sigma\sigma.
 \ee
Here $\sigma_{xy}=\sigma_{yx}=\pm 1$ is a $\mathbb{Z}_2$ gauge field living on the links of the lattice,
and ${S_x=\pm 1}$ is a Higgs field living on the sites.
We have written the last term schematically: it
  is a sum over the square plaquettes of the lattice, and $\sigma\sigma\sigma\sigma$ is the product of the gauge fields on the four links of a given plaquette.
  If ${\sigma\sigma\sigma\sigma=-1}$ for a plaquette, this plaquette is said to host a nonzero gauge flux.
  
  This theory \cite{wegner1971duality,
fradkin1979phase,
kogut1979introduction}  is sometimes referred to as a $\mathbb{Z}_2$ gauge--Higgs model.
 It has the gauge redundancy
${S_x \rightarrow \chi_x}$, 
 ${\sigma_{xy}\rightarrow\chi_x\chi_y \sigma_{xy}}$
 (where $\chi_x=\pm 1$ are 
 arbitrary gauge transformation parameters).
That is, configurations that are related by such a gauge transformation are regarded as identical.\footnote{We are free to pick a gauge in which $\sigma_{xy}=1$, 
but below it will be more convenient not to do this.}
This gauge redundancy should not be confused with the local symmetry $\mathbb{Z}_2^\text{local}$ which we encountered in the paramagnet--measurement problem (Sec.~\ref{sec:paramagnetsymmetry}), which related physically distinct configurations.

Configurations are most easily visualized on the dual lattice. 
In outline, the values of the gauge-invariant degrees of freedom $\{\sigma_{xy}S_xS_y\}$ 
can be represented in 2D by a configuration of ``strings'', made up of links on the dual lattice, 
and in 3D by a collection of ``membranes'' made up of plaquettes of the dual lattice.
See footnote for details.\footnote{In 2D, a link $\<xy\>$ of the original square lattice pierces an orthogonal link of the dual square lattice. By defining the dual link to be occupied if 
 ${\sigma_{xy}S_xS_y = -1}$  we map the configuration to a configuration of occupied links on the dual lattice. 
 We refer to these occupied links colloquially as making up strings (this terminology does not imply any restriction on the topology of the configuration --- the ``strings'' can intersect). 
  In 3D a link $\<xy\>$ pierces a plaquette of the dual lattice, and 
 a similar mapping gives a configuration of membranes, which just means a collection of occupied plaquettes on the dual lattice. A ``closed'' string configuration (a configuration of loops) means one where each site of the dual lattice is incident on an even number of occupied links. Similarly, a closed membrane configuration means one where each link of the dual lattice is adjacent to an even number of occupied plaquettes.}
 
 In 2D, a gauge flux (a plaquette with ${\sigma\sigma\sigma\sigma=-1}$) is the endpoint of a string.
 In 3D, the gauge fluxes form ``loops'' on the dual lattice and these loops are the boundaries of membranes.
 Therefore when $K=\infty$, so that there are no gauge fluxes,  
  the hypersurfaces (strings/membranes) are closed.
 
 In this limit, the model is closely related to the Ising model (with coupling $J$) \cite{wegner1971duality,
fradkin1979phase,
kogut1979introduction}. Essentially, the closed loops/membranes can be mapped to Ising domain walls.\footnote{Alternately: if $K=\infty$ then in any simply connected region we can choose the gauge $\sigma=1$, so that  the first term in (\ref{eq:LGT}) has the form of an Ising Hamiltonian. The model is an orbifold of the Ising model, meaning just that we have gauged the Ising model with a ``flat'' gauge field (a gauge field for which the gauge flux $\sigma\sigma\sigma\sigma$ is nonzero). One effect of this gauging is that there is no longer a local observable corresponding to the Ising spin. ($S_x$ is not gauge-invariant so is not a local observable.)} 

When $K=\infty$ and $J$ is small enough, the model is in a nontrivial  ``deconfined'' state.
This regime maps to the \textit{paramagnetic} phase of Ising in the above mapping,
but it is a thermodynamically nontrivial phase in the gauge theory.
This nontriviality can be diagnosed from the properties of the  topological line operators mentioned in  the preface to Sec.~\ref{sec:nishimoricloserelatives}, 
which are simply Wilson loops or open Wilson lines \cite{kogut1979introduction,gaiotto2015generalized,nussinov2009symmetry}.
Formally, the deconfined state is nontrivial because it has a spontaneously broken 
${(d-2)}$---form $\mathbb{Z}_2$ symmetry \cite{gaiotto2015generalized} that is associated with these topological operators.\footnote{In 2D this is a 0-form symmetry. Usually we say that a   $0$-form symmetry is just a conventional global symmetry. However in  App.~\ref{app:symmetrynote} we note a subtlety about the interpretation of such symmetries in classical, as opposed to quantum, models.
Classical statistical mechanics models come equipped with a preferred choice of basis for the transfer matrix, and this must be taken into account when we classify states according to their symmetries. Models which are ``equivalent'' 
in the formal quantum field theory sense 
(because they are related by a change of basis in the transfer matrix) 
may be inequivalent as classical statistical mechanics problems.
In the present context, the 0-form symmetry is not equivalent to what we usually mean by a global symmetry in a classical model (instead it is a constraint on the classical configurations).}

In 2D the deconfined state is unstable and the model becomes thermodynamically trivial  (confined) for  ${K<\infty}$. 
In 3D the deconfined state is stable and occupies an open region in the $(J,K)$ phase diagram for small enough $J$ and large enough $K$ 
\cite{wegner1971duality,kogut1979introduction,fradkin1979phase,TupitsynTopological,vidal2009low,somoza2021self}.

 \subsubsection{Measuring a $\mathbb{Z}_2$ deconfined state}
 \label{sec:measuringZ2deconfined}

 First we consider measurements in the model with ${J=0}$ and  ${K=\infty}$
 (the extreme limit of the deconfined regime/phase).
Specifically we consider Gaussian measurements of the basic gauge-invariant observable, i.e. $\sigma_{xy}S_xS_y$ on bonds.
This corresponds to measuring the string density in 2D  or the membrane density in 3D: see Figs.~\ref{fig:N3types2D}.

The general recipe  (Sec.~\ref{sec:generalities}) gives the replica Hamiltonian
\be\label{eq:Hreplicagaugetheory1}
\mathcal{H}_N =
- \lambda \sum_{a\neq b} \sum_{\<xy\>} \sigma^{ab}_{xy} X^{ab}_x X^{ab}_y
 - K \sum_a \sum_\square (\sigma\sigma\sigma\sigma)^a .
\ee
In the first term, $X^{ab}_x = S^a_x S^b_x$ (compare Eq.~\ref{eq:nishimorilandau}). 
We have similarly defined 
\be
\sigma^{ab}_{xy} = \sigma^a_{xy} \sigma^b_{xy}.
\ee
Again we have simplified the notation in the second term of (\ref{eq:Hreplicagaugetheory1}), where the replica index $a$ is to be understood to be carried by each $\sigma$.

Since $K=\infty$, the gauge fields $\sigma^a$ are deconfined for each $a$.
Neglecting boundary-condition effects, we can choose the gauge $\sigma_{xy}^a=1$. 
We then see that we formally  recover the 
replica theory for the standard Nishimori inference problem in Eq.~\ref{eq:nishimorilandau},
with a phase transition at some $\lambda_c$ 
at which $X^{ab}$ ``orders'',
with Nishimori exponents.

However, the physical interpretation here is slightly different.\footnote{In Eq.~\ref{eq:Hreplicagaugetheory1}, the replica theory for the Nishimori inference problem has been gauged with a 
flat ($K=\infty$) gauge field. 
We can say that we have an orbifold of the usual Nishimori critical point.
In another terminology we can say that we have a  ``starred''  
\cite{lammert1993topology,senthil2002microscopic,grover2010quantum,misguich2008quantum,moessner2001ising,slagle2014quantum, serna2024worldsheet} version
of the usual Nishimori replica theory.}
The condensation of $X$ is a Higgs transition rather than a standard ordering transition, since $X^{ab}$ is not gauge-invariant.

Physically, this is because the weak and strong measurement phases must be distinguished using Wilson loop operators 
rather than  local operators. 
As discussed around  Eq.~\ref{eq:VPexp},  we can distinguish the phases using 
$\mathbb{E}_M \< V_P\>_M^2$ for an open path $P$. 
Standard ideas \cite{kogut1979introduction} 
show that this expectation value becomes nontrivial when $X$ condenses.\footnote{In more detail: 
if $P$ is a path from $x$ to $y$, then $V_P=S_x (\prod_{\<wz\>\in P}\sigma_{wz}) S_y$.
Note that, in the replica theory, 
${\<V_P^a V_P^b\>} = 
{\< X^{ab}_x 
(\prod_{\<wz\>\in P}\sigma^{ab}_{wz})
X^{ab}_x\>}$. 
In the gauge  $\sigma^a=1$ 
(valid in the infinite system for $K=\infty$)
this becomes $\<X^{ab}_x X^{ab}_y\>$.}

The case with large but finite ${K}$  can also be understood using standard results for  discrete gauge theories \cite{kogut1979introduction}. 
In 2D, the weak and strong measurement phases immediately collapse to a single phase once ${K<\infty}$ 
(i.e. once the loops in Fig.~\ref{fig:N3types2D}, Top Center, become open strings).
All correlators of $V_P$ decay exponentially in the resulting regime, and this operator is no longer topological (invariant under deformation of $P$).

In 3D,  $1/K$ is an irrelevant perturbation at the critical point, and the distinction between the two phases survives.
While the ``bare''  operator $V_P$ is no longer topological, it can be replaced with a ``dressed'' Wilson line operator~\cite{serna2024worldsheet} which distinguishes the phases in the same manner as for ${K=\infty}$.
The geometrical interpretation is that, although the membranes are not strictly closed (so do not map to Ising domain walls at the microscopic scale) the holes in them are of a finite size and disappear under RG (so that the mapping to Ising domain walls, and Nishimori inference, is recovered in the IR).

\subsubsection{Measurement of fluxes in the confined state}
\label{sec:measuremembraneboundary}

In the previous section it was important that the gauge fields coupling to 
$X^{ab}$ were deconfined. 
But note that $X$ couples only to the ``interreplica'' gauge fields $\sigma^a \sigma^b$. Therefore, so long as these inter-replica fluctuations are suppressed,
we can have a nontrivial Higgs transition for $X^{ab}$, even if a single gauge field $\sigma^a$ is strongly fluctuating.
This setting is relevant to various error correction problems \cite{dennis2002topological,weinstein2024computational}, as summarized in the preface to Sec.~\ref{sec:nishimoricloserelatives}.

Consider the setup of Ref.~\cite{weinstein2024computational}, generalized to arbitrary $d$. 
This is an inference problem in which the information we are given is
the positions of the gauge \textit{fluxes}, i.e.\ the values of $\sigma\sigma\sigma\sigma$ on plaquettes.
We consider the regime where the gauge field is confining:
in 2D this just means that $K$ is finite rather than infinite.
Assume that the information about the gauge fluxes is  perfectly reliable, which corresponds to measuring these operators with a measurement strength ${\lambda_\square =\infty}$,  the replica Hamiltonian is
\ba\notag
\mathcal{H}_N  = & \,
- J \sum_{a} \sum_{\<xy\>} \sigma^a_{xy} S^a_x S^a_y
\\
& - K \sum_a \sum_\square (\sigma\sigma\sigma\sigma)^a
- \lambda_\square  \sum_{a\neq b}  \sum_\square (\sigma\sigma\sigma\sigma)^{ab}
\label{eq:Hreplicagaugetheory2}
\end{align}
in the limit $\lambda_\square \to \infty$.
The $\lambda_\square$ term involves the product of $\sigma^{ab}=\sigma^a \sigma^b$ over the four links of a plaquette.
The $K=0$ case can be mapped to the 2D toric code (single-round) error correction problem \cite{dennis2002topological} by viewing occupied links as errors.

The transition will be driven by varying $J$. Here we consider only the phase-diagram structure and assignment of universality classes; the operational meaning of these kinds of transition is discussed in Refs.~\cite{weinstein2024computational,dennis2002topological}.
In outline, if $J$ is sufficiently large, then we can estimate $V_P$ using the flux measurement information.\footnote{Unlike in the previous section, $V_P$ is not a topological operator (its value is not invariant under deformation of $P$). However, the perfect measurement of fluxes means that the product $V_P^1 V_P^2$ over two replicas is a topological operator.}

Let us separate out the replica-uniform and replica-asymmetric gauge fluctuations by writing
\be
\sigma^a_{xy} = \sigma_{xy} \tau^a_{xy},
\ee
where, to avoid redundancy, $\tau^1_{xy}=1$.
Since $\lambda_\square =\infty$, 
we have $(\tau\tau\tau\tau)^a=1$ for all $a$, and the nontrivial terms in the Hamiltonian are then\footnote{As usual we  may set $N\to 1$ in the coefficients of the Hamiltonian for the purposes of understanding phase diagrams.}
\ba\notag
\mathcal{H}_N  =
- J \sum_{a} \sum_{\<xy\>} \sigma_{xy} \lf \tau_{xy}^a S^a_x S^a_y \ri
 - K \sum_a \sum_\square (\sigma\sigma\sigma\sigma)^a.
\label{eq:Hreplicagaugetheory2prime}
\end{align}

We have assumed that we are  at finite $K$ in 2D, or small enough $K$ in 3D,
such that the gauge field $\sigma$ is confining.
This means that at large scales we can work with an effective Hamiltonian that only includes fields that do not carry gauge charge under $\sigma$. This is simplest to see when $K=0$ and $J$ is small, when 
\ba\notag
\mathcal{H}_N  \simeq 
- \f{J^2}{2} \sum_{a\neq b} \sum_{\<xy\>} \tau_{xy}^{ab} X^{ab}_x X^{ab}_y  - \lambda_\square \sum_{a\neq b} \sum_\square (\tau\tau\tau\tau)^{ab}
\end{align}
as we see by doing the sum over $\sigma$ separately for each link.
We have restored the $\lambda_\square$ term explicitly to emphasise that this is a gauge-Higgs model for the ``replica-asymmetric'' fields, in the limit $\lambda_\square\to\infty$ where  replica-asymmetric gauge fluctuations are suppressed.

When $K$ is nonzero the effective Hamiltonian is not strictly short range.
However,  this should not matter for the universal physics.\footnote{For example it is straightforward to check that making $K$ nonzero does not lead to any new relevant perturbations at the critical point.}
By increasing $J$, we expect to be able to drive a Higgs transition for $X$.

Ref.~\cite{weinstein2024computational} noted that in the special case $K=0$ a relation with the Nishimori line 
 could be established directly, i.e. without using replicas (see also \cite{dennis2002topological}), 
 while in the limit $K=\infty$ a conventional Ising transition was obtained.
The generic universality class of the transition was left open.
 Here we find that 
 the universal behaviour in the ``inter-replica'' sector
is identical to that in Sec.~\ref{sec:measuringZ2deconfined}
(despite the fact that the control parameter is now a physical coupling, $J$, rather than a measurement strength).
That is, we expect Nishimori exponents to apply generically for this transition 
in 2D, for any finite $K$.

In 2D this transition only exists in the limit of   infinite $\lambda_\square$, 
as otherwise the inter-replica gauge field $\tau^{ab}$ becomes confined.

In the analogous 3D problem, however, (i.e.\ at small enough $K$) 
the ``gauged Nishimori''  transition extends to finite $\lambda_\square$, 
since the deconfined phase for $\tau^{ab}$ is stable.

\subsubsection{More general phase diagrams in 3D}
\label{sec:replicatedGTmoregeneral}

To end the discussion of inference in $\mathbb{Z}_2$ gauge theory, 
we briefly note that  in three dimensions the full phase diagram includes more types of transition than those discussed above.

For simplicity we restrict here to the simplest possibility for the initial ensemble, where both $J$ and $K$ are zero,
\be
\mathcal{H}=0,
\ee
but we expect the universality classes of transitions that arise are representative of a broader class of ``paramagnetic'' initial states.

The terms in the replica Hamiltonian then come entirely from measurement. 
We combine the  bond measurements of Sec.~\ref{sec:measuringZ2deconfined} with the flux  measurements on plaquettes of Sec.~\ref{sec:measuremembraneboundary} with arbitrary strengths $(\lambda, \lambda_\square)$:
\be\label{eq:Hreplicagaugetwolambdas}
\mathcal{H}_N = 
- \lambda \sum_{a\neq b} \sum_{\<xy\>} \sigma^{ab}_{xy} X^{ab}_x X^{ab}_y
- \lambda_\square  \sum_{a\neq b} \sum_\square (\sigma\sigma\sigma\sigma)^{ab}.
\ee
This is a gauge-Higgs theory with gauge group\footnote{We write $(\mathbb{Z}_2)^{N-1}$ rather than $(\mathbb{Z}_2)^N$ because the replica-uniform gauge fluctuations have dropped out of the Hamiltonian and only the inter-replica gauge fluctuations appear.} $(\mathbb{Z}_2)^{N-1}$.

By analogy with simpler gauge theories,
 we   expect (in 3D)  two absolutely stable phases:
 a  deconfined phase 
 (in the corner of the phase diagram where $\lambda_\square$ is sufficiently large and $\lambda$ is sufficiently small)
 and a trivial phase.
Again by analogy we would expect distinct  Higgs and confinement transitions out of this phase (perhaps meeting at a multicritical point).
The Higgs transition is driven by condensation of $X$ (e.g.\ by increasing $\lambda$) and the confinement transition is driven by fluctuations of the gauge field.

We leave the analysis of this 2D phase diagram to the future. 
Here, to make contact with Refs.~\cite{dennis2002topological,wang2003confinement}, we comment briefly on the  ${\lambda=0}$ axis, where 
(\ref{eq:Hreplicagaugetwolambdas}) becomes a pure gauge theory,
\be\label{eq:Hreplicapuregauge}
\mathcal{H}_N = 
- \lambda_\square  \sum_{a\neq b} \sum_\square (\sigma\sigma\sigma\sigma)^{ab}.
\ee
We expect the confinement transition to occur at some $(\lambda_\square)_c$.\footnote{This is the limit $\lambda\to 0$ of the confinement transition line in the more general $(\lambda, \lambda_\square)$ phase diagram. The universality class at $\lambda=0$ presumably matches that for small nonzero $\lambda$.  We could also restore a small ``physical'' gauge coupling $K$ without affecting the following discussion.}
The confinement transition in this  3D effective field theory is 
relevant to a \textit{dynamical} process, in two spatial dimensions,  describing error correction in the toric code, which Refs.~\cite{dennis2002topological,wang2003confinement}
related to a Nishimori line in a disordered gauge theory. We expect that model to have the same exponents as (\ref{eq:Hreplicapuregauge}).

The pure gauge theory in Eq.~\ref{eq:Hreplicapuregauge} describes a problem in which we imperfectly measure the density of unoriented ``loops'' 
(more precisely, we measure the links of a $\mathbb{Z}_2$ one-chain \cite{dennis2002topological}) in 3D. 
We may discard the Higgs fields, which do not appear in the Hamiltonian or in the measured operators. 
The gauge-invariant observables are then only the gauge fluxes $\sigma\sigma\sigma\sigma$ on plaquettes, which map to a loop configuration on the dual lattice. 
Measurements are made of loop density on the dual links.
In the error-correction interpretation, the loops are (heuristically) worldlines of excited defects (anyons), which are imperfectly measured.
As in the 2D case, this problem
--- involving  measurement of  the string density for  \textit{closed} loops --- is also closely related to a problem involving measurement of the density of \textit{open} loops (App.~\ref{app:gaugedual}).

In 3D, pure discrete gauge theories are dual to theories without any gauge fields \cite{wegner1971duality}. For the case of  
Eq.~\ref{eq:Hreplicapuregauge},
the dual order parameter is a spin ${(T^1,\ldots,T^N)\in \{+1,-1\}^N}$ satisfying ${T_1T_2\cdots T_N=1}$, as we discuss in App.~\ref{app:gaugedual}. 
This description may make it possible to study the confinement transition using Landau theory techniques.

\subsection{Continuous symmetry and other generalizations}
\label{sec:ctsnishimori}

So far we have discussed Nishimori-Inference-like problems with discrete symmetries ---
we briefly mention an example with continuous symmetry
 \cite{abbe2018group,garban2022continuous}  which it may be interesting to investigate with field theory.
 
 Each site hosts an  independent  Haar-random element from a Lie  group  such as $\mathrm{SU}(q)$ or $\mathrm{SO}(q)$.
 Measurements are of the group elements (or basis changes) that relate matrices on adjacent sites \cite{abbe2018group}. This problem is known as group synchronization and the $\mathrm{SO}(q)$ version has many  applications in image processing \cite{wang2013exact,govindu2001combining,martinec2007robust}.

Restricting to $\mathrm{SU}(q)$, we could measure two quantities on links
\ba
W_{xy} & = V_x^\dag V_y, & 
\overline{W}_{xy} & = V_x  V_y^\dag
\end{align}
(note that the bar does not denote conjugation; $W_{xy}^\dag = W_{yx}$).
Perfect knowledge of either $\{ W\}$ or $\{\overline{W}\}$ is sufficient to ``reconstruct'' the matrices (up to a global transformation).
For example, given $V_0$, any other matrix $V_x$ can be obtained using a product of  $\overline{W}$ matrices along a path or a product of ${W}$ matrices along a path.
Rigorous results (for the case where $W$  is measured) show that in 3D  there is a stable phase at sufficiently accurate measurement where reconstruction is possible \cite{abbe2018group} and that the ``corresponding''  disordered system
(Sec.~\ref{sec:rbimmapping})
has a long-range-ordered phase \cite{garban2022continuous}.

In the replica formalism for Gaussian measurements,
\ba
\mathcal{H}_N & = - \sum_{\<xy\>} \sum_{a\neq b} 
\lf 
\lambda \, \tr \, W_{xy}^a W_{yx}^b
+
\overline{\lambda} \, \tr \, \overline{W}_{xy}^a \overline{W}_{yx}^b
\ri
\\
& = -  \sum_{\<xy\>} \sum_{a\neq b} 
\lf 
\lambda \, \tr \, X_{x}^{ab} X_y^{ba}
+
\overline{\lambda} \, \tr \, \overline{X}_x^{ab} \overline{X}_y^{ba}
\label{eq:matsyncreplicaH}
\ri,
\end{align}
showing that the natural order parameters are collections of $\mathrm{SU}(q)$ matrices
\ba
X^{ab}_x & = V_x^{a} V_x^{b\dag}, 
& 
 \overline{X}^{ab}_x & = V_x^{a\dag} V_x^{b}. 
\end{align}

The symmetry structure is richer than in   Sec.~\ref{sec:paramagnetsymmetry}. 
The $V_x$  are Haar-random and uncorrelated (${\mathcal{H}=0}$) so  the initial ensemble is invariant under  \textit{local} ${\mathrm{SU}(q)_L \times \mathrm{SU}(q)_R}$ transformations,\footnote{More precisely the local symmetry is $[\mathrm{SU}(q)_L \times \mathrm{SU}(q)_R]/\mathbb{Z}_q$ since  transformations in the centre of $\mathrm{SU}(q)$ can be implemented via either $L$ or $R$.}
\ba
V_x & \rightarrow L_x V_x R_x^\dag, 
&
L_x, \, R_x & \in \mathrm{SU}(q).
\end{align}
The $W$ measurements are invariant under $\mathrm{SU}(q)^\text{global}_L$, and covariant under $\mathrm{SU}(q)^\text{local}_R$
(see Sec.~\ref{sec:symmetry}), and the converse is true for the $\overline{W}$ measurements. As a result, the two terms in (\ref{eq:matsyncreplicaH}) preserve different symmetries
\ba
\text{$\lambda$ term:} \quad\,  &[\mathrm{SU}(q)^\text{global}_L]^N\times \mathrm{SU}(q)_R^\text{local},
\\
\text{$\overline{\lambda}$ term:} \quad\, & \mathrm{SU}(q)^\text{local}_L \times [\mathrm{SU}(q)_R^\text{global} ]^N.
\end{align}
It will be interesting to explore this structure, and possible emergent symmetries in the IR, further.

The simplest case is where only one measurement strength, say $\lambda$, is nonzero. Then it is natural to formulate a continuum nonlinear sigma model for the order parameter $X$,
${\mathcal{H} = \f{1}{g}\int \sum_{a\neq b} \tr (\nabla X^{ab})(\nabla X^{ba})}$.
In the nontrivial phase, we expect $X$ to develop an expectation value that breaks ${[\mathrm{SU}(q)_L^\text{global}]^N}$ down to the diagonal subgroup (compare Sec.~\ref{sec:paramagnetsymmetry}).

Standard dimensional analysis shows that in 3D the sigma model has such a long-range ordered phase (i.e a phase where reconstruction is possible --- see Sec.~\ref{sec:paramagnetsymmetry}). This agrees with the rigorous result \cite{abbe2018group,garban2022continuous}.
Two dimensions is the marginal dimensionality for continuous symmetry breaking, so we must examine the beta function for the sigma model in more detail to determine the topology of the 2D phase diagram --- we will discuss this elsewhere. 

Another class of toy models that may be tractable (using the sigma model in the present case, or domain-wall RG in  cases with   discrete symmetries \cite{anderson1970exact}) is obtained by 
allowing measurements of $W_{xy}$ for distant site pairs in \textit{one} dimension, giving a long-range replica theory.

\section{Imaging polymers and cluster connectivities}
\label{sec:imagingpolymers}

In this section we consider measuring, imperfectly, the spatial density of a polymer chain. We also consider imaging percolation configurations.

These setting provides simple examples in which the weak-measurement regime and the strong-measurement limit can both be understood. 
It is possible to understand these examples without using field theory, but we put them in field-theory language because they nicely illustrate  that the field-theory degrees of freedom appropriate for the strong-measurement limit may be different from those appropriate for the weak-measurement limit (similar structures may be relevant to other problems where the phase diagram is not so easy to guess).

Unlike the problems in the preceding sections, many of the natural observables in this Section are nonlocal, relating to the way in which local segments of the polymer connect up, or to the connectivity of distant percolation sites.
However, the formalism is contiguous with that of the previous Sections since, as is well known (see e.g.\ \cite{cardy1996scaling}) many of these geometrical observables may be related to standard field-theory obervables by invoking a separate  limit in an additional ``integer'' parameter $m$ (not to be confused with the replica limit for $N$).

\subsection{Polymers}

There are many variants of the polymer problem, depending on the dimensionality and the polymer's interactions. 
First we will consider a polymer with excluded volume interactions 
(the universality class of the ``self-avoiding walk'' or of a polymer in a good solvent \cite{cardy1996scaling}). 
Here we see that weak density measurements are enough to reveal the coarse-grained geometry with essentially perfect precision. 
Next we consider a polymer at the $\Theta$ point in 3D
(where the leading interaction term is tuned to zero),
which is essentially Brownian at large scales. 
In this case, weak measurements are still RG relevant, but they do not fully reveal the polymer's coarse-grained geometry.

The partition function for a self-avoiding walk (of some length $
\ell$) on a finite square or cubic lattice is given simply by summing over all possible configurations of the walk that do not visit any site more than once. 
At large scales this is a random fractal: in two dimensions it has  fractal dimension  $d_f = 4/3$, while in three dimensions $d_f$ is close to $5/3$ \cite{cardy1996scaling}. 
Each bond is  visited  either once or zero times, so the configuration can be characterized by densities $\rho_x=0,1$,
where for convenience we take $x$ to run over bonds here. 
We imagine a Gaussian measurement of the local densities with strength (inverse variance)~$\lambda$.

A well-known framework  relates  long self-avoiding walks to the critical $O(m)$ field theory,\footnote{More precisely, the field theory at a fixed mass maps to a partition function for a polymer with a fixed fugacity for the length rather than a fixed length \cite{cardy1996scaling}. 
The typical polymer length diverges
as the mass$^2$ tends to zero from above.} for a field $\phi_\alpha$, with ${\alpha = 1,\ldots, m}$,  in the limit $m\to 0$ \cite{cardy1996scaling,degennes1972exponents}:
\be\label{eq:Om}
\mathcal{H} =  \f{1}{2} \int \sum_\alpha (\nabla \phi_\alpha)^2 + g  \int\lf \sum_\alpha \phi_\alpha^2\ri^2.
\ee
The polymer can be thought of as a ``wordline'' of the field  (or, in a lattice formulation, as a graph appearing in a high-temperature expansion). 
When $m$ is an integer greater than zero, this worldline carries a ``flavor'' index $\alpha$ which runs over $m$ values.
Closed loops therefore have a weight $m$, and the limit $m\to 0$ may be used to isolate a single polymer.
The lattice density $\rho_x$ maps onto the continuum operator  $\vec\phi^2$ (up to less relevant terms), and in the limit $m\to 0$ the scaling dimension of this operator is  ${x_{\phi^2}= d-d_f}$.
Note that $m$ should not be confused with the replica index $N$ (which has not yet been introduced  in Eq.~\ref{eq:Om} since that theory describes the unmeasured ensemble).

The usual result for the RG eigenvalue of the measurement strength in terms of the scaling dimension of the measured operator (Sec.~\ref{sec:rgsmalllambda})
 gives ${y_\lambda=2 d_f - d}$, i.e. 
 ${y_\lambda = 2/3}$ in two dimensions and ${y_\lambda \simeq 1/3}$ in three dimensions.
 Therefore weak measurement is relevant.

Where does this RG flow lead? 
We argue that the fixed point at infinite measurement strength is stable, 
and that weak measurement ($\lambda\ll 1$) 
flows to infinite measurement strength ($\lambda=\infty$). 
So, while measurement noise may prevent us from inferring the position of every monomer, the coarse-grained geometry is fully determined. More concretely, there is a lengthscale $L_* \sim \lambda^{-1/y_\lambda}$ beyond which measurements become informative.\footnote{The renormalized measurement strength at scale $L$ is $\sim\lambda L^{y_\lambda}$, leading to a lengthscale
$L_* \sim  \lambda^{-1/y_\lambda}$
at which the coarse-grained measurement strength is of order 1. 
We interpret this as a characteristic scale above which the spatial structure of the polymer can be reliably inferred.
For example, consider a  polymer chain of length $\ell>L_*$.
What is the probability that the geometry of this chain is ambiguous even on the largest scale $\ell$?
It will be ambiguous if the chain, \textit{after coarse-graining to this scale}, has a self-approach (see the discussion later in the section). 
In the field theory language, this
probability is determined by the (negative) RG eigenvalue $y_4$ of the four-leg operator: $P_\text{ambiguity}\sim (\ell/L_*)^{-|y_4|}$.}

In heuristic terms, the ${\lambda=\infty}$ fixed point is stable for a simple reason.
When measurements are strong but not perfect (${1\ll \lambda< \infty}$), 
the main cause of ambiguity in inferring the structure of the polymer arises from self-approaches, 
where the polymer approaches the vicinity of the same point twice. 
(Fig.~\ref{fig:percolationmeasurement} below illustrates a closely related phenomenon in a different geometrical model.)
But a well-known property of the polymer is that such self-approaches become more and more rare under coarse-graining. 
Formally, there is an RG eigenvalue $y_4$ associated with such events, which  is negative. 

Let us describe this in field theory terms.
We will initially  restrict to the 3D case, because the 2D case has an additional subtlety arising from the fact that the polymer does not cross itself.

First, when we have a single replica ($N=1$), we have a single walk.  When we view this as a worldline of the $O(m)$ model, it acquires a flavor index $\alpha$ that runs over $m$  values.

Now consider the replica theory.
When $\lambda=0$, we have $N$ independent walks, 
carrying flavor indices $\alpha_1, \ldots, \alpha_N$.

When  ${\lambda\to\infty}$ the spatial conformations of these walks become identical:
so we are back to a partition function for a single walk. 
However, this walk still carries the flavor indices from all the replicas, i.e.\ it is labelled by a multi-index $(\alpha_1, \ldots, \alpha_N)$, which can take $\mathcal{M}=m^N$ values.
The ``replica-locked'' walk may therefore be described by an $O(\mathcal{M})$ model with Hamiltonian density 
\be\label{eq:OmN}
\f{1}{2}\sum_{\alpha_1, \ldots, \alpha_N}(\nabla \Phi_{\alpha_1, \ldots, \alpha_N})^2
+ g \lf \sum_{\alpha_1, \ldots, \alpha_N} ( \Phi_{\alpha_1, \ldots, \alpha_N})^2 \ri^2
\ee
(neglecting the mass, which will tend to zero).

\begin{figure}
    \centering
\includegraphics[width=0.35\columnwidth]{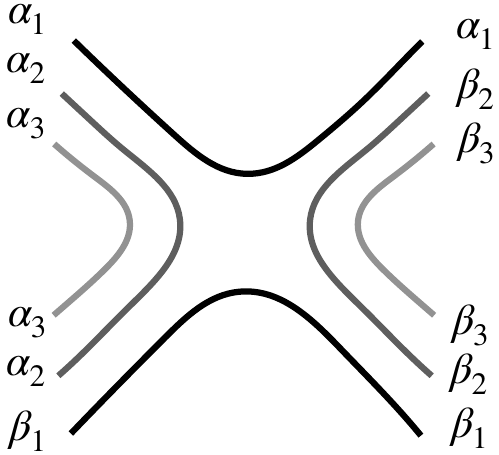}
    \caption{Schematic of the local reconnection event discussed in main text, for the case ${N=3}$.}
\label{fig:reconnection}
\end{figure}

Now consider reducing $\lambda$ away from $\infty$. 
This allows local events with the topology  in Fig.~\ref{fig:reconnection}, where one of the replicas reconnects differently from the others.
This diagram can be thought of a vertex in the field theory, and it corresponds to a perturbation
\be\label{eq:walkexchangeperturbation}
\Phi_{\alpha_1 \alpha_2 \ldots \alpha_N}
\Phi_{\beta_1 \beta_2 \ldots \beta_N}
\Phi_{\beta_1 \alpha_2 \ldots \alpha_N}
\Phi_{\alpha_1 \beta_2 \ldots \beta_N}
+ \cdots.
\ee
(The ellipses represent symmetrization with respect to $S_N$ replica permutations and possible  subtraction of an $O(\mathcal{M})$ singlet term.)
Further, for arbitrary $N$ there are arbitrarily many further quartic terms, representing more complex kinds of reconnection.\footnote{In a given replica there are three ways of connecting the four outgoing polymer strands. We must assign each replica to one of the three patterns.
In three dimensions, the topologically distinct kinds of assignment 
(i.e., after identifying patterns that differ by permuting replicas or by permuting the different connectivities)
are labelled by integers $(n_1, n_2, n_3)$ with $N\geq n_1 \geq n_2 \geq n_3 \geq 0$ and $n_1+n_2+n_3=N$. } 
However, these quartic terms are all of the form $\Phi_A \Phi_B\Phi_C\Phi_D$ where $A,B,C,D$ are
 \textit{distinct} values for the multi-index.
 These operators are known as four-leg operators in the polymer language, and are irrelevant \cite{duplantier1989statistical}.\footnote{They belong to the four-index symmetric tensor representation of $\mathrm{O}(\mathcal{M})$, which is RG-irrelevant when $\mathcal{M}\to 0$. (This is  the limit of interest, since $\mathcal{M}=m^N$ and  $m\to 0$, $N\to 1$.)}
 
 As a result, the infinite-measurement fixed point is stable, and the likely situation is that weak measurement flows to this fixed point.  The 2D case may be discussed similarly. Again the irrelevance of the four-leg operator guarantees the stability of the strong-measurement limit.

 To see that the RG flows are not always of this type, consider the case where we turn on physical interactions for the  polymer to tune it to the ``$\Theta$ point'' (in three dimensions). This  means that the (renormalized) coefficient $g$ in Eq.~\ref{eq:Om} is tuned to zero. 
 In 3D the sextic term is marginally irrelevant, so the theory is free in the infra-red: i.e.\ the walk becomes essentially a Brownian path. (We will neglect corrections due to the marginally irrelevant coupling.)
 
The fractal dimension is $d_f=2$, so weak  density measurements are strongly relevant with $y_\lambda=1$.

What about the strong-measurement fixed point?
We must consider the quartic perturbation (\ref{eq:walkexchangeperturbation}) for the case where $\Phi_{\alpha_1,\ldots, \alpha_N}$ is simply a free field. 
Unlike the previous case, this perturbation is strongly relevant in three dimensions.

An interpretation is that weak measurements reveal the coarse-grained \textit{density} of the polymer, but they cannot reveal its coarse-grained \textit{geometry}, because there are many ambiguities about how the polymer strands are locally connected up.

\subsection{Percolation}
\label{sec:percolation}

\begin{figure}
    \centering
\includegraphics[width=0.93\columnwidth]{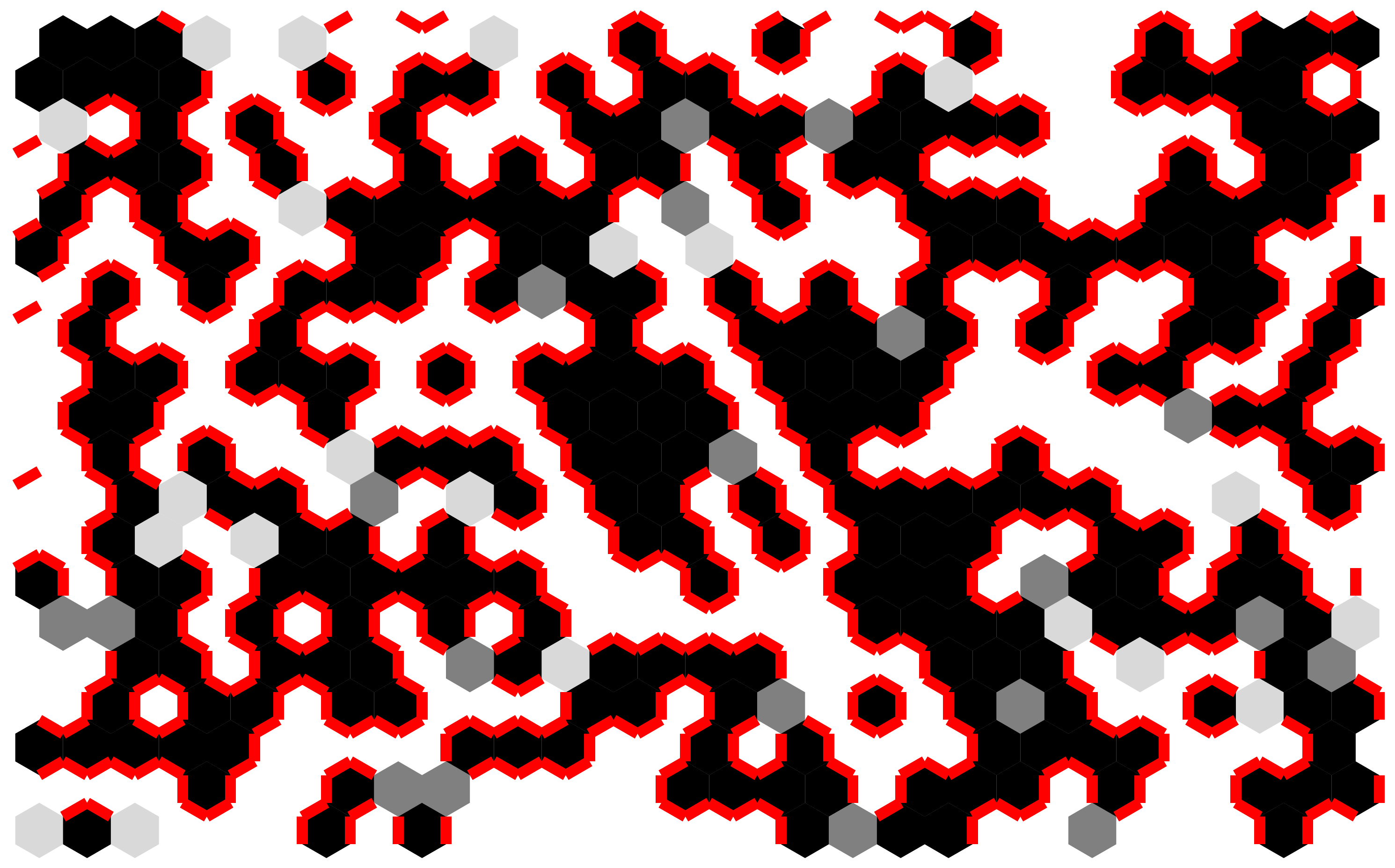}
    \caption{Critical site-percolation configuration as described in the main text.}
\label{fig:percolationmeasurement}
\end{figure}

As a final example of field theory for measurement of  geometrical configurations, 
we consider percolation.
Fig.~\ref{fig:percolationmeasurement} illustrates a measurement problem in which we measure the site occupancies in triangular-lattice site percolation and attempt to infer long-distance connectivities.

To simplify the picture, we show a protocol where some sites are known to be measured with perfect accuracy (black $=$ occupied, white $=$ unoccupied) 
while for some sites (gray) we have no information.
(Since we know which sites we lack information about, this is like ``heralded errors'' in error correction.)
For the latter sites, the true occupancy,
unknown to the measurer,
is indicated via dark vs.\ light gray.
Red lines show the known cluster boundaries.

The picture illustrates the fact that, when errors are rare,
the main cause of ambiguity in the long-distance connectivity arises from sites where four segments of cluster boundary meet.
It is easy to argue, from standard properties of percolation configurations (essentially the fact that such sites have a positive fractal dimension, $d_4 = \tfrac34$)
that any finite error rate leaves us ignorant of the connectivities at large scales.
Nevertheless, it is interesting to formalize the problem using field theory.
One formulation uses a nonlinear sigma model representation of 2D percolation and  is described in App.~\ref{app:geometricalstrongmeasurement}.
In addition to the replica number $N$, this theory has a second replica-like number $m$ ($m\to 1$) which is required, even in the unmeasured theory, in order to express nonlocal geometrical correlation functions.
If we start at very strong but finite measurement, the RG flow goes from a single $CP^{\mathcal{M}-1}$ sigma model in the UV to a tensor product of $N$ copies of the $CP^{m-1}$ model in the IR.
Here ${\mathcal{M}=m^N}$.
This RG flow may have analogs in other replica sigma models.

\section{Other physical applications of the conditioned ensembles}
\label{sec:otherinterpretations}

\subsection{Partial quench}
\label{sec:partialquench}

Let us give an alternative interpretation of the  formulas  in Sec.~\ref{sec:generalities}, which does not require the language of measurements, and which is   useful for  Monte Carlo simulations of the conditioned ensembles.

First consider the special case where $\Delta\to 0$, so that the observable  $\measO^a_x$ is equal for all replicas (giving a constraint in the replica partition sum).
One way to think about the replicas is then as follows. 

We first equilibrate a sample $S^1$ using the original physical Hamiltonian. 
We then freeze the degrees of freedom $\{\measO_x\}_x$ to the values they take in configuration $S^1$. 
As an example, if our model contained two different species of spins, then this could mean freezing one of the two species.

We then rethermalize the remainder of the degrees of freedom (still using the original Hamiltonian) to get a new sample $S^2$. The correlations between $S^1$ and $S^2$ are the same as those of the replicas above.
This can be continued to larger numbers of replicas.
The replica correlation functions give information on the post-quench distribution.

For example, the efficacy of the quench in freezing the local spin fluctuations will be reflected in the correlator in Eq.~\ref{eq:spinuncertaintyreplica}.

An alternative quench interpretation of the formulas in Sec.~\ref{sec:gaussianmeasurements}, which does not require the limit ${\Delta\to 0}$, is to promote the measurement outcomes $\{M_x\}_x$ to additional physical degrees of freedom, i.e.\ to interpret 
$\mathcal{H}_\text{meas}(S,M)$ in  
 Eq.~\ref{eq:Hmeasdefn}
as a \textit{physical} Hamiltonian for coupled degrees of freedom $S$ and $M$. 
The quench is then the quench of the $M$ degrees of freedom, leaving the $S$ degrees of freedom to evolve.

Remarkably, the vulcanization process for rubber, 
i.e.\ its transformation from a liquid to a solid by the formation of chemical bonds that connect different polymer segments, is a physical example of a ``partial quenching process'', at least in an idealized limit where cross-links are formed instantaneously \cite{deam1976theory,goldbart2004sam,goldbart2000random}. 
The statistical mechanics of the vulcanized state was addressed using an ${N\to 1}$ replica limit in  Refs.~\cite{deam1976theory,goldbart2004sam,goldbart2000random, castillo1994distribution,
goldbart1996randomly,
peng2000renormalization,
janssen2001relevance,
peng2001connecting,xing2013generalized}.
A difference between that problem and the partial quenches discussed immediately above is that the constraints in the rubber problem are not associated with spatially localized degrees of freedom. As a result, the replica theory has a very different structure: for example, the relevant order parameter has its spatial argument replicated, rather than having the field itself replicated \cite{goldbart2000random}.

\subsubsection{Measuring $c_\text{eff}$}
\label{sec:meas_ceff}

Related ideas can be used to  measure the effective central charge of Eq.~\ref{eq:ceffectivemeasurement}
(the non-trivial derivative term $c'(N)$ in the replica limit $N \to 1$) numerically.

The central charge is notoriously difficult to obtain in Monte Carlo simulations, where it requires constructing a lattice discretisation of the stress-energy tensor. While this can be done in simple models \cite{BastiaansenKnops1998} it does not look promising in the case at hand.

By contrast, $c$ is simply related to the finite-size scaling of the free energy $f_L$ per unit area on a semi-infinite cylinder of circumference $L$ \cite{Affleck1986,BloteCardyNightingale1986}:
\ba\label{eq:FSS}
 f_L = f_\infty - \frac{\pi c}{6 L^2} + o(L^{-2}).
\end{align}
The latter is readily obtained from the leading eigenvalue $\Lambda_0$ of the corresponding row-to-row transfer matrix, as $f_L = -\tfrac{1}{L} \ln \Lambda_0$.
In the case of quenched disorder the derivative of the replica free energy is $\left. f_L'(N) \right|_{N\to0} = -\tfrac{1}{L} \ln \Xi_0$, where $\Xi_0$ is now the leading Lyapunov exponent of a product of random matrices, still describing the transfer in the cylinder geometry, but now depending on the  quenched randomness. For instance, for a random-bond problem the transfer matrix would depend on the random realisation of coupling constants for the corresponding row of the lattice. The finite-size scaling of $\left. f_L'(N) \right|$ then produces $c_\text{eff} = c'(0)$, as in Eq.~\ref{eq:FSS}. The feasibility of this protocol as a numerical scheme was demonstrated in \cite{CardyJacobsen1997,JacobsenCardy1998}.

The issue of $c_\text{eff}$ for a critical point in a problem of imperfect measurements is slightly more involved, because of the need to sample from the nontrivial correlated distribution of measurement outcomes.
This has been addressed for the quantum measurement phase transition in Ref.~\cite{zabalo2022operator}.

 In the classical system, we suggest first making a Monte Carlo simulation of the system on a cylinder of size $L \times L_\infty$, with $L_\infty \gg L$. 
Performing the measurements gives outcomes $\{M_x\}_x$.
We can then  compute the leading Lyapunov exponent $\Xi_0$ for the corresponding  system,
using the transfer matrix for the physical  degrees of freedom $S$, with the values  $\{M_x\}_x$   quenched to their previously measured values. This computation can be made via exact diagonalisation (transfer matrices), or alternatively by an approximate method based on linear operators that evolve the system in imaginary time along the $L_\infty$ direction, such as DMRG. 

\subsection{Real space RG and RG-breaking transitions}\label{sec:RSRG}

We now describe a connection between conditioned ensembles and the real-space renormalization group (RSRG). 

In the  RSRG for the Ising model \cite{kadanoff2000statistical}, the spins $S_x$ are grouped into blocks (labelled by $X$) for which  block spins $S_X'$ are defined.
Block spins are determined from microscopic spins either via a deterministic rule, such as the  the majority rule, or more generally by a probabilistic one: 
$S'_X$ is drawn from a distribution $P(S'_X|S_{X,1},\ldots, S_{X,b^d})$, where $b^d$ is the number of spins in the block.
We denote the product (over blocks) of these conditional probabilities by $P(S'|S)$.
Then renormalized Hamiltonian $\mathcal{H}'$ 
is defined by a  partition sum  for $S$ of the form:
\ba\label{eq:blockH}
e^{-\mathcal{H}'[S']}
= 
\sum_{S} e^{-\mathcal{H}[S]} P(S'|S).
\end{align}
(In the special case where the block spins are deterministic functions of the microscopic spins, then the right-hand side is just the constrained sum over $S$, with the block spins $S'$ held fixed.)
Summing over $S'$ shows that the original partition function is preserved.

Note that the right-hand side determines an ensemble for $S$ which depends on the values of $S'$. For a given choice of $S'$, this ensemble is precisely of the form defined by $\mathcal{H}_\text{meas}[S,M]$ in Eq.~\ref{eq:Hmeasgeneral}, 
if we identify the block spin values $S'$ with the measurements $M$. 
(This identification is legitimate since $S'_X$ only depends  on the spins in a local region, i.e.\ $S'_X$ corresponds formally to a local measurement.)
What is the relation between the universal properties of the conditioned ensemble and the properties of the RSRG?

In order for the RSRG to be useful, the coarse-grained Hamiltonian $\mathcal{H}'[S']$ must remain \textit{quasilocal} \cite{kennedy1993some,ould1997effect}:
in some sense, the amplitude of long-range couplings should decrease exponentially with distance.
Heuristically, we expect $\mathcal{H}'[S']$ to be quasilocal if the conditioned ensemble in Eq.~\ref{eq:blockH} is in a trivial  phase with exponentially decaying correlations \cite{kennedy1993some,ould1997effect}.
For the majority-rule transformation of the square lattice Ising model, 
this has been established for some specific configurations of $S'$ in Refs \cite{kennedy1993some,ould1997effect}.

A strict requirement on the RG transformation 
would be that the conditioned ensemble for $S$  has exponentially decaying correlations for \textit{any} configuration of the block spins $S'$. 
However, it is natural to expect that a weaker requirement is sufficient for many purposes, namely that the conditioned ensemble for $S$ has exponentially decaying correlations for a \textit{typical} configuration $S'$.
Then we are in precisely the situation considered in this paper, where we investigate the properties of the spins conditioned on typical measurement outcomes.
Let us define an RSRG transformation that obeys the above weaker requirement, for a given initial Hamiltonian, to be a ``valid'' RSRG rule for that Hamiltonian.

In standard RG schemes, the block spins do not correspond to ``weak'' measurements. Instead we are in a stronger-measurement regime.
Nevertheless, we suggest that 
phase transitions in the conditioned ensemble can be relevant to classifying RG transformations.

Since the RG transformation is to be applied iteratively, we would like the RSRG transformation to remain valid all the way along the flow (and in particular for the fixed-point Hamiltonian). For simplicity, here we consider only the validity of the transformation 
after a \textit{finite} number of   coarse-graining steps, arguing that this can already show a nontrivial transition.
Without loss of generality, we can then, in fact, consider just the first coarse-graining step, since the composition of several steps can be regarded as a single step with a larger scale factor $b$.
In the future it will be interesting to consider the action along the full flow, and at the fixed point $\mathcal{H}_*$.\footnote{For self-consistency, the RG transformation must be valid when applied to $\mathcal{H}_*$.
The difference from the case of a single RG step is that now  the Hamiltonian that the RG transformation acts on, and the transformation kernel ${P(S'_X|S_{X,1},\ldots, S_{X,b^d})}$, both depend on $\kappa$.}

In general, there is freedom to choose the kernel $P(S'_X|S_{X,1},\ldots, S_{X,b^d})$
defining the RG transformation
(for example, this freedom may be used to optimize the transformation for numerical efficiency).
Therefore let us imagine that the kernel depends on additional parameters $\{\kappa\}$ that can be smoothly varied. 
Let us say that $\kappa=0$ represents a valid RG transformation rule. It is easy to show (see below) that as we vary $\kappa$ we can encounter an ``RSRG--breaking'' phase transition at which the validity of the RG step breaks down.
In the simple setting of a single RG step, this transition is of precisely the kind described in this paper.

The properties of the conditioned ensemble show that, for some models, 
any sensible RSRG rule is valid, at least in the sense of preserving locality after a finite number of RG steps\footnote{A well-known fact is that the decimation transformation \cite{kadanoff2000statistical}  does not lead to a valid representation of the Ising RG fixed point in $d>1$. The  argument above indicates that it gives a quasilocal renormalized Hamiltonian after any finite number of RG steps; however, the range of this Hamiltonian diverges as the RG time tends to infinity.} (locality could still fail as RG time $\to \infty$).
However   this is not the case in all  models.

For example, consider the critical Ising model. 
The spin $S_x$ provides a local scaling operator 
that is sufficiently relevant (i.e.\ $x_S<d/2$,  Sec.~\ref{sec:generalitiesweak}) that weak measurements are relevant.
We expect  that the measured ensemble is in a trivial replica-locked phase even for small  measurement strength.
A generic block spin transformation in which the block spin $S'$ is odd under Ising symmetry is equivalent, 
at the level of symmetry (e.g.\ in the replica formalism) to measurement of $S$. Since the RG fate of this model is independent of the measurement strength,
we therefore expect that ``quenching'' the block spins is sufficient to trivialize the microscopic spins. This  is our assumed criterion for the validity of the transformation.

The self-avoiding walk gives a less trivial example in which any reasonable RG transformation 
will be valid (at least for a finite number of RG steps).
Here we assume the coarse-grained variable couples to the monomer density.
The RG flows in  Sec.~\ref{sec:imagingpolymers} 
show that any generic such rule lies in the strong measurement (valid) phase.
(But that this is not true for the $\Theta$-point polymer, where the RG flows are different.)

In other models, on the other hand, it is not possible to lock the replicas with weak measurement. 
For example, we may lack a local operator with a small enough scaling dimension.
Another possibility is that an operator with a small scaling dimension exists, but that the RG flow induced by weak measurement does not lead to a trivial replica-locked state.

In these cases, the RG transformation ${P(S'_X|S_{X,1},\ldots, S_{X,b^d})}$  may fail if it does not correspond to a strong enough measurement of the microscopic degrees of freedom.
On the other hand, it may be possible to recover a valid RG rule by changing the kernel ${P(S'_X|S_{X,1},\ldots, S_{X,b^d})}$ so that it corresponds to a stronger measurement.

For a simple example of the first mode of failure, we can take a vortex-free 2D XY model\footnote{We suppress vortices in order to be able to access the critical state at small $K$, which would otherwise be destabilized by vortices}
at a small enough stiffness $K$ that weak measurement of $\vec S$ is irrelevant. 
As a proof of principle  we can imagine a transformation in which the block spin $\vec S'_X$ is formally like a noisy measurement of the magnetization of block $X$. When the noise variance is large, this is like the weak measurement problem, showing that the replicas are not locked.

An even simpler example is a deconfined state of a 3D discrete gauge theory (Sec.~\ref{sec:nishimoricloserelatives}). This state is not critical, but it is nevertheless thermodynamically nontrivial, i.e.\ it flows to a distinct RG fixed point from the paramagnet.
Since it is non-critical, there are no relevant operators to couple to, and weak measurement is irrelevant, so there is a stable ``phase''  in which the RG rule is invalid.\footnote{The properties of this phase are easily appreciated from the extreme limit in which $S'$ is completely decoupled from $S$. The original Hamiltonian for $\mathcal{H}[S]$ is in the deconfined phase. The renormalized Hamiltonian $\mathcal{H'}[S']$ has the schematic form ${\mathcal{H'}[S'] =\mathcal{H'}_\text{paramagnet}[S'] - \ln \mathcal{N}}$, where $\mathcal{H'}$ is a trivial paramagnetic Hamiltonian for $S'$ and $\mathcal{N}$ is the topological degeneracy of the classical state (e.g. 8 for $\mathbb{Z}_2$ gauge theory on a 3D torus, corresponding  to the 8 possible values for the winding Wilson loops in each of the three directions). Note that the nonlocal topological quantity
$\mathcal{N}$ spoils the quasilocality of the renormalized Hamiltonian. Even if we discard this term, $\mathcal{H'}_\text{paramagnet}$ is not in the same phase as the initial Hamiltonian.}  The results in Sec.~\ref{sec:measuringZ2deconfined} show that there is a ``gauged Nishimori inference'' transition  into a phase where the RG rule is valid.

For an example of the second mode of failure,  consider either the Higgs transition  or the confinement transition in the 3D 
$\mathbb{Z}_2$ gauge-Higgs model.
Both of these transitions are ``starred'' Ising transitions (orbifolds of the Ising critical point \cite{wegner1971duality,lammert1993topology,senthil2002microscopic,grover2010quantum,misguich2008quantum,moessner2001ising,slagle2014quantum, serna2024worldsheet}): 
the spectrum of local operators is obtained 
by eliminating $\mathbb{Z}_2$-odd operators from the spectrum of the Ising critical point.
Weak measurement of a generic lattice operator in the gauge theory
will map to weak measurement of the Ising energy operator.  Weak measurements of this operator are relevant, but as shown in Sec.~\ref{sec:isingddims}, the resulting flow does not lead to the locked state.
Therefore we expect that an RSRG rule for the gauge theory transition can either be valid or invalid depending on the parameters of the transformation. 
Another example, where the IR theory is less well understood, is given by the multicritical point in 3D $\mathbb{Z}_2$ gauge-Higgs theory \cite{TupitsynTopological,vidal2009low,somoza2021self,bonati2022multicritical}.\footnote{At this critical point there is an operator $A$ with dimension close to $1.2$ and an operator $S$ with dimension close to $1.5$ \cite{somoza2021self}. Weak measurement is relevant at least for $A$, but it seems unlikely that the RG flow is to a fully replica-locked state.}

In practise, of course, an RSRG rule will only be useful if it lies in the interior of the ``valid'' phase (and not too close to the phase boundary), so it is an open question whether the critical properties of the RSRG-breaking transitions have relevance to applications. 
Nevertheless, it would be interesting to understand the structure of these transitions better, taking into account the entire flow and not only a finite number of steps.

\section{Outlook}

We end by noting some promising directions for the future, and then some general lessons from the models we have examined. 

There are both  specific open questions about the phase diagrams of the models and  general directions to pursue.

\smallskip

{\bf Questions about specific phase transitions.} Let us start with the former. We have for example made conjectures about the phase diagram for bond-energy measurement in the critical Ising model, in general dimension $d$, that it would be interesting to test. 
Similarly it would  be interesting to resolve the question of how the fixed points evolve, when $d$ is continuously varied.

For the 2D Ising and Potts models it would be interesting to study the conjectured intermediate--$\lambda$ critical point,
denoted $\fpU$ in Sec.~\ref{sec:pottsglobal}, either numerically  or analytically. (For example, can it be addressed using a fermionic representation for the Ising case?)  It would be worthwhile to measure $c_\text{eff}$ for both $\fpW$ and $\fpU$, using the method outlined in Sec.~\ref{sec:meas_ceff}, and establish the precise locations and properties of both critical points akin to what was done in \cite{JacobsenCardy1998,JacobsenPicco2000}.

Sec.~\ref{sec:3dfluxes}
introduced a measurement problem for flux lines in 3D and suggested a replica-Higgs description: it would be interesting to analyze the critical point in this system. 
(We can also ask whether the resulting  critical exponents have a connection with 2+1D charge sharpening, Sec.~\ref{sec:chargesharpeninghigherd}.)

Sec.~\ref{sec:replicatedGTmoregeneral} described a two-parameter phase diagram for a 3D discrete gauge theory describing inference in a paramagnetic state: what is the structure of this phase diagram? For example, does it have a confinement transition and a Higgs transition that meet at a multicritical point?

All of these models are numerically accessible via a Monte-Carlo scheme that mimics the   ``partial quench'' (Sec.~\ref{sec:partialquench}).

Next let us turn to broader directions.

\smallskip

{\bf Monitored  noisy classical dynamics.}
We have proposed a general continuum formulation for 
 monitored classical stochastic processes
or fluctuating hydrodynamics, via a replicated  Martin-Siggia-Rose action (Secs.~\ref{sec:chargesharpening},~\ref{sec:othermarkov}}). 
We used this formalism to analyse monitored dynamics of 
classical particles. 
It is clear in our approach that  charge sharpening is a generic classical phenomenon that relies only on universal 
properties of fluctuating hydrodynamics
(and does not require quantum effects).

This formalism could be applied to many other stochastic or chaotic systems: it will be interesting both to explore examples and to try to formulate general heuristics for phase diagrams. 

\smallskip

{\bf Measuring free field models.}
The models studied in Secs.~\ref{sec:freefieldandchargesharpening},~Sec.~\ref{eq:interactingfixedlinefreefield}, while physically varied, 
were all described in the IR by  free-field theory (prior to introducing measurements, which yielded interactions in the replica theory).
Even for models described by (some number of) measured free fields,
there is in principle an arbitrarily large space of models to explore and perhaps partially classify.
Different measurement processes are natural depending on the physical interpretation of the field.

\smallskip

{\bf Measuring interacting conformal field theories.}
The Ising and Potts models in Sec.~\ref{sec:potts2d} (and the self-avoiding walk in Sec.~\ref{sec:imagingpolymers}) 
are examples where the pre-measurement ensemble is an interacting conformal field theory.
(Formally these cases differ from the free-field examples, and unlike many of the latter, the  replica-symmetric mode cannot be decoupled.)
These examples showed some generic kinds of phase transition and  RG topology, 
but it is natural to ask whether similar things happen for other celebrated  models of classical critical systems.
In addition, in this paper we have only explored spatially homogeneous measurement for translationally invariant models: as in standard critical phenomena,
various elaborations may lead to  interesting criticality --- for example quenched disorder\footnote{There are distinct problems depending on whether the disorder realization is known to the experimentalist. The natural situation in the condensed matter context is where it is not known. By contrast  we have assumed throughout that in the \textit{translationally invariant} setting   the Hamiltonian is known to the experimentalist. The reason for this is that, in the infinite system limit, the parameters of the Hamiltonian can anyway be inferred with perfect precision by the experimentalist (if translation invariance, and some restriction on the range of the Hamiltonian, are assumed) since a single infinite sample supplies an infinite amount of data with which to fix the effectively finite number of parameters in the Hamiltonian.} or critical phenomena associated with boundaries/defects/impurities (which may allow connections with \cite{garratt2023measurements}).

\smallskip

{\bf Discrete gauge theories.}
The discrete gauge-theory problems in Sec.~\ref{sec:nishimoricloserelatives} also permit many generalizations.
On the formal side, phase transitions in gauge theories can give interesting examples of critical points that are beyond standard Landau-theory techniques. It will also be interesting to explore whether there are applications to error correction.
Separately, other  RG approaches could be applied to  more general ``Nishimori-like'' critical points (Bayesian inference for non-critical states) even for simple order parameters: e.g. $2+\epsilon$ expansions using nonlinear sigma models in cases with continuous symmetry, or 1D long-range toy models (Sec.~\ref{sec:ctsnishimori}). 

\smallskip

{\bf Fine-tuned structures and algorithmic questions.}
In Sec.~\ref{sec:FKmeasurement} we described a (fine-tuned) measurement process that mapped to measurement of FK clusters.
This process  involves a pair of ``mutually refining'' observables, with FK  clusters carrying less information than spin clusters \cite{Wu1988,VasseurJacobsen2012,GLJRprep}. Other examples with a similar structure could be set up, where we measure the finest observable and acquire imperfect knowledge of the coarsest. These examples are closely related to Monte Carlo cluster algorithms \cite{ZhangMichelElciDeng2020}.
We can  also ask about algorithmic aspects for more general measurement processes.
One possible  direction here would make contact with the disordered complex-systems literature \cite{zdeborova2016statistical,nishimori2001statistical}. 
Here we are interested in directly simulating the conditioned ensemble. 
(Another important question in many inference problems is about how to efficiently estimate observables without a brute-force simulation, but that is not our focus here.
Note that it is remarkable that replica approach can give results about Bayesian inference that are algorithm-independent.)
Recall that highly efficient cluster algorithms are available for many of the workhorse models of classical statistical mechanics in the translationally invariant setting \cite{newman1999monte}. In what cases are there efficient nonlocal algorithms for simulation of the conditioned ensemble? How does the measurement process affect the dynamical exponent of cluster-based algorithms?

\smallskip

{\bf RSRG-breaking transitions.} The practical relevance of the RSRG-breaking transitions in Sec.~\ref{sec:RSRG}
remains to be investigated, but the point that models fall into different classes depending on how delicate it is to coarse-grain them seems intriguing.
The most famous applications of RSRG are to systems with an order parameter, where it is ``easy'' to formulate a reasonable RG rule. But there  also exist topological phase transitions which lack  an order parameter,
and where we also lack a convenient continuum field theory\footnote{See e.g.\ discussion in ~\cite{somoza2021self} and appendices of \cite{serna2024worldsheet}.}: in these cases an efficient  RSRG scheme might  be useful.

Finally we summarize a few general points.

\smallskip

{\bf Measurement vs.\ quenched disorder.} First, consider the analogies and differences with problems of quenched disorder. 
There is a formal analogy between measured systems and disordered ones, and in various settings it is possible to do perturbation theory for both cases in parallel (e.g.\ Sec.~\ref{sec:pottsweak}). 
However, we have also seen that this can obscure differences between the two kinds of problems. Measured systems obey additional constraints on the RG flows,
and the behavior in the strong-measurement limit is typically very different from the behavior of the disordered system in the strong-disorder limit.
Even in weak-coupling examples (e.g.\ Sec.~\ref{sec:isingddims}) the flows can have a  different topology at $N\to 0$ and at $N\to 1$, but at strong coupling there can be fundamental differences that are formally to do with differences in replica group theory and physically to do with the difference between a glassy phase and a simple ``replica-locked'' phase where measurement trivializes the conditioned ensemble 
(see e.g.\ Secs~\ref{eq:2Dheightlocking},~\ref{sec:3DdimerEFT}).
(There are also qualitative differences between an $N\to 1$ limit representing quantum measurements, and an $N\to 0$ limit
representing a certain postselected measurement process, in the quantum setting \cite{nahum2023renormalization,nahum2021measurement,jian2023measurement,fava2023nonlinear,poboiko2023theory}.)

\smallskip

{\bf Nontrivial Bayesian critical points in trivial models.}
The examples in 3D gauge theory demonstrate that  nontrivial measurement phase transitions can occur 
even when the degrees of freedom in the initial ensemble are  not only short-range correlated, but also completely structureless --- with no  imposed microscopic symmetries, constraints or conservation laws at all.
In those examples, the  nontrivial state was associated with 
a stable deconfined phase for interreplica fluctuations
(with corresponding 
emergent one-form symmetries in inter-replica space). 
It would be interesting to look for other  mechanisms for nontrivial transitions that do not require microscopic symmetries 
(or significant fine-tuning of RG-relevant couplings). 
For example, in 4D it is possible to have stable deconfined phases for continuous gauge groups, and we could ask if there is any connection to dynamical processes (e.g.\ in quantum circuits) in 3+1D.

Having seen that the conditioned ensemble 
can have more structure than the 
 pre-measurement  ensemble
 (e.g.\ can have nontrivial phase transitions even when the pre-measurement ensemble is trivial), it is clear that this is relevant to monitored stochastic systems too.  
It is possible to construct proof-of-principle examples based on Sec.~\ref{sec:nishimoricloserelatives}, but it  would be interesting to explore physically natural examples.

\smallskip

{\bf Measurement-enforced symmetries and constraints.}
Strong (accurate) measurement  can enforce  constraints on the conditioned ensemble.
In some cases these constraints are formally equivalent to a \textit{symmetry}
or higher-form symmetry in the post-measurement ensemble and in the replica theory. Sec.~\ref{sec:nishimoricloserelatives} gave examples of a zero-form symmetry and one-form symmetries emerging in this way. 

(Note, however, that in the context of classical statistical mechanics,
{\bf symmetries} which would be naturally regarded as equivalent in a quantum partition function should be regarded as distinct, see App.~\ref{app:symmetrynote}.)

In other examples 
(e.g.\ for level lines in  Sec.~\ref{eq:interactingfixedlinefreefield} or percolation configurations in Sec.~\ref{sec:percolation}) strong measurement imposes a geometrical constraint on configurations.
In some cases this can restructure the field content of the replica theory, organizing the replicated field $\phi^a_{\alpha_a}$ of the original theory into a ``superfield'' $\Phi_{\alpha_1,\ldots,\alpha_N}$ that inherits an index from every replica.

\smallskip

{\bf Quantum analogies.} The classical problems here may yield insights into quantum analogs.
To give one example, they suggest examining a wider range of quantum measurement transitions.
In the examples of measuring a classical paramagnet, the natural replica order parameter was an overlap field of the schematic form $X^{ab}$ carrying two replica indices.
Overlap order parameters with two replica indices also arise in quantum measurement problems \cite{nahum2023renormalization}, albeit with a  different replica group theory structure/ordering pattern and a very different  physical interpretation.
In many of the examples of  measurement of a  critical classical state, it was natural instead to work with replicas of the ``elementary'' order parameter, say $\phi^a$, i.e.\ fields with a single replica index.
Quantum models where measurement modifies existing critical order parameter correlations also exist  \cite{garratt2023measurements}.
But one of the  classical critical points we have discussed in Sec.~\ref{sec:isingddims}
shows that for some
classical measurement transitions it is natural to retain both ``elementary'' and ``overlap'' fields in the critical Langrangian.  One can also imagine settings in  quantum models  where both elementary and overlap fields should be retained in the effective Lagrangian. We hope to discuss this elsewhere.

\acknowledgments 
 We thank 
Denis Bernard,
Patrick Draper,
Paul Goldbart,
Timothy Hsieh,
Jorge Kurchan,
Sarang Gopalakrishnan,
Stephen Powell, 
Slava Rychkov, 
Guilhem Semerjian, 
Brian Skinner, 
T. Senthil, 
Zack Weinstein and
Kay Wiese for useful discussions and references. We are particularly grateful to Sarang Gopalakrishnan for valuable encouragement to write up these results.
This work was 
supported by the French Agence Nationale de la Recherche (ANR) under grant ANR-21-CE40-0003 (project CONFICA).

\appendix

\section{Replica identity}
\label{app:replicaidentity}

Let $P_k(S^1,\ldots,S^k;M)$, for $k\in \mathbb{N}^+$, be the joint probability distribution of the measurement outcomes, together with $k$ samples, that are conditioned to have these outcomes. 
We may write this as 
\ba\notag
P_k(S^1,\ldots,S^k; M) 
& \equiv P(M) P(S^1|M)\cdots P(S^k|M) \\
& =
\f{Z(M)}{Z} \f{1}{Z(M)^k}
e^{
-\sum_{\alpha=1}^k \mathcal{H}_{\rm meas}[S^\alpha, M] 
}\label{eq:PkSSSSMfirst}
\end{align}
(see Eq.~\ref{eq:PSgivenM}). 
We use the replica trick to formally promote factors from the denominator to the numerator:
\ba
P_k(S^1,\ldots,S^k; M)
& =
\lim_{N\to 1}  \f { Z(M)^{N-k}}{Z}
e^{
-\sum_{\alpha=1}^k \mathcal{H}_{\rm meas}[S^\alpha, M] 
}.\notag
\end{align}
Writing $Z(M)$ explicitly as an integral (Eq.~\ref{eq:Mmarginal}),
\ba
P_k(S^1,\ldots,S^k; M)
& =
\lim_{N\to 1}  \f {1}{Z}
\int_{S^{k+1}, \ldots, S^{N}}
e^{
-\sum_{\alpha=1}^N \mathcal{H}_{\rm meas}[S^\alpha, M] 
}.\label{eq:PreplicaAsintegral}
\end{align}
Using Eq.~\ref{eq:PreplicaAsintegral}, the  expectation value of a quantity $\bullet$ that depends on $S^1, \ldots, S^k$, and potentially also on $M$, is 
\be\label{eq:EkMandS}
\mathbb{E}^{(k)} \left[ \bullet \right]
=\f{1}{Z}
\lim_{N\rightarrow 1}
\int_{M, \,S^1,\ldots, S^N}
e^{
-\sum_{\alpha=1}^N \mathcal{H}_{\rm meas}[S^\alpha, M] 
} \,  \left( \bullet \right).
\ee
The integral on the right-hand side (over $M$ together with $S^1, \ldots, S^N$) defines the replica partition function for a given value of $N$, which we may denote $Z_N$:
\ba\label{eq:ZNintMS}
Z_N & = \int_{M, \,S^1,\ldots, S^N}
e^{
-\sum_{\alpha=1}^N \mathcal{H}_{\rm meas}[S^\alpha, M] 
} 
\end{align}
Since $M$ appears quadratically in Eq.~\ref{eq:Hmeasdefn} we may integrate it out:
\ba\label{eq:ZNintS}
Z_N & = \int_{S^1,\ldots, S^N}
e^{- \mathcal{H}_N[S^1, \ldots, S^N]},
\end{align}
with 
\ba  \notag
\mathcal{H}_N[S^1, \ldots, S^N]=  &
 \sum_{a=1}^N \mathcal{H}[S^a]
 + \f{1}{2\Delta^2 N} \sum_{a<b} (\mathcal{O}^a - \mathcal{O}^b)^2 
 \\
& - (N-1)\ln d + \f{1}{2} \ln N . \label{eq:replicaHappendix}
\end{align}
Note that the limit of the replica partition function is simply the physical one, $\lim_{N\to 1} Z_N = Z$.
This allows us to replace the factor of $1/Z$   outside the limit in Eq.~\ref{eq:EkMandS} with a factor of $Z_N^{-1}$ inside the limit. 
This finally gives:
\be
\mathbb{E}^{(k)} \left[ \bullet \right]
=
\lim_{N\rightarrow 1}
\< \bullet \>_{N},
\ee
as written in the main text (see Eq.~\ref{eq:firstreplicaidentity}),
where $\< \bullet \>_{N}$ denotes an expectation value for the $N$-replica problem, taken with the Hamiltonian $\mathcal{H}_N$ in Eq.~\ref{eq:replicaHappendix}.
The right-hand side must in principle be computed for $N\geq k$ and analytically continued to $N=1$. The terms in the second line of Eq.~\ref{eq:replicaHappendix} vanish for $N \to 1$, leading to Eq.~\ref{eq:Hreplica} in the main text.

For many purposes we may take the limit ${N\to 1}$ directly in the coefficients in $\mathcal{H}$, giving the simplified forms stated in the main text (see Eq.~\ref{eq:replicaHsimplified}).

Note also that one of the replicas appearing on the right-hand side of Eq.~\ref{eq:PreplicaAsintegral} has been used to generate the factor of $P(M)$ in Eq.~\ref{eq:PkSSSSMfirst}.
Physically this replica (which we can take to be, for example, $S^N$) 
can be interpreted as the actual measured sample, $S^\text{ref}$ \cite{zdeborova2016statistical}.

Finally, let us derive the identity
\be\label{eq:shannonfreeenergy}
S_\text{meas} = F'(1) - F(1)
\ee
used in the main text (see Eq.~\ref{eq:Fprimeidentity}), where $F(N)$ is the free energy for $N$ replicas, i.e. $Z_N = e^{-F(N)}$.
The replica partition function may be written
\be
Z_N = \int_{M}
\left(
\int_S
e^{-\mathcal{H}[S]}
P(M|S)
\right)^N.
\ee
Up to a factor of $Z=Z_1$, the factor inside the parentheses is just the probability $P(M)$ for a given set of measurement outcomes, so
\be
Z^{-N} Z_N = 
\int_M P(M)^N.
\ee
The right-hand side appears in the definition of the $N$th R\'enyi entropy:
\be\label{eq:renyidefn} 
\int_M P(M)^N
=
\exp\left[{-(N-1) S^{N}_\text{R\'enyi}}\right].
\ee
The $N\to 1$ limit of the R\'enyi entropy is the Shannon entropy. This gives Eq.~\ref{eq:shannonfreeenergy}.

As an illustration, consider two extreme limits. For concreteness, let us consider the Gaussian measurement protocol in App.~\ref{app:replicaidentity}, as above, and let us assume the spins are discrete variables.

First, consider the case in which the measurements reveal no information about the system. This can be achieved by taking the measured operator $\mathcal{O}_x$ to be a trivial constant, e.g. $\mathcal{O}_x= 0$.
The replicas are then decoupled from each other. 
As a result, the free energy of $N$ replicas is just $N$ times the free energy of a single replica, except for trivial terms that come from the Gaussian integrals over local measurement outcomes,\footnote{\be
F(N) = NF(1) + \f{1}{2} \sum_x \lf
\ln N + (N-1) \ln (2\pi \Delta^2)
\ri.
\ee} 
giving 
\be
S_\text{meas} =  S_\text{triv}(\Delta),
\ee
where $S_\text{triv}(\Delta)=\f{V}{2}\ln \lf 2\pi \Delta^2 e \ri$ is the  entropy of $V$ Gaussian variables with variance $\Delta^2$, where $V$ is the number of measured sites.

Next, consider the case where the measured operator is the spin itself, and the measurements are very precise ($\Delta\to 0$). Then the configurations of the replicas become identical, and the replica Hamiltonian is effectively equivalent to the initial Hamiltonian, but at a temperature that is reduced by a factor of $N$.\footnote{That is, the nontrivial part of the replica Hamiltonian is $N\mathcal{H}[S] = \mathcal{H}[S] + (N-1) \mathcal{H}[S]$. Treating the second term as a quantity to be averaged leads to 
\ba
F(N)  = 
\,  &  F(1)
- \ln \< e^{-(N-1) \mathcal{H}[S]}\> 
\\
& +
 \f{1}{2} \sum_x \lf
\ln N + (N-1) \ln (2\pi \Delta^2)
\ri
\end{align}}
Expanding around $N=1$ gives 
\be
S_\text{meas} = S_\text{thermo} + S_\text{triv}(\Delta),
\ee
where ${S_\text{thermo} = \<\mathcal{H}\> - F(1)}$ is the thermodynamic entropy of the original model.

\section{Alternative measurement protocols}
\label{app:binarymeasurements}

\subsection{Diluted binary measurements}

Consider a binary  observable  $\measO_x=\pm 1$ 
that gets measured with a probability $p_{\rm meas}$.
If it is measured, the probability of an error is $p_{\rm err}$.
We can represent the absence of a measurement by the value $M_x=0$.
The above probabilities then specify the distribution
\be
P(M|S) = e^{- K[M,S]}
\ee
for $M_x=-1,0,+1$.
Note  that the case where ${p_{\rm meas} <1}$ but ${p_{\rm err} = 0}$ is similar to the case of ``heralded'' errors in error correction: while we do not have information about all sites, we know that the information that we do have is trustworthy.

The replica Hamiltonian, after integrating out $M$, is given by 
(to simplify notation we consider a single site and omit the site index)
\ba
\mathcal{H}_N[S^1,\ldots,S^N]
= \sum_a \mathcal{H}[S^a]
+ \mathcal{H}_{\rm coupling}[S^1,\ldots,S^N]
\end{align}
with 
\ba
e^{-\mathcal{H}_{\rm coupling}}
& = 
\sum_{M=-1,0,+1}
\prod_a P(M|S_a)
\\
& = 
(1-p_{\rm meas})^N
+ 
p_{\rm meas}^N
\sum_{\sigma= \pm 1}
(1-p_e)^{N_\sigma} p_e^{N_{-\sigma}},
\end{align}
where $N_\sigma$ is the number of replicas with $\measO_x = \sigma$.

First consider the case where $p_{\rm meas}=1$ (all sites are measured) but the measurements are very imprecise: $p_{\rm err} = \f{1}{2} - \epsilon$ with $\epsilon \ll 1$. Then
\ba
\mathcal{H}_{\rm meas} & = - 2\epsilon^2 \lf \sum_{ab} \measO^a\measO^b  -N \ri + (N-1)\ln 2
\\
& = - 2\epsilon^2  \sum_{a\neq b} \measO^a\measO^b   + (N-1)\ln 2.
\label{eq:appweakbinarymeasurementHeff}
\end{align}
We see that in this weak measurement limit ($\epsilon\ll 1$) we recover the same kind of replica coupling that we found in the Gaussian measurement case. 

If we consider very imprecise measurements that are also diluted, $0<p_{\rm meas}<1$, then 
(apart from a  change to the additive constant, which anyway vanishes at $N=1$) the only effect is to change the coefficient $2\epsilon^2$ in Eq.~\ref{eq:appweakbinarymeasurementHeff} to $2\epsilon^2 p_{\rm meas}$.

Finally,  consider the opposite limit where $p_e = 0$, so that the measurements are perfectly precise. Then
\ba
e^{-\mathcal{H}_{\rm coupling}}
& = 
(1-p_{\rm meas})^N
+ 
p_{\rm meas}^N
\delta_{\measO_x^1 = \measO_x^2 =\ldots = \measO_x^N},
\end{align}
where the delta function enforces equality of $\measO$ in all the replicas. 
Taking the $N\to 1$ limit for the coefficient and dropping a constant,
\be
\mathcal{H}_{\rm coupling} = 
\lf 1- \delta_{\measO_x^1 = \measO_x^2 =\ldots = \measO_x^N} \ri \ln (1- p_{\rm meas})^{-1}.
\ee
If the initial theory is critical, and if  $p_\text{meas}$ is small (so that $\mathcal{H}_{\rm coupling}$ is small), then at large scales we should decompose the perturbation in terms of the scaling operators of the continuum theory, and the leading term is again of the type discussed in the Gaussian case.

\subsection{Global symmetries in  replicated theory}

We comment on the symmetry of the replicated theory (Sec.~\ref{sec:symmetry}) in the specific example of a $\mathbb{Z}_2$ (Ising-like) symmetry of the initial ensemble that acts as ${S\to - S}$.

Let us consider various cases. The obvious ones are:

{\bf (a)} We measure a $\mathbb{Z}_2$-even observable $\measO$ using the prescription in the main text.
The replica Hamiltonian preserves the $\mathbb{Z}_2$ symmetry within each replica, so that there is a $\mathbb{Z}_2^N$ symmetry.
Together with replica permutations, this gives the global symmetry group $\mathbb{Z}_2^N \rtimes S_N$.\footnote{The notation indicates that $\mathbb{Z}_2^N$ is a normal subgroup of  $\mathbb{Z}_2^N \rtimes S_N$.}

{\bf (b)} We measure a $\mathbb{Z}_2$-odd observable $\measO$ using the prescription in the main text.
Then the replica Hamiltonian does not preserve a separate $\mathbb{Z}_2$ for each replica (because of the terms $S^a S^b$) but it does preserve a single global $\mathbb{Z}_2$ that acts simultaneously on all replicas. 
This gives the symmetry $\mathbb{Z}_2\times S_N$.

However, the measurement protocol in the main text is not fully general, even if we assume that the $P(M|S)$ is Gaussian.
We may also allow the standard deviation $\Delta$ of the measurement error to become a function of a local operator, rather than being a constant. 
As an example, we could take
\ba\label{eq:Deltavarying}
\Delta_x^{-2} =  \Delta^{-2} \lf 1 + \epsilon \, S_x \ri,
\end{align}
so that the precision of the measurements depends on the local $\mathbb{Z}_2$-odd observable $S_x$.
In other words, it is possible to introduce symmetry-breaking not only via the observable that is measured, but also via the measurement strength.

We will consider the effect of this in the two cases above, in which the measured operator $\measO$ is even/odd respectively.
First, let us write the more general form of the replica Hamiltonian that results.
The term induced by measurements in the replica Hamiltonian, at a given site, is 
\ba
\delta\mathcal{H} = 
 \f{1}{4\sum_c \Delta_c^{-2} }
\sum_{a\neq b} \f{(\measO^a - \measO^b)^2}{\Delta_a^2 \Delta_b^2},
\end{align}
where we suppress the $x$ index, and write
${\Delta_a^{-2} \equiv \Delta^{-2}(1+\epsilon S^a)}$.
For concreteness, let us assume that $\epsilon$ is small, so that we can expand: 
\ba\notag
\delta\mathcal{H} = & \, 
\f{N}{4\Delta^2}
\sum_{a\neq b} (\measO^a - \measO^b)^2 
 + 
\f{\epsilon}{2\Delta^2} \sum_{\substack{abc \\ \text{all distinct}}} S^a \measO^b \measO^c
\\
&
+ 
\f{(2-N)\epsilon}{2\Delta^2}
\sum_{a\neq b} S^a [\measO^b]^2
+ \ldots
\label{eq:replicaHsoo}
\end{align}
We have dropped terms whose coefficients vanish when $N=1$.
(In order to perform RG,  we would expand $[\measO^b]^2$ in terms of scaling operators.)

We consider the analogs of the cases (a) and (b) above:

{\bf (a')} The measured operator $\measO$ is $\mathbb{Z}_2$ even, but the measurement error variance depends on a $\mathbb{Z}_2$-odd operator. 

We claim that this  is really case (b) in disguise, 
because the measurement outcomes do in fact carry information about $\mathbb{Z}_2$-odd observables.

First, this can be seen  heuristically:
consider a long-wavelength spatial variation
in $S_x$. 
In the regions where $S_x$ is smaller, the 
measurement outcomes will fluctuate more strongly, by (\ref{eq:Deltavarying}).
Therefore we can obtain coarse-grained information about $S$ from the  fluctuations in $M$.

This can be seen more formally from the replica Hamiltonian in Eq.~\ref{eq:replicaHsoo}.
This ``bare'' Hamiltonian does not contain the term $\sum_{a\neq b} S^a S^b$ which would be present if we directly measured the operator $S$. 
However, this term will be generated under RG from the terms that are present, with an \textit{effective} measurement strength of order $\epsilon^2/\Delta^4$. (Details in footnote\footnote{To see this it is sufficient to consider the final term in (\ref{eq:replicaHsoo}). $[\measO^b]^2$ can be decomposed into local $\mathbb{Z}_2$ even scaling operators. Let the leading nontrivial operator  be ${\widetilde\measO}^b$ 
(generically $\widetilde\measO=\measO$;
the contribution from the identity operator vanishes, because the sum over $b$ gives a coefficient ${N-1\to 0}$).
Therefore we have a term 
$\widetilde \lambda \sum_{a\neq b} S^a \widetilde\measO^b$ in the bare Hamiltonian, with $\widetilde \lambda=\epsilon/\Delta^2$.
By considering the OPE  \cite{cardy1996scaling} of this term with itself, we see that the term $\sum_{a\neq b}S^a S^b$ is generated under RG. If the coupling for the latter term is denoted $\lambda'$, then the beta function for $\lambda'$ contains a term proportional to $\widetilde \lambda^2$.
After coarse-graining for an order-1 amount of RG time,  a $\lambda'$ of order $\epsilon^2/\Delta^4$ is generated.}.)
That is, roughly speaking, after some coarse-graining the situation is similar to one in which we directly measure $S$ with an error variance $\sim \Delta^4/\epsilon^2$.

In many cases the effective measurement of the $\mathbb{Z}_2$-odd operator $S$ will be more RG relevant than the direct measurement of the  $\mathbb{Z}_2$-even operator $\mathcal{O}$.
However, if $\Delta^2\ll 1$, then the effective measurement strength for $S$ is much weaker than that for $\measO$ (in the present setup), 
and both facts need to be taken into account in determining the fate of the system in the IR.
Our main point is that, as far as symmetry is concerned, case (a') is equivalent to case (b).

{\bf (b')} The measured operator $\measO$ is $\mathbb{Z}_2$ odd, and the  measurement error variance depends on a $\mathbb{Z}_2$-odd operator. 

In general, the replica symmetry is then different from both case (a) and case (b) above. 
Whereas in case (b) the replicated theory has a single $\mathbb{Z}_2$ symmetry for any $N$,
in case (b') this $\mathbb{Z}_2$ 
symmetry is absent
except when we set $N=1$.
For general $N$ the symmetry is therefore just $S_N$. 
However, the fact that the $N=1$ model has a $\mathbb{Z}_2$ symmetry is important: this implies that the terms breaking $\mathbb{Z}_2$ involve at least two distinct replica indices, making them less RG-relevant.

It is certainly possible to construct examples where the difference in symmetry between (b) and (b') reflects a genuine  difference in universal properties.\footnote{A contrived example is a system with spontaneously broken $\mathbb{Z}_2$ coexisting with critical $\mathbb{Z}_2$-even degrees of freedom.  
After picking one of the two equivalent symmetry-breaking macrostates,
we are left with an \textit{effective}
problem of type (a) for the critical $\mathbb{Z}_2$-even degrees of freedom.
(Coarse-graining effectively generates measurements of $\mathbb{Z}_2$-even operators.)
If the original process is of type  (b), the two macrostates give rise to equivalent effective measurement problems for the critical degrees of freedom.
On the other hand in case (b') the  measurement strength in the effective problem differs  for the two macrostates.}
However, we expect that for many ensembles in the weak measurement regime,  (b) and (b') will be equivalent up to RG-irrelevant terms.
In other words, the ``missing'' $\mathbb{Z}_2$ symmetry can emerge under RG.

To see this, consider the case where
 $\measO=S$. 
We may expand
$[\measO^b]^2=[S^b]^2$ in Eq.~\ref{eq:replicaHsoo} as a sum of $\mathbb{Z}_2$-even scaling operators. 
Letting the nontrivial leading  operator be denoted $\mathcal{E}^b$, and setting $N\to 1$ in the coefficients,
\ba\notag
\delta\mathcal{H} = & \, 
- \f{1}{2\Delta^2}
\sum_{a\neq b} S^a S^b
 + 
\f{\epsilon}{2\Delta^2} \sum_{\substack{abc \\ \text{all distinct}}} S^a S^b S^c
\\
&
+ 
\f{\epsilon\times \text{const}}{2\Delta^2}
\sum_{a\neq b} S^a \mathcal{E}^b
+ \ldots
\label{eq:replicaHsoo2}
\end{align}
If $\mathcal{E}$ is less relevant than $S$, then the leading term is the first one, 
$\sum_{a\neq b} S^a S^b$,
just as in case (b).
This leading term preserves a single $\mathbb{Z}_2$ symmetry [as in case (b)].
If the terms proportional to $\epsilon$ are  irrelevant at the IR fixed point, then the $\mathbb{Z}_2$ symmetry is \textit{emergent} under the RG flow, and we are back to case~(b).

(For concrete examples of this emergent symmetry we could probably adapt the problems in Secs.~\ref{eq:2Dheightlocking},~\ref{sec:chargesharpening}  by adding  variations in the measurement strength that introduce irrelevant cosine terms which break the $h\to - h$ symmetry.)

\section{RG-instability of the FK-cluster measurement protocol}
\label{app:FKdetails}

We expand further on the instability of the FK-cluster-measurement fixed point $\fpFK$ discussed in Sec.~\ref{sec:FKmeasurement}.
It is convenient to use a Landau-Ginzburg notation, although the basic points depend only on symmetry.

Consider a critical Potts model with $\mathcal{Q}$ spin states and $S_\mathcal{Q}$ symmetry. For the present application, this is the fine-tuned replica theory of Sec.~\ref{sec:FKmeasurement}, which gives a Potts model with $\mathcal{Q}=Q^N$ states.

We can represent the Potts spin with a Landau-Ginzburg field $\phi_\mathcal{S}$,  
where $\mathcal{S}=1,\ldots, \mathcal{Q}$ runs over the possible spin values. 
In the present case, there is a correspondence between the values of   $\mathcal{S}$ and the values of the vector $\vec{\sigma}=(\sigma_1,\ldots,\sigma_N)$ introduced in the main text.
That is, picking a value for the spin $\mathcal{S}$ corresponds to picking a spin value for each of the $N$ replicas.
We also impose $\sum_{\mathcal{S}=1}^\mathcal{Q} \phi_\mathcal{S} = 0$. The remaining $\mathcal{Q}-1$ independent components then form an irreducible representation of $S_\mathcal{Q}$ symmetry.
We may think of $\phi_\mathcal{S}$ as the density of sites with spin value $\mathcal{S}$ (up to a constant shift).

The mass term, which is an $S_\mathcal{Q}$ scalar, and which tunes the Potts model to criticality, is $\sum_{\mathcal{S}} \phi_\mathcal{S}\phi_\mathcal{S}$.
The coefficient of this term is fixed by the fact that the model is critical. 
In our interpretation, this coefficient is fixed by the fact that the original physical model that we are measuring is the \textit{critical} Potts model. (In other words, the theory that remains when we set $N=1$ is critical.)

Consider instead the operators of the form
\be
A_{\mathcal{S}, \mathcal{S}'} = 
\phi_\mathcal{S}\phi_{\mathcal{S}'} - \f{1}{\mathcal{Q}}  \sum_{\mathcal{S}''} \phi_{\mathcal{S}''} \phi_{\mathcal{S}''},
\ee
which form a nontrivial irrep.
These operators are ``two cluster'' operators, in the geometrical interpretation of the FK cluster model, and have scaling dimension $x = 2g-(g-1)^2/(2g)$ for $g=\f{1}{\pi}\arccos(-\sqrt{Q}/2)$. They are relevant if $Q<4$ and marginal when $Q=4$.

When we have $S_\mathcal{Q}$ symmetry, $A_{\mathcal{S}, \mathcal{S}'}$ is forbidden by symmetry from appearing in the action  (since it is not an $S_{\mathcal{Q}}$ singlet). However, when we break $S_\mathcal{Q}$  down to $G_{Q,N}= (S_Q  \times \cdots \times S_Q )\rtimes S_N$,
various components of $A$ can be added to the action.
We do not enter into the group theory here, but nevertheless it is clear that, when we treat $N$ as arbitrary, there are an infinite number of $G_{Q,N}$-allowed relevant terms, because
\be
\sum_{\substack{{\vec\sigma, \vec \sigma'} \\ {\text{($\vec\sigma, \vec \sigma'$ differ in $k$ replicas)}}}}
A_{\vec\sigma, \vec \sigma'}
\ee
is invariant for any $k$.
(We have used $\vec \sigma$, as a label, since this is equivalent to $\mathcal{S}$.)

From a lattice perspective, note that in the fine-tuned model, the interaction between the ``superspins'' $\mathcal{S}$, i.e. between the vectors $\vec{\sigma}$, is of the form $\delta_{\vec \sigma_x, \vec \sigma_y}$. When the measurements  of the FK clusters are imperfect, more general interactions are allowed, which respect $G_{Q,N}$ but may not respect $S_{\mathcal{Q}}$.
A basis for these interactions is given by functions $\{\eta_k(\vec\sigma_x, \vec \sigma_y)\}_{k=0}^{N-1}$,
where 
$\eta_k(\vec\sigma_x, \vec \sigma_y)$ is 1 if $\vec \sigma_y$ differs from 
$\vec\sigma_x$ in $k$ of the $N$ replicas, but agrees in the other $N-k$ replicas. Only $\eta_0(\vec\sigma_x, \vec \sigma_y)=\delta_{\vec\sigma_x, \vec \sigma_y}$ respects $S_{\mathcal{Q}}$ symmetry.

There may be further constraints on the interactions beyond those given by symmetry. Nevertheless the natural expectation is that a generic small change to the measurement protocol (away from the FK-measurement protocol) will induce an infinite number of these relevant perturbations.

Finally, note that the measurement protocol discussed above, which effectively amounts to measuring FK  occupation numbers, 
allows an extension to any real $Q>0$ in which all the probabilities are positive. 
(The Gibbs weight for an FK cluster configuration is proportional to $Q^\text{no. clusters}$, which is positive for $Q>0$.)
By contrast, if we wish to measure the energy operator $\delta_{\sigma_x,\sigma_y}$,
then we must use a formulation of the model in which this is a local observable. 
This can be done, for arbitrary $Q$, by writing the partition function in terms of Potts domain wall configurations weighted by a chromatic polynomial (obtained by summing over all spin configurations consistent with the given domain wall locations). However, in general this weight can be negative when $Q$ is noninteger.

\section{Square-lattice dimer model}
\label{app:dimer}

For convenience, we first recall \cite{fradkin2004bipartite}  the formulas for the lattice occupation numbers in the square lattice dimer model, in the case where the partition function is the equally-weighted sum over all fully-packed configurations. A constructive derivation was given recently in Ref.~\cite{wilkins2023derivation}.
If $d_x(x,y)$ [resp. $d_y(x,y)$] is the occupation number of the bond connecting $(x,y)$ to its east [resp. north] neighbor,
\ba\label{eq:dx1}
d_x & \to \f{(-)^{x+y+1}}{2\pi}\partial_y h +\f{1}{\pi a} (-)^x \cos h + \ldots
\\
d_y & \to \f{(-)^{x+y}}{2\pi}\partial_x h +\f{1}{\pi a} (-)^y \sin h + \ldots,
\label{eq:dx2}
\end{align}
where $a$ depends on the regularization of the field theory~\cite{wilkins2023derivation}.

Second, we show 
that ideas from deconfined criticality 
\cite{levin2004deconfined,freedman2011weakly} give a
quick  way of rationalizing the field theory for the square lattice dimer model without the detailed height-field analysis.

We orient all the links of the square lattice such that each of the four columnar dimer configurations corresponds to occupying only links of a single orientation,
with each of the four compass directions represented at each site.

For each site $(x,y)$ we then define a vector $\vec S(x,y)$ that 
is parallel to the \textit{occupied}  oriented  link touching that site.
Since $\vec S(x,y)$ on a single site can only point in four directions, we view this as an unconventional lattice regularization of an XY model with strong fourfold anisotropy.

Next note that, 
as for the valence-bond solid order parameter in a 2D antiferromagnet \cite{levin2004deconfined},  
the full-packing constraint prevents $\vec S$ from having vortices. 
For example, it is easy to check that a monomer (unoccupied site) corresponds to a vortex or an antivortex, depending on the sublattice the monomer sits on (see Fig.~4 in Ref.~\cite{levin2004deconfined}).

Finally, we write a continuum Hamiltonian for the coarse-grained phase $h$ of ${\vec S = (\cos h, \sin h)}$.  We include $\cos (4 h)$ in the action since the microscopic regularization has fourfold anisotropy, but we do not allow vortices in~$h$:
\be
\mathcal{H}
=
\f{K}{2} (\nabla h)^2 
- g \cos 4h + \ldots
\ee
This heuristic argument does not fix the couplings, 
but it turns out that for equally weighted dimer configurations ${K=\f{1}{4\pi}}$ is small enough that $g$ is irrelevant.

Finally, consider the operator identifications for the dimer occupancies.
The presence of the $\cos h$ and $\sin h$ terms in Eqs.~\ref{eq:dx1},~\ref{eq:dx2},
with their alternating sign factors, 
follows immediately from the microscopic relation between $\vec S$ and bond occupations. For example, 
\be
2\sum_{x,y} \lf (-)^x d_x(x,y), (-)^y d_y(x,y)\ri =  \sum_{x,y} \vec S(x,y)
\ee
(up to boundary terms, and for some choice of origin).
Therefore we expect ${(-)^x S_x=(-)^x \cos h}$ to appear in the continuum expression for $d_x$, etc.

The presence of derivative terms in 
Eqs.~\ref{eq:dx1},~\ref{eq:dx2}
can be understood from the constraints imposed on $\vec S$ by its relation with a dimer configuration.
For example, if $S_y(x,y)=\pm 1$,
then the phase increment 
${h(x+1,y)-h(x,y)}$ can be
$(-)^{x} \f{\pi}{2} S_y(x,y)$, but it cannot be $-(-)^{x}\f{\pi}{2}S_y(x,y)$.
Therefore, microscopically,
$-(-)^{x} S_y$ is positively correlated with $\partial_x h$.
Equivalently
$d_y$ is positively correlated with $-(-)^{x+y} \partial_x h$.
Therefore in the continuum we expect
$(-)^{x+y} \partial_x h$ also to appear 
with a positive 
coefficient in the expression for~$d_y$.

\section{Field theory for noisy diffusion}
\label{app:diffusion}

\subsection{Martin-Siggia-Rose action}

Starting with (\ref{eq:countingfield}), the Martin-Siggia-Rose approach gives a functional integral
\ba\notag
Z & = \int \mathcal{D}\eta \mathcal{D}h 
\delta\lf\partial_t h  - D \partial_x^2 h - 2\pi \eta\ri
\exp\lf{-\int\dd t \dd x \f{\eta^2}{2\kappa}}\ri
\\
& = 
\int  \mathcal{D}h \exp\lf{- \f{1}{8\pi^2\kappa}}\int\dd t \dd x 
\lf 
\partial_t h  - D \partial_x^2
\ri^2
\ri.
\end{align}
When we expand the square we encounter the total derivative term ($\dot h = \partial_t h$, $h' = \partial_x h$):
\be\label{eq:writeastotalderiv}
\dot h  h'' = 
\partial_x ( \dot h h' ) -\f{1}{2} \partial_t
(h')^2.
\ee
In writing $Z$ we were vague about boundary conditions. We can impose boundary conditions on a finite domain $x\in [0,L]$ such that $h(0,t)=h(L,t)=0$. The integral of the first term in 
Eq.~\ref{eq:writeastotalderiv} then vanishes, but the second term gives a contribution on the temporal boundaries.
If time ranges from $0$ to $T$, then
$Z = \int \mathcal{D} h e^{-\mathcal{H}}$ with
\ba\label{eq:MSRaction}
\mathcal{H} & = 
\f{1}{8\pi^2 \kappa} \int \dd x \dd t \left[
(\partial_t h)^2 + D^2 (\partial_x^2 h)^2
\right]
+ \mathcal{H}_B,
\\
\mathcal{H}_B & = 
\f{D}{8\pi^2 \kappa}
\int \dd x \left[
\rho(x,T)^2 - 
\rho(x,0)^2
\right].
\end{align}
We have written the boundary term $\mathcal{H}_B$ in terms of ${\rho = \partial_x h}$ in order to emphasize that it is an integral of a local observable.
Since it is localized at the temporal boundaries, it cannot affect the bulk RG flow. 
It is  the boundary term that ``remembers'' the direction of time in Eq.~\ref{eq:noisydiffusion} 
 ($D\to -D$ changes the sign only of $\mathcal{H}_B$).

\subsection{Lattice and continuum operators}

In this Appendix we review a standard construction from the sine-Gordon description of 1D quantum fluids  \cite{haldane1981effective}.
Define a microscopic version of the counting field, $h_\text{micro}(x,t)$, using the microscopic density:
\be
h_\text{micro}(x,t) = 2\pi \int^{x+0^+}_{0^-} n_\text{micro}(x,t) .
\ee
In continuous space, the density (at a given time)  is a sum of delta functions, and 
$h_\text{micro}$ jumps by $2\pi$ every time that a particle is crossed.
In a 1D lattice model,  $h_\text{micro}$ is naturally assigned to bonds of the lattice, and again jumps by $2\pi$ every time a particle is crossed. 
$h_\text{micro}$ will be coarse-grained  to obtain a continuum field $\widetilde h$.
Finally, the field that appears in the continuum action is obtained by subtracting the uniform part of the ``tilt'' in $\widetilde h$, which is proportional to the background density $n_0$:
\be\label{eq:hshift}
h(x,t) = \widetilde h(x,t) - 2\pi n_0  x.
\ee

Note that the introduction or removal of particles very far away from $x$ (to the left)
can change $h_\text{micro}$, and therefore $\widetilde h$, by multiples of $2\pi$.
This implies that when local microscopic observables are written in terms of $h$, the expressions must be  invariant under $\widetilde h\to \widetilde h \pm 2\pi$.
In addition, parity symmetry (spatial reflection) acts as 
\ba
x& \to L-x, 
& 
n_\text{micro} & \to n_\text{micro},
& 
\widetilde h &\to 2\pi N - \widetilde h.
\end{align}
Note that $\nabla \widetilde h$ and
$\cos( \widetilde h)$ are invariant under $\widetilde h\to \widetilde h \pm 2\pi$ 
and are even under parity, like $n_\text{micro}$.

While $n_\text{micro}$ can be written simply as\footnote{If we are considering a lattice model, then $\nabla$ is the lattice derivative and we take the lattice spacing $a=1$ for now.} ${n_\text{micro}=\f{1}{2\pi}\nabla h_\text{micro}}$, the right hand side is not simply equal to ${\f{1}{2\pi} \nabla \widetilde h}$.
In general, determining an exact relation between correlators of ${\f{1}{2\pi}\nabla h_\text{micro}}$ and those of the coarse-grained $\widetilde h$ would require a nontrivial renormalization group calculation starting at the lattice scale,\footnote{In momentum shell RG, the continuum field $\widetilde h$ is determined by a linear operation on the lattice field $h_\text{latt}$, but note that this does not mean that correlators of $h_\text{latt}$ can be 
rewritten as correlators of $\widetilde h$ by a simple linear substitution. The nontrivial integral over fast modes renormalizes the operator insertions.} or  matching against exact results for correlators. 
Haldane instead gave a heuristic argument \cite{haldane1981effective} determining the leading terms which we now attempt to summarize.

$h_\text{micro}$ is a sequence of steps between values in $2\pi \mathbb{Z}$. Imagine that  $\widetilde h$ has been obtained by performing a smoothing operation on an intermediate lengthscale 
in such a way that ${\widetilde h\in 2\pi (\mathbb{Z}+1/2)}$ at the locations of the particles $x_j$ (where $j$ indexes particles). 
Then the density may be written (we consider the spatial continuum)  
\ba
n_\text{micro}(x) & = \sum_j \delta(x-x_j) 
\\
& = 
(\partial_x \widetilde h) 
\sum_{k=-\infty}^\infty
\delta\lf  \widetilde h - (2k+1)\pi\ri 
\\
& = (\partial_x \widetilde h) 
\sum_{m=-\infty}^\infty e^{i m (\widetilde h-\pi)}.
\end{align}
This argument does not fix coefficients, but it suggests that in the IR we should have (keeping the most relevant terms)
\be
n_\text{micro}(x,t) = \f{1}{2\pi} \partial_x \widetilde h(x,t) + B \cos  \widetilde h(x,t) +  \ldots,
\ee
where $B$ is a nonuniversal constant. We may argue (by considering the integral of $n_\text{micro}$, see below) that the coefficient of the first term is the naive one.

Making the shift in Eq.~\ref{eq:hshift}, 
\be
n_\text{micro} \simeq n_0 + \f{1}{2\pi} \partial_x  h + B \cos  \big(  h + 2\pi n_0 x \big).
\ee
The coefficient $\f{1}{2\pi}$ is fixed by considering adding $\delta n_0 \times L$ additional particles to the system, so that the integral of ${n_\text{micro}}$
is increased by 
$\delta n_0 L$. Therefore 
$\f{1}{2\pi}(h_\text{micro}(L)-h_\text{micro}(0))$ is increased by the same amount. Finally,  we can perform coarse-graining in such a way that this also matches the increase of $\f{1}{2\pi}(h(L)-h(0))$.
This fixes the coefficient.

\section{Solvable model of level-line measurement}
\label{sec:looporientationexample}

Consider a standard model of a discrete lattice height field, $h\in {2\pi} \mathbb{Z}$, that is defined on the sites of the triangular lattice, with the constraint that adjacent heights are equal, or differ by $\pm 2\pi$ in which case an energy cost is incurred. If this energy cost is not too large, then
in the mid-IR this model is  described \cite{nienhuis1982exact} by a sine-Gordon model 
\be
{\cal H} = \f{K}{2} \int (\nabla h)^2 - g \int \cos h,
\ee
where the $g$ term is the remnant of the discreteness of the microscopic field.
In the deep IR, the model becomes a free field with  $g=0$ and $K=\f{1}{8\pi}$.\footnote{With appropriate boundary conditions \cite{PasquierSaleur1990}, the partition function can be mapped to the partition function of a model with an $\mathrm{SU}(2)$ symmetry. This symmetry is one way to understand why the IR value of $K$ is forced to be equal to $1/8\pi$.}

In this model we may naturally define level lines on the dual honeycomb lattice.
These are nonintersecting loops that separate differing values of $h$. (A given level line is thus naturally associated with a field value in ${2\pi [\mathbb{Z}+1/2]}$.)
The measured operator lives on a bond $\<ij\>$ and can be taken to be
\be
\measO = 1 - \delta_{h_i, h_j},
\ee
detecting the presence of a level line.

For \text{weak} measurement,  $\measO$ may be decomposed into scaling operators, and the most relevant ones that appear 
($\cos h$, and the components of the tensor $\nabla_\mu h \nabla_\nu h$) have dimension $2$. Therefore weak (highly noisy) measurement of the level lines is irrelevant for the above value of $K$: all the replicas are independent in the IR.

The limit of strong measurement is more interesting. 
In this limit, all the replicas share the same level-line configuration.
But they can differ in the \textit{orientation} of the level lines.

As is well known \cite{nienhuis1982exact}, a configuration of the height field $ h$ maps (modulo to a global shift $ h\to h+\text{const}$) to a configuration of oriented loops (level lines) on the dual (honeycomb) lattice.
For an appropriate choice of boundary conditions,  the partition function of a single replica is 
\be\label{eq:Zloopnequals2}
Z = \sum_{C} 2^{\text{\#loops}} x^{\text{loop length}}.
\ee
Here $C$ is a configuration of unoriented loops, and the factors of 2 are from summing over orientations.
The weight $x$ is determined by the energy cost for height differences. For $x\geq 1/\sqrt{2}$  the model is in the massless phase \cite{nienhuis1982exact}.

Now consider $N$ replicas with perfect measurement. The measurements determine $C$, but they do not determine the loop orientations. Therefore
\be
Z_N = \sum_{C} (2^N)^{\text{\#loops}} (x^N)^{\text{loop length}}.
\ee
This is the partition function for a loop model with a modified loop fugacity $n=2^N$.

For large enough $x$, this model is critical for $n \le 2$, but not for $n>2$.
The central charge, as a function of $N = \log_2 n$, for $N\leq 1$, is \cite{DifrancescoSaleurZuber1987}
\ba
c(N) & = 1- 6 \f{(1-g(N))^2}{g(N)},
& 
g(N) & = \f{1}{\pi} {\arccos}( - 2^{N-1}).
\end{align}
How should we interpret the effective central charge (cf. Eq.~\ref{eq:ceffectivemeasurement}),
\be\label{eq:ceffrepeat}
c_\text{eff} = c'(1),
\ee
in this example? 
This is not immediately obvious, because the replicated theory flows to a nontrivial CFT for $N\leq 1$, but flows to a trivial fixed point (with only short loops) for $N>1$.
This means that $c(N)$ is not even continuous at $N=1$.
However, it has a left derivative there, and we conjecture that this is the correct definition to use in Eq.~\ref{eq:ceffrepeat}.\footnote{The derivation of the entropy in Sec.~\ref{sec:entropies} makes clear that the correct order of limits is to take the derivative in $N$ first, and then the thermodynamic limit. 
Since it is written in terms of the derivative of $c(N)$ (which is a quantity defined through the thermodynamic limit),  
Eq.~\ref{eq:ceffrepeat} assumes these limits can be commuted.
When $c(N)$ is singular this may not be guaranteed. In principle, we should consider the RG flow of a finite-size estimate, $\partial_N c(N,L)|_{N=1}$. We expect that here this converges to the one-sided derivative of $c(N)$.}
This yields:
\be
c_\text{eff} = \f{12 \ln 2}{\pi^2}
\simeq 1 - 0.157.
\ee

The fact that the replicated theory is nontrivial can be seen more directly from correlation functions.

First consider the height difference 
${ h(x) -  h(0)}$ between two points. 
In fact measurements give us only subleading information about this quantity, but they give us more information about other quantities.

For a single replica we have as usual 
\be\label{eq:2point8}
\< ( h(x) -  h(0) )^2 \> \sim
 \f{1}{\pi K} 
\ln |x| 
 \sim
8 
\ln |x|.
\ee
By symmetry ${\<  h(x) -  h(0)\>_M}$ vanishes even after conditioning on measurements, but  we can ask whether measurement information improves our estimate of ${[ h(x) -  h(0)]^2}$.
Let 
\be
V_{0,x}(M) \equiv 
\< ( h(x) -  h(0) )^2 \>_M
\ee
be the estimate of this quantity for a given set of measurement outcomes.
We will show that 
$V_{0,x}(M)$ differs from the unconditioned estimate (\ref{eq:2point8}) 
only by a subleading (though universal) term.

Note that the value of $ h(x) -  h(0)$ is given by the signed number of oriented level lines that are traversed in connecting $0$ and $x$ 
(see Ref.~\cite{cao2021level} for an application of this fact to the percolation of level sets).
We can write this as 
\be
 h(x) -  h(0) 
=
2\pi \sum_{i = 1}^{\mathcal{N}_{0,x}^C} \chi_i,
\ee
where $\mathcal{N}_{0,x}^C$ is the number of unoriented level lines separating $0$ and $x$, and $\chi_i =\pm 1$ distinguishes the two orientations of the level lines.
After conditioning on $C$ (i.e. on the measurements),
the orientations  $\chi_i$ are uniformly random, so that
\be
V_{0,x}(M)
=(2\pi)^2 \mathcal{N}_{0,x}^C.
\ee
The distribution of $\mathcal{N}_{0,x}^C$ is in principle accessible using Coulomb gas 
 (or rigorous \cite{miller2016extreme})
techniques. More simplistically, a standard coarse-graining argument indicates that 
$\mathcal{N}_{0,x}^C$ is asymptotically Gaussian, with a mean $\< \mathcal{N}_{0,x}^C\>=\f{2}{\pi^2}\ln|x|$ that is fixed by (\ref{eq:2point8}), and a variance of order $\ln|x|$.
Therefore 
\ba
\operatorname{var} V_{0,x}(M)
\sim 
\ln |x|.
\end{align}
In other words, conditioning on measurement outcomes only changes the two-point function by a subleading amount:
\be
\<( h(0)- h(x))^2\>_M 
= 
\<( h(0)- h(x))^2\>
+ \Delta V_{0,x}(M),
\ee
where $\Delta V_{0,x}(M)$ is a Gaussian random variable whose order of magnitude is $\sqrt{\ln |x|}$, and which is therefore much smaller than the leading term $\<( h(0)- h(x))^2\>\simeq 8 \ln |x|$.

Next consider correlations of $e^{i\alpha  h}$, which are more strongly affected by conditioning.
In our model with a discrete height field we cannot distinguish $e^{i\alpha  h}$ from $e^{i(\alpha+1)  h}$, so we restrict to $\alpha \in [-1/2,1/2]$.
In terms of the level lines,
\ba
\< e^{i\alpha  h(0)}
e^{-i\alpha  h(x)}\>_M
& = 
\< \exp \lf 
2\pi i \alpha \sum_i^{\mathcal{N}_{0,x}^C} \chi_i
\ri \>
\\
& = 
\lf \cos (2\pi \alpha)\ri^{\mathcal{N}_{0,x}^C}.
\end{align}
Let us consider the typical absolute value\footnote{For $|\alpha|>1/4$ it is possible for the conditioned correlator to be negative, hence the use of the absolute value.}  of the correlator, obtained by averaging the logarithm of $|\<\ldots\>_M|$ over measurement outcomes, and then re-exponentiating. (Recall that the average over measurement outcomes is equivalent to the average over $\mathcal{N}_{0,x}^C$ in the loop ensemble, Eq.~\ref{eq:Zloopnequals2}.)
\ba
\left|\< e^{i\alpha  h(0)}
e^{-i\alpha  h(x)}\>_M \right|_\text{typ}
& = 
|\cos(2\pi \alpha)|^{ \< \mathcal{N}_{0,x}^C\> }.
\end{align}
This gives the typical value 
quoted in the main text (Eq.~\ref{eq:typicalcorrelatorlevellinestrongmeasurement}), with a scaling dimension
${x_\text{typ}(\alpha)=-\f{1}{\pi^2} \ln \cos |2\pi \alpha|}$, to be compared with the exponent  ${x_\text{av}(\alpha)=2\alpha^2}$ governing the standard expectation  value 
${\< e^{i\alpha  h(0)}
e^{-i\alpha  h(x)}\>}$.
The Taylor development of the typical exponent  reads $x_\text{typ}(\alpha)= 2\alpha^2(1+\tfrac{2\pi^2}{3} \alpha^2 + \ldots)$, so
the two exponents agree to order $\alpha^2$, but the typical correlator decays faster than the mean correlator.
This is because the standard correlation function 
${\< e^{i\alpha  h(0)}
e^{-i\alpha  h(x)}\>}
=
{{\mathbb{E}_M\< e^{i\alpha  h(0)}
e^{-i\alpha  h(x)}\>_M}}$
is dominated by 
rare loop configurations
(i.e.\ rare measurement outcomes $M$)
with anomalously small $\mathcal{N}^C_{0,x}$.

\section{Note on symmetry in classical vs. quantum models}
\label{app:symmetrynote}

The following comment is a digression from our main topic --- 
the aim is to point out that specifying the symmetry structure of a classical system requires  more information than specifying that of a quantum system.

We use the concrete example of the the gauge-Higgs theory at $K=\infty$ in 2D (Sec.~\ref{sec:Z2gtreview}), but the basic point is more general.

The gauge-Higgs theory at $K=\infty$ 
has a  $(d-2)$-form $\mathbb{Z}_2$ symmetry, 
which loosely speaking means that there are nontrivial topological  operators $V_P$ supported on closed paths $P$ (here they are Wilson loops), with $V_P^2=1$ \cite{gaiotto2015generalized}.
In 2D this is a 0-form symmetry.

In the usual field theory terminology, a 0-form symmetry is equivalent to a conventional global symmetry.
In the present context this is most easily understood in the transfer matrix language, i.e.\ via the classical-to-quantum mapping \cite{cardy1996scaling}.

In this mapping the 2D gauge theory at $K=\infty$ is 
mapped  to a  one-dimensional quantum  gauge theory \cite{kogut1979introduction}.
But after a local change of basis in the Hilbert space,  
we find that this spin chain is equivalent to a conventional \textit{non}-gauged Ising model with quantum Hamiltonian $\hat H_\text{QM}^\text{Ising}$, with order parameter $\hat \tau^3_j$ 
(where $\hat\tau^3_j$ is the third Pauli matrix at site $j$ of the new model).
(Note that this  Ising model should not be confused with the Ising model discussed in Sec.~\ref{sec:nishimoricloserelatives},~\ref{sec:Z2gtreview}. The two are related by Kramers-Wanner duality.)

Before the basis rotation, a Wilson line in the quantum gauge theory, which runs along all the bonds of the chain, gives a 0-form symmetry operator. 
After the basis change this becomes  the operator $\prod_j \hat \tau^1_j$ implementing the conventional  symmetry transformation of $\hat H_\text{QM}^\text{Ising}$, i.e. a flip of all the quantum spins to the symmetry-related state.  
Note that this is an off-diagonal operator in the basis where the Ising order parameter $\hat\tau^3_j$ is diagonal.

In the 1D quantum setting (or if we are doing formal quantum field theory) we would say that the $K=\infty$ gauge theory and the Ising model  are equivalent systems viewed in different bases.

Returning to the classical case, however, it is clear that the gauge theory and the Ising model should no longer be viewed as equivalent. This is  because a classical statistical mechanics problem comes with a preferred choice of basis.

In the classical gauge theory, the symmetry operator  associated with the  symmetry is a standard classical observable: that is, the Wilson line is 
a function of the gauge-invariant classical degrees of freedom $\sigma_{xy}S_xS_y$. 
Correspondingly, it is a diagonal operator in the quantum formulation if we use the natural ``classical'' basis.

In the classical Ising model, the symmetry operator is not a diagonal operator. This is apparent from the fact that, in the quantum formulation, it is an off-diagonal operator when we use the $\hat \tau^3$ basis mentioned above.
Instead, it is a ``defect line'' in the language of classical statistical mechanics.

The above gauge theory example falls into a larger class of examples involving classical models with local constraints, including e.g.\ classical dimer models with full-packing constraints.

Yet another example is the   3D XY model. As a quantum field theory, this is equivalent by duality \cite{wen2004quantum} to a model of a U(1) gauge theory interacting with a complex Higgs field, and we would say that both models have U(1) global symmetry. As classical problems these are distinct. The first is a model of spins that are acted on by U(1) transformations and the latter is essentially a model of  fluxes that satisfy a conservation law.

These points are straightforward, but they indicate that the classification of phases by symmetry (and higher symmetry) in classical systems should take into account the  preferred choice of local basis that the classical statistical mechanics problem comes equipped with. This could be investigated further.

These considerations are relevant to classical inference problems, 
because the measurements are necessarily of classical observables
(corresponding to diagonal operators in the classical basis).

\section{Dual representation of 3D replica gauge theory}
\label{app:gaugedual}

Here we relate the pure gauge theory in Eq.~\ref{eq:Hreplicapuregauge} to a dual order-parameter theory.

We write the partition function of the gauge theory as 
\be\label{eq:Hreplicapuregaugeexpanded}
Z \propto \sum_{\{\sigma^a\}} \prod_\square \prod_{a<b}
\lf 1 + x \, (\sigma\sigma\sigma\sigma)_\square^{ab}\ri
\ee
for $x=\tanh 2\lambda_\square$,
and expand the product over plaquettes.
For each plaquette, and for each pair of replicas $(ab)$, we must choose either the 1 term or the $x$ term.
We visualize the latter as an ``occupied'' plaquette of type~$(ab)$.

The sum over $\{\sigma^a_{xy}\}$ on the links ensures that, for each replica index $a$, and for each link, 
an even number of occupied plaquettes that involve index $a$ meet at the link. 
This ensures that
(neglecting boundary condition effects)
we may map the plaquette configurations to 
domain wall configurations for an order parameter 
\be\label{eq:dualspinTconstraint}
(T^1,\ldots, T^N)\in \{+1,-1\}^N
\quad \text{with} \quad  \prod_{a=1}^N T^a=1.
\ee
The dual spin $\vec T$ lives at the centers of lattice cubes (i.e. on sites of the dual lattice). To illustrate the basic idea, let us assume that $x$ is small, so occupied plaquettes are rare, and neglect doubly-occupied plaquettes.
Then a single occupied plaquette of type $(ab)$
is interpreted as a domain wall where spin components $T^a$ and $T^b$ change sign. 
The cost $x$ for such a domain wall is reproduced by an energy $\widetilde J \vec T_i \cdot \vec T_j$ for the corresponding bond $\<ij\>$ of the dual lattice, with $x=e^{-4\widetilde J}$.

The full dual Hamiltonian (beyond this small-$x$ approximation) is more complicated,\footnote{On each plaquette 
we must sum up configurations of occupied plaquettes that correspond to the same domain wall configuration. For example an occupied plaquette of type $(12)$ and a pair of occupied plaquettes of types $(13)$ \& $(23)$ correspond to equivalent domain wall plaquettes.}
but the above is sufficient to see the global symmetry of the dual theory, which allows replica permutations and  $\mathbb{Z}_2^{N-1}$ transformations  (note that if
we take an element of $\mathbb{Z}_2^{N-1}$  to assign signs to the first $N-1$ components of $\vec T$, the 
the sign  of the last component  is fixed by Eq.~\ref{eq:dualspinTconstraint}).

The model in Eq.~\ref{eq:Hreplicapuregauge} and Eq.~\ref{eq:Hreplicapuregaugeexpanded} can be viewed as arising from measuring the occupied link density  for a configuration of closed loops.
Now we discuss another measurement problem which we expect to give a transition with the same exponents, and where the effective Hamiltonian can be derived more simply.

In this problem the physical degrees of freedom are a collection of occupied bonds of the cubic lattice. 
Heuristically we think of this as defining a collection of --- possibly open --- strings. The relation between this problem and the previous one is therefore analogous to the relation between the 2D problems in the central and right panels of Fig.~\ref{fig:N3types2D}.

Each bond $\<xy\>$ is occupied with some probability $p$; we define $\widetilde J$ by $\tanh \widetilde J=p/(1-p)$. Such an ensemble of occupied bonds arises from the ``high-temperature'' expansion \cite{cardy1996scaling} of an  Ising model in the limit of an infinitely strong magnetic field,\footnote{We expect the universal behavior would be the same in a finite field (the latter would correspond to introducing an additional probability cost for string endpoints).} which can be written
\be
Z = \sum_{\{T\}} \prod_{\<xy\>} \lf 1 + (\tanh \widetilde J) \, T_x T_y\ri
\prod_x \lf 1  + T_x\ri.
\ee
As usual, the expansion of the product over $\<xy\>$ generates diagrams made up of occupied links.
These diagrams may be referred to as 1-chains or more colloquially as string diagrams.
A given string diagram determines a collection of string endpoints (endpoints are  sites adjacent to an odd number of occupied links, sites in the boundary of the 1-chain).

Now we   measure the positions of the endpoints with (for simplicity) perfect precision.
In the replica description, we obtain a partition function for $N$ replicas that are forced to have the same endpoints. 
The resulting ensemble matches the high-temperature expansion of
\be\label{eq:TreplicahighT}
Z = \sum_{\{T^a\}} \prod_{\<xy\>}\prod_a \lf 1 + (\tanh \widetilde J) \, T_x^a T_y^a \ri
\prod_x \lf 1  + T_x^1 \cdots T_x^N \ri.
\ee
Equivalently,
\be\label{eq:TconstrainedH}
Z \propto \sum'_{\{\vec T\}}\, \exp \bigg( \widetilde  J \sum_a \sum_{\<xy\>} \vec T_x \cdot \vec T_y \bigg),
\ee
where the prime on the sum indicates that for each site we sum over a vector $\vec T$ in $\{+1,-1\}^N$ obeying the constraint in Eq.~\ref{eq:dualspinTconstraint}
(which is enforced by the final factor in Eq.~\ref{eq:TreplicahighT}).

Note that, as in the first model in this appendix, the spins $\vec T$ are ``dual'' degrees of freedom used to express the partition function and are not local physical observables. 
(We could also obtain Eq.~\ref{eq:TconstrainedH} by starting with a gauge theory representation of the loops and performing duality.)

The replica Hamiltonian (\ref{eq:TconstrainedH}) is slightly different from that of the previous example, and the physical parameter driving the transition is also different
(it is the string ``tension'' in the pre-measurement ensemble, rather than a measurement strength). However the natural expectation is that there is a transition in the same universality class in both models.

\section{Strong measurement regime for percolation}
\label{app:geometricalstrongmeasurement}

In Sec.~\ref{sec:imagingpolymers} we discussed measurement of a single polymer. 
In this Appendix we consider a related problem where we measure the local occupancy in 2D percolation.
The physical conclusion here is almost trivial and easily guessed without field theory: unless measurement is perfect, it reveals essentially nothing about percolation connectivities at large scales.
However,  it is interesting to see how this is interpreted in field theory. 
Starting with a weak perturbation of a nonlinear sigma model,
we end up in the IR with $N$ independent nonlinear sigma models.

To be concrete, consider two-dimensional bond percolation. 
There is a well-known mapping of configurations to configurations of completely packed loops on the ``medial'' lattice \cite{BaxterKellandWu1976}. These loops are (modulo some small loops) just the percolation cluster boundaries, 
so local measurement of the bond occupancies in the percolation language is equivalent to measurement of the local geometry of the loops.
Similarly, knowing the large-scale geometry of the clusters is equivalent to knowing the large-scale geometry of the loops.
We focus first on the loops since this connects to the discussion in Sec.~\ref{sec:imagingpolymers}.

We would expect that any finite measurement strength flows to zero. Heuristically, this is because (for example) a spanning percolation cluster in a system of size $L$ has $O(L^{3/4})$  ``red bonds'' --- breaking any one of these bonds makes the cluster non-spanning \cite{coniglio1982cluster}. Therefore once the probability of a measurement error reaches $O(L^{-3/4})$ we lose the ability to accurately determine whether the configuration has a spanning cluster.

We focus on the limit of almost perfect measurement, and relate it to field theory.

First consider a single replica.
A loop may be viewed \cite{candu2010universality,nahum2012universal,nahum2013loop,nahum2014critical}  as a ``worldline'' of a complex field $z_\alpha$, with $\alpha=1,\ldots, m$ and $|z|^2=1$.
Here $\alpha$ is a fictitious flavor index for the loops,
introduced to allow correlators to be expressed, 
and the relevant limit is $m\to 1$.
(If we consider $m\neq 1$, then there is an additional weight $m^\text{no. loops}$ in the partition function; by a standard mapping \cite{BaxterKellandWu1976}, this corresponds to boundaries of FK clusters for  $Q$-state Potts with $Q=m^2$.) 
The field theory for $z_\alpha$ is the $CP^{m-1}$ nonlinear sigma model,
\be\label{eq:CPmm1}
\mathcal{H} = \f{1}{2g} |(\partial - i a)z_\alpha|^2 
+ \f{i\theta}{2\pi}  \epsilon_{\mu \nu} \partial_\mu a_\nu,
\ee
where $a_\mu=\f{i}{2}((\partial_\mu z)^\dag z - z^\dag \partial_\mu z)$, and $\theta=\pi$ if the percolation model is critical. (The sum on $\alpha$ is implied.)

In order to consider measurements we  introduce replicas and the  additional  limit $N\to 1$.
When measurements are perfect, the loops in all replicas are locked. 
As in Sec.~\ref{sec:imagingpolymers}, this means that we go back to the theory  of a single loop, except that now the loop carries not one flavor index but $N$ of them. The corresponding field also carries multiple indices:
\be\label{eq:CPMm1}
\mathcal{H} = \f{1}{2g} |(\partial - i A)Z_{\alpha_1,\ldots,\alpha_N}|^2 
+ \f{i\theta}{2\pi}  \epsilon_{\mu \nu} \partial_\mu A_\nu.
\ee
This is the $CP^{\mathcal{M}-1}$ model, with $\mathcal{M}=m^N$.
Note that the relevant limit is again $\mathcal{M}\to 1$  (just as for the theory in Eq.~\ref{eq:CPmm1} it was $m\to 1$) because taking $m\to 1$ and $N\to 1$ gives $\mathcal{M}\to 1$.

Now we slightly weaken the measurements. By a logic similar to that in Sec.~\ref{sec:imagingpolymers} this induces terms such as 
\be
Z^*_{\alpha_1 \alpha_2\ldots \alpha_N}
Z^*_{\beta_1 \beta_2\ldots \beta_N}
Z_{\beta_1 \alpha_2\ldots \alpha_N}
Z_{\alpha_1 \beta_2\ldots \beta_N} + \ldots,
\ee
corresponding to locations where loops in the first replica reconnect differently to those in other replicas, as well as similar terms with different numbers of differently-reconnecting replicas.
These terms are RG-relevant, with RG eigenvalue $y_4 =3/4$ 
(they are ``four-leg'' operators \cite{saleur1986new,saleur1987exact}).
In principle we could compute the coefficients of the corresponding perturbations explicitly in a lattice version of the field theory \cite{nahum2014critical} that maps to the original percolation model but we do not do this here.

Instead we conjecture that the universal effect of these terms, beyond the lengthscale where they become significant,
 can be captured by pretending that they simply impose a potential which restricts $Z$ to a submanifold of $CP^{\mathcal{M}-1}$, of the form ${CP^{m-1}\times \cdots CP^{m-1}}$,\footnote{We expect that a somewhat analogous collapse of complex projective space to a submanifold, induced by quartic terms, gives the RG flow from percolation, viewed as a fine-tuned limit of the integer quantum Hall transition, to the generic universality class of the latter transition \cite{ANunpublished2013}.} parameterized as
\be
Z_{\alpha_1 \alpha_2\ldots \alpha_N}
=z^{1}_{\alpha_1}
z^{2}_{\alpha_2}
\cdots
z^{N}_{\alpha_N},
\ee
where each of the $z^{a}$ is a unit vector. 
Substituting this into Eq.~\ref{eq:CPMm1} we find that it becomes a sum of $N$ $CP^{m-1}$ model Hamiltonians, one for each of the $z^{a}$.
This is the zero-measurement fixed point, where the replicas are independent, and each one is described by a theory of the form (\ref{eq:CPmm1}).

If we consider $m>1$ instead of $m\to 1$ then the above corresponds to measurement of FK clusters, as discussed in App.~\ref{app:FKdetails} (and the main text),
for the Potts model with $Q=m^2$ states.
For $Q<2$, i.e.\ $m<\sqrt{2}$, weak measurement is RG-irrelevant, and the phenomenology above carries over.

We can also phrase the RG flow in terms of the  Landau theory for the Potts model, as discussed in 
App.~\ref{app:FKdetails}.
In the strong measurement limit we work with  a field $\Phi_{\sigma_1, \ldots, \sigma_N}$ with 
$\sum_{\sigma_1,\ldots,\sigma_N}\Phi_{\sigma_1, \ldots, \sigma_N}=0$
(App.~\ref{app:FKdetails}).
For $Q<2$, we can think of this field as fragmenting, during  the RG flow to weak measurement, into $N$ fields $\phi^a_\alpha$
(with $\sum_{\sigma}\phi_{\sigma}^a=0$),
one for each replica $a$, via
\be
\Phi_{\sigma_1, \ldots, \sigma_N} 
= \phi^1_{\sigma_1}+\ldots+\phi^N_{\sigma_N}.
\ee

\bibliography{refs}

\end{document}